\documentclass[12pt,epsf]{article}

\usepackage{amsmath}
\usepackage{amssymb}
\usepackage{amsthm}
\usepackage{epsfig}
\usepackage{color}

\newcommand{\prt}{\partial}

\def\ni{{\noindent}}

\def\be{\begin{equation}}
\def\ee{\end{equation}}

\usepackage{color}
\usepackage{epstopdf}

\theoremstyle{definition}

\begin{document}
\bibliographystyle{plain}

\title{{\Large\bf A universal asymptotic regime in the hyperbolic nonlinear Schr\"odinger equation}}
\author{Mark~J.~Ablowitz, Yi-Ping~Ma, Igor~Rumanov\footnote{corresponding author; e-mail: igor.rumanov@colorado.edu} \\
{\small Department of Applied Mathematics, University of Colorado, Boulder, CO 80309 USA} }

\maketitle

\bigskip

\begin{abstract}
The appearance of a fundamental  long-time asymptotic regime in the two space one time dimensional hyperbolic nonlinear Schr\"odinger (HNLS) equation is discussed. Based on analytical and extensive numerical simulations an approximate self-similar solution is found for a wide range of initial conditions -- essentially for  initial lumps of small to moderate energy. Even relatively large initial amplitudes, which imply strong nonlinear effects, eventually lead to local structures  resembling those of the self-similar solution,
with appropriate small  modifications. These modifications are important in order to properly capture the behavior of the phase of the solution.
This solution has aspects that suggest it is a universal attractor emanating from wide ranges of initial data.

\end{abstract}

\newpage

\section{Introduction}

The (2+1)-dimensional hyperbolic nonlinear Schr\"odinger (HNLS) equation,

$$
i\prt_Z\Phi + (\prt_{xx}-\prt_{yy})\Phi + |\Phi|^2\Phi = 0,   \eqno(1.1)
$$

\ni describes many types of physical phenomena including electromagnetic pulse propagation in optical waveguides, e.g.~plasmons propagating along a flat dielectric/metal interface~\cite{AbBu14},  cyclotron waves in plasmas~\cite{PerSenBer78, MyrLiu80, SulSul99} and surface waves in  deep water ~\cite{AbSeg79, SulSul99, ZahKuz12}. Recently it has been suggested  that the HNLS equation may explain the existence of rogue waves~\cite{KharPel03, Rub15}. In the plasmon application the variable $Z$ corresponds to propagation distance, $x$ to the transverse spatial direction, $y$ to a retarded time and $\Phi$  to the complex amplitude of the electromagnetic field~\cite{AbBu14}. In water wave applications $Z$ typically is related to physical time shifted by the group velocity  while $x$ and $y$ correspond to the horizontal dimensions; see e.g.~\cite{AbSeg79}. For other applications $x$, $y$ and $Z$ (e.g.~plasma waves~\cite{SulSul99}) are related to the three physical spatial axes. It is important in all these applications to study the various long-time propagation regimes in HNLS equation.

\par An exact similarity solution of eq.~(1.1) was found in~\cite{SKJB78} and independently in~\cite{AbSeg79}. It has the form

$$
\Phi = \frac{\Lambda_0}{Z}\exp\left(i\left[s + \theta_0 - \frac{\Lambda_0^2}{Z}\right]\right),  \qquad  s = \frac{x^2-y^2}{4Z},   \eqno(1.2)
$$

\ni where $s$ defined above is the similarity variable.
\par The main result of this paper is to show that the solution eq.~(1.2),  with a remarkably small modification, appears at long times (i.e. large $Z$) in the central zone of the $(x,y)$-plane for a wide range of initial conditions. This is basically the case for all initial conditions such that some lump of energy is initially present in the neighborhood of the origin in the $(x,y)$-plane; hence we term this solution as universal. The form of the modified similarity solution is found to be

$$
\Phi = \frac{\Lambda_0}{Z}\exp\left(i\left[s + \theta_0 + \frac{ \eta_0-\Lambda_0^2}{Z}\right]\right),   \eqno(1.2A)
$$

\ni where $\theta_0, \Lambda_0, \eta_0$ depend on the initial conditions. This is more fully discussed in the sections below. We only remark that we do not consider either rapidly varying  or large initial data; this is consistent with the derivation of the HNLS from physical principles where all terms are of the same order.

\par Apart from the linear and nonlinear stability analysis cf.~\cite{ZahRub73, RasRyp86, Ber98, Pel01, PelEtAl14} (and also numerous references therein) which are usually  dedicated to the stability  of exact one-dimensional solutions of the 1D-NLS equation (e.g. solitons)  (and thus also exact solutions of the HNLS eq.),  there are relatively few analytical results  in the literature regarding  the general behavior of solutions of the  HNLS equation, though some results can be found in ~\cite{MyrLiu80, LitEtAl83, Ber94, GidSaut96, KuzBerRas96, BerEtAlChirp96, KeNaZe11, Rub15}.
\par Studies conducted in the 1970s-1990s are summarized in the reviews~\cite{RasRyp86, Ber98} and the book~\cite{SulSul99} where both elliptic NLS and HNLS in various dimensions are considered.

\par Since the seminal work~\cite{ZahRub73} it is known that one-dimensional solitons in multidimensional NLS equations are unstable with respect to transversal perturbations, see also e.g.~\cite{AbSeg81}.  Recently there has been more research regarding the types of instability and the growth rates of various  instabilities in the HNLS eq.; see e.g.~\cite{PelEtAl14, Pel01} and their associated experimental demonstration~\cite{GorEtAl09,  GorEtAl11, GorEtAl12}.

\par It is known that nonelliptic NLS equations do not admit localized traveling wave solutions~\cite{GidSaut96}. This is consistent with our observations that lumps of energy in the HNLS eq.~eventually disperse, and in turn, lead to the universal asymptotic solution described here. The underlying structure can consist of many localized hyperbolas. Sometimes we observe a number of such hyperbolic structures with different centers, partly superimposed. These may be preceded by more intricate structures at intermediate ``times" $Z$.

\par Apart from analyzing the development of instabilities of one-dimensional solutions~\cite{GorEtAl09, GorEtAl11, GorEtAl12}, there has been some numerical and experimental research on the HNLS equation~\cite{LitEtAl83, RyAg95, Pietr97, LitEtAl00, BerEtAl02, MoshEtAl09, KeNaZe11}. There has also been a number of studies of the (3+1)-D HNLS eq., see e.g.~\cite{RaSchGa96, Ber98, BerEtAl01, BerEtAl02, CouaironEtAl06} and references therein. The 3D case has attracted researchers due to a wide variety of applications ranging from short (femtosecond) laser beams in condensed media~\cite{RaSchGa96, BerEtAl01, ContiEtAl03, CouaironEtAl06, SkuBer06}, cyclotron waves in plasma~\cite{PerSenBer78, SKJB78, MyrLiu80, BerRasMultispl96, BerEtAl01, BerEtAl02} and high energy nonlinear electromagnetic phenomena~\cite{ShuklaEtAl04}.

\par Given its numerous applications (surface waves in deep water, optical pulses in planar waveguides etc.) and its fundamental role as an intrinsically simple (2+1)-dimensional nonlinear equation, the HNLS eq.~is a laboratory for novel types of phenomena and its behavior can shed light on related problems such as the more complicated (3+1)-D HNLS eq. Indeed many processes associated with the (3+1)-D HNLS eq. such as pulse splitting, multi-filamentation, fragmentation, so-called snake and neck instabilities and nonlinear X-waves, have qualitative counterparts in the  HNLS eq.~\cite{LitEtAl00, BerEtAl01, BerEtAl02, GorEtAl09, KeNaZe11, GorEtAl11, GorEtAl12}.

\par Interestingly, there is still controversy as to whether there exists a catastrophic collapse, or blowup with infinite singularity formation in the (3+1)-D HNLS eq.~\cite{NewellEtAl94, FiMaPa95, BerEtAl01, LitEtAl02, LitEtAl03, BerEtAl02}. In this respect, the situation with the HNLS eq.~is clearer. There is a simple convincing, though non-rigorous, argument regarding the absence of catastrophic collapse in this case, see e.g.~\cite{FiMaPa95}. The HNLS eq.~without the second-derivative defocusing term ($\prt_{yy}\Phi$ in eq.~(1.1)) is the integrable 1D NLS eq.~which is known to have no collapse (and possess multisoliton solutions). Adding the second-derivative term which causes defocusing in the transversal direction should only improve the situation, further dispersing the energy. There are some rigorous quantitative arguments~\cite{KuzBerRas96}, based on virial inequalities (i.e.~inequalities for second moments or variances and their $Z$-derivatives), which show that total collapse, i.e.~finite energy concentration on sets of measure zero (points or lines) is impossible for the (2+1)-D HNLS eq. This does not rigorously rule out a blowup singularity formation but, together with the qualitative argument above, makes it much less plausible.

\par Virial-based arguments like those in~\cite{KuzBerRas96} suggest that the HNLS eq.~favors structures stretched along the defocusing direction leading to hyperbolas in the $(x,y)$-plane. This is similar to X-wave phenomena cf.~\cite{ContiEtAl03, Conti04, KeNaZe11} which exhibit characteristic X-shapes in the $(x,y)$-plane formed by the lines $x=\pm y$ and characteristic hyperbolas asymptotic to these lines. While such existing exact solutions have infinite energy (i.e.~infinite $L^2$-norm $\int\int|\Phi|^2dxdy$), their finite-energy  counterparts have been observed in experiments on electromagnetic beam propagation; they appear in the central cores of the beams and split at sufficiently large propagation distance $Z$, see e.g.~\cite{ContiEtAl03, Conti04}. When the phenomenon is described by the HNLS equation~\cite{BerEtAl01}, the end result of their splitting must be the hyperbolic structure which we observe numerically and describe analytically below.

\par Recently some exact X-wave solutions of the HNLS eq.~and the (3+1)-D HNLS equation with an additional supporting potential have been found~\cite{EfSiChri09} but they still have infinite energy. Also, some infinite energy standing wave solutions were proven to exist in~\cite{ Lu15}.
The existence of certain types of bounded and continuous hyperbolically radial standing and self-similar waves was established in~\cite{KeNaZe11} where the asymptotics of such solutions at large hyperbolic radius, i.e. large $|x^2-y^2|$, are computed. The region we consider here is different: $x^2 \pm y^2 \lesssim 4Z$ i.e.~the center zone of the pulse. The  scenario, also confirmed  by our numerical studies is that initial localized lumps asymptotically tend to a similarity  solution valid in the above central zone and falling off sharply (exponentially) beyond it. Although the similarity solution has infinite energy, other small amplitude solutions, which can be obtained by WKB methods and matched to these similarity solutions, exist in regimes away from the core; cf.~e.g.~\cite{AbSeg81}.

\par A number of exact solutions based on symmetry reductions using Lie group invariance methods, have been constructed for both the NLS and HNLS eq.~\cite{Taj83, GagWin89, GagEtAl89, Gag90, Clark92, OzGun06}; this was also recently revisited in application to HNLS eq. in~\cite{GuoLin10}. We summarize some of this work in Appendix A. Some of these similarity reductions and exact solutions may be relevant at intermediate stages before reaching the long-time regime. They may allow for a better quantitative description of various phenomena like self-focusing, splitting, pattern formation. The descriptions available so far in literature~\cite{Ber94, BerRasMultispl96, Ber98, BerEtAl01} are only approximate and an improved understanding might be reached by obtaining more sophisticated exact solutions. This question deserves further, more systematic study and we plan to consider it elsewhere. We emphasize, however, that our current results are of importance for these questions since they allow one to select among the many complicated transient solutions those which are asymptotically close to the universal regime described here.
\par Of the  approximate methods applied to all types of NLS equations, variational methods, though not rigorous, have proven to be very popular. They were used to construct approximate solutions for both the HNLS~\cite{Ber94, Rub15} and the (3+1)-D HNLS equations ~\cite{BerRasMultispl96, BerEtAl01, BerEtAl02}. They were used in~\cite{Ber94, BerRasMultispl96} to quantitatively understand the self-focusing and pulse-splitting phenomena and in~\cite{Rub15} to investigate possible mechanisms of generating rogue waves in HNLS. While the range of validity of a variational ansatz remains to be rigorously established, the results of using the Gaussian ansatz of~\cite{Ber94, Rub15} for HNLS can be obtained from a usual approximate solution where the validity and the precision of the approximation are completely clarified, see Appendix B. Thus, the variational approximation can be useful e.g.~at long times in $Z$, and can be related to the universal solution eq.~(1.2A), e.g.~the amplitude there also may decay as the inverse of time or propagation distance $Z$. However, some limitations of this approach are exposed when we compare the phases in Appendix B.

\par It is well-known  that similarity solutions play a crucial role in the long time asymptotic solution of certain integrable nonlinear dispersive wave equations \cite{AbSeg81, AbCl91}. Equations which are not known to be integrable, such as the HNLS equation have been less intensively studied from this point of view. For example, the one dimensional integrable NLS equation

\[ iu_z+ u_{xx} + \sigma |u|^2u =0  \eqno(1.3) \]
has the similarity solution 
\[u(x,z)= \frac{A}{z^{1/2}} \exp (i \theta)  ~~\mbox{where} ~~ \theta= \frac{x^2}{4z}+\sigma A^2 \log z +\theta_0  ~~~\eqno(1.4) \]
In 1976 Manakov showed that for $\sigma =-1$ as $z \rightarrow \infty$ the solution tended to the above similarity solution in the central core region. Ablowitz and Segur showed how to include suitable perturbations and solitons (when $\sigma=+1$) cf.~\cite{AbSeg81}.

\par It was also shown that similarity solutions played key roles in the long time limit of other well-known integrable PDEs, e.g. the Korteweg-deVries (KdV) and modified KdV (mKdV) equations~\cite{AbSeg77,SeAb81}. In the case of the mKdV equation

\[u_t-u^2u_x+u_{xxx}=0 ~~~  \eqno(1.5) \]
with decaying initial data, it has the similarity reduction: $u(x,t)= w(\eta)/(3t^{1/3})$ where $w$ satisfies the 2nd Painlev\'e equation

\[ w''-\eta w-2 w^3=0  \eqno(1.6)  \]
Here the solution $u$ tends (up to the factor $3t^{1/3}$) to a solution of the Painlev\'e equation (1.6) (in the case of KdV, a related ODE) in the long time limit. Indeed, Ablowitz, Kruskal and Segur \cite{AKS79} showed that the decaying solution of mKdV equation had the following property. Corresponding to the boundary condition

\[ w(\eta) \sim r_0 Ai(\eta),  \mbox{as} ~~ \eta \rightarrow +\infty  \eqno(1.7) \]
where Ai($\eta$) is the well-known Airy function, there were three types of behavior as $\eta \rightarrow -\infty$.

\par i) For $|r_0|<1$ (subcritical),
\[ w(\eta) \sim d_0(-\eta)^{1/4}  \sin\left( \frac{2}{3}(-\eta)^{3/2}-\frac{3}{4}d_0^2\log(-\eta) +\theta_0\right)  \eqno(1.8)  \]
where $d_0^2= -\frac{1}{\pi}\log(1-r_0^2)$; the formula for $\theta_0=\theta_0(r_0)$ is more complicated; see \cite{SeAb81}.

\par ii) For $|r_0|=1$ (critical),
\[ w(\eta) \sim Sgn(r_0) \left( (-\eta/2)^{1/2}-\frac{(-\eta)^{-5/2}}{2^{7/2}}+ O((-\eta)^{-11/2}) \right)  \eqno(1.9) \]

\par iii) When $|r_0|>1$ (overcritical),

\[ w(\eta) \sim Sgn(r_0)  \left( \frac{1}{\eta-\eta_0} - \frac{\eta_0}{6}(\eta-\eta_0)+O((\eta-\eta_0)^2) \right)  \eqno(1.10) \]
where $\eta_0=\eta_0(r_0)$. Subsequently, Hastings and McLeod \cite{HM80} studied case (ii) in detail.

\par What is clear from the above is the important role similarity solutions play in long time evolution of nonlinear dispersive wave equations.

\par The plan of the paper is the following. In section 2 we compute the perturbative solution of the HNLS equation for the Gaussian lump initial condition keeping first order in nonlinearity and determine the form of corrections to its exact solution eq.~(1.2) which is relevant for long-time (large $Z$) asymptotics. Section 2 also shows how focusing and defocusing can be described for moderate amplitudes in the the HNLS equation. In section 3 we present the general large $Z$ asymptotics for both the linear and nonlinear equation, assuming in the last case that the solution falls off as $1/Z$ in the central region. Section 4 presents extensive numerical results demonstrating the appearance of the solution eq.~(1.2A) with the corrections discussed in sections 2 and 3. Section 5 is dedicated to the discussion of the results. In appendix A we present some exact similarity reductions of the HNLS equation which may be useful for understanding the complicated intermediate dynamics of the HNLS eq.~prior to the long-time regime. Appendix B shows what kind of approximation underlies  the variational approach subject to a Gaussian ansatz; cf.~\cite{Ber94, Rub15}, and how it relates to the solutions (1.2-1.2A).

\section{A perturbative calculation}

It is instructive to compute the first order corrections due to the nonlinearity to the exact solution of the linearized HNLS equation (see also \cite{AbBu14}) i.e.~to consider

$$
\Phi \approx \Phi_L + \Phi_n,   \eqno(2.1)
$$

\ni where $\Phi_L$ satisfies

$$
i\prt_Z\Phi_L + (\prt_{xx}-\prt_{yy})\Phi_L = 0,   \eqno(2.2)
$$

\ni and $\Phi_n$ is found as the first order perturbation to $\Phi_L$ from

$$
i\prt_Z\Phi_n + (\prt_{xx}-\prt_{yy})\Phi_n = -|\Phi_L|^2\Phi_L.   \eqno(2.3)
$$

\ni If one takes the Gaussian initial condition $\Phi(x,y, Z=0) = \Phi_L(x,y,Z=0) = A_0e^{-x^2-y^2}$, $A_0$ constant, then the exact solution of eq.~(2.2) is

$$
\Phi_L(x,y,Z) = \frac{A_0}{\sqrt{16Z^2+1}}e^{-\frac{x^2+y^2}{16Z^2+1}}\cdot e^{\frac{4iZ(x^2-y^2)}{16Z^2+1}} = A_L(x,y,Z)e^{i\theta_L(x,y,Z)}.   \eqno(2.4)
$$

\ni Using Fourier transform techniques, one obtains the exact formula for $\Phi_n$ from eq.~(2.3) and eq.~(2.4) with $\Phi_n(x,y,Z=0)=0$

$$
\Phi_n(x,y,Z) = \frac{iA_0^3}{4}\int_0^Z\frac{dZ'}{\sqrt{(16Z'^2+1)(16Z'^2+9)(R^2+J^2)}}\cdot e^{-\frac{R(x^2+y^2)}{4(R^2+J^2)}}\cdot e^{\frac{iJ(x^2-y^2)}{4(R^2+J^2)}},   \eqno(2.5)
$$

\ni where we denote

$$
R = R(Z') = \frac{3(16Z'^2+1)}{4(16Z'^2+9)},   \qquad     J = J(Z',Z) = Z - \frac{8Z'}{16Z'^2+9}.     \eqno(2.6)
$$

\par Next we consider the asymptotics of the exact perturbative expression eq.~(2.5) as $Z$ becomes large. It is then convenient to rewrite eq.~(2.5) as

$$
\Phi_n(x,y,Z) = I_1-I_2,
$$

\ni where

$$
I_1 = \frac{iA_0^3}{4}\int_0^\infty\frac{dZ'}{\sqrt{(16Z'^2+1)(16Z'^2+9)(R^2+J^2)}}\cdot e^{-\frac{R(x^2+y^2)}{4(R^2+J^2)}}\cdot e^{\frac{iJ(x^2-y^2)}{4(R^2+J^2)}}   \eqno(2.7)
$$

\ni and

$$
I_2 = \frac{iA_0^3}{16Z^2}\int_4^\infty\frac{du}{\sqrt{(u^2+1/Z^2)(u^2+9/Z^2)((R/Z)^2+(J/Z)^2)}}\cdot e^{-\frac{R(x^2+y^2)}{4(R^2+J^2)}}\cdot e^{\frac{iJ(x^2-y^2)}{4(R^2+J^2)}},  \eqno(2.8)
$$

\ni where $u=4Z'/Z$. In the first integral $I_1$, we change integration variable to

$$
\zeta = \frac{16Z'^2+1}{16Z'^2+9},   \qquad     4Z' = \sqrt\frac{9\zeta-1}{1-\zeta}.   \eqno(2.9)
$$

\ni Then $I_1$ can be rewritten as

$$
I_1 = \frac{iA_0^3e^{\frac{i(x^2-y^2)}{4Z}}}{32Z}\int_{1/9}^1\frac{d\zeta}{\sqrt{\zeta(1-\zeta)(9\zeta-1)}}\cdot\frac{e^{-\frac{3(x^2+y^2)\zeta}{16Z^2g(\zeta,Z)} + \frac{i(x^2-y^2)(u_1-u_2/Z)}{4Z^2g(\zeta,Z)}}}{\sqrt{g(\zeta,Z)}},   \eqno(2.10)
$$

$$
g(\zeta,Z) = 1 - \frac{2u_1}{Z} + \frac{u_2}{Z^2},  \qquad u_1 = u_1(\zeta) = \frac{\sqrt{(9\zeta-1)(1-\zeta)}}{4},  \qquad u_2 = u_1^2 + \frac{9\zeta^2}{16}.
$$

\ni The last formula is convenient to expand in inverse powers of $Z$. Restricting the consideration to the central zone $x^2+y^2 \lesssim 4Z$, we find

$$
I_1 = \frac{iA_0^3e^{\frac{i(x^2-y^2)}{4Z}}}{32Z}\int_{1/9}^1\frac{d\zeta}{\sqrt{\zeta(1-\zeta)(9\zeta-1)}}\cdot
$$

$$
\cdot\left(1 + \frac{u_1}{Z} - \frac{3(x^2+y^2)\zeta}{16Z^2} + \frac{i(x^2-y^2)u_1}{4Z^2} + \frac{3u_1^2-u_2}{2Z^2} + O\left(\frac{1}{Z^3}\right) \right).  \eqno(2.11)
$$

\ni As for the second integral $I_2$, since

$$
J = Z\left(1 - \frac{2}{Z^2u(1+9/(Z^2u^2))}\right),  \qquad R = \frac{3}{4}\frac{\left(1+\frac{1}{Z^2u^2}\right)}{\left(1+\frac{9}{Z^2u^2}\right)},
$$

\ni it is easy to see that, for large $Z$, $I_2$ is expanded as

$$
I_2 = \frac{iA_0^3e^{\frac{i(x^2-y^2)}{4Z}}}{16Z^2}\int_4^\infty\frac{du}{u^2}\left(1 + O\left(\frac{1}{Z^2}\right) \right) = \frac{iA_0^3e^{\frac{i(x^2-y^2)}{4Z}}}{64Z^2}\left(1 + O\left(\frac{1}{Z^2}\right) \right).  \eqno(2.12)
$$

\ni Thus, gathering the contributions of $I_1$ and $I_2$, we obtain

$$
\Phi_n = \frac{iA_0^3e^{\frac{i(x^2-y^2)}{4Z}}}{32Z}\left(C_1 - \frac{1}{6Z} - \frac{3C_2(x^2+y^2)}{16Z^2} + \frac{i(x^2-y^2)}{12Z^2} + \frac{C_3}{Z^2} + O\left(\frac{1}{Z^3}\right) \right),  \eqno(2.13)
$$

\ni where $C_1$, $C_2$ and $C_3$ are numerical constants given by

$$
C_1 = \int_{1/9}^1\frac{d\zeta}{\sqrt{\zeta(1-\zeta)(9\zeta-1)}} \approx 1.68575, \qquad C_2 = \int_{1/9}^1\frac{\sqrt\zeta d\zeta}{\sqrt{(1-\zeta)(9\zeta-1)}} \approx 0.742494,
$$

$$
C_3 = \frac{1}{16}\int_{1/9}^1\frac{d\zeta}{\sqrt{\zeta(1-\zeta)(9\zeta-1)}}\left((9\zeta-1)(1-\zeta) - \frac{9\zeta^2}{2}\right) \approx -0.05268.   \eqno(2.14)
$$

\ni Similarly, expanding the exact solution eq.~(2.4) of the linearized equation in powers of $1/Z$ for $Z\gg1$, we find

$$
\Phi_L = \frac{A_0e^{\frac{i(x^2-y^2)}{4Z}}}{4Z}\left(1 - \frac{x^2+y^2}{16Z^2} - \frac{1}{32Z^2} - \frac{i(x^2-y^2)}{64Z^3} + O\left(\frac{1}{Z^4}\right) \right).  \eqno(2.15)
$$

\ni Adding eqs.~(2.15) and (2.13), we obtain the large $Z$ asymptotics of $\Phi$ for the Gaussian initial condition,

$$
\Phi \approx \Phi_L + \Phi_n = \frac{A_0e^{\frac{i(x^2-y^2)}{4Z}}}{4Z}\left(1 + \frac{iC_1A_0^2}{8} - \frac{iA_0^2}{48Z} -  \right.
$$

$$
\left. - \frac{x^2+y^2}{16Z^2}\left(1+\frac{3C_2iA_0^2}{8}\right) - \frac{A_0^2(x^2-y^2)}{96Z^2} - \frac{1-4C_3iA_0^2}{32Z^2} + O\left(\frac{1}{Z^3}\right) \right).  \eqno(2.16)
$$

\ni It is illuminating to present the factor in the parentheses as

$$
1 + \mu + i\nu = \rho e^{i\sigma},   \eqno(2.17)
$$

\ni where

$$
\mu = - \frac{x^2+y^2}{16Z^2} -  \frac{A_0^2(x^2-y^2)}{96Z^2} - \frac{1}{32Z^2} + O\left(\frac{1}{Z^3}\right),  \eqno(2.18)
$$

$$
\nu =  \frac{C_1A_0^2}{8} - \frac{A_0^2}{48Z} -  \frac{3C_2A_0^2(x^2+y^2)}{128Z^2} + \frac{C_3iA_0^2}{8Z^2} + O\left(\frac{1}{Z^3}\right).  \eqno(2.19)
$$

\ni Then

$$
\rho^2=(1+\mu)^2+\nu^2,  \qquad  \sigma = \arctan\frac{\nu}{1+\mu}.   \eqno(2.20)
$$

\ni Keeping the terms up to order $\sim1/Z^2$ in $\rho^2$ and $\sigma$, we find

$$
\rho^2 = 1 + \frac{C_1^2A_0^4}{64} - \frac{C_1A_0^4}{192Z} -
$$

$$
-\frac{x^2+y^2}{8Z^2}\left(1+\frac{3C_1C_2A_0^4}{64}\right) - \frac{A_0^2(x^2-y^2)}{48Z^2} - \frac{1}{16Z^2}\left(1 - \frac{(72C_1C_3+1)A_0^4}{144}\right) + O\left(\frac{1}{Z^3}\right),  \eqno(2.21)
$$

$$
\sigma = \arctan\frac{C_1A_0^2}{8} + \frac{\left( -\frac{A_0^2}{48Z} + \frac{(C_1-3C_2)A_0^2(x^2+y^2)}{128Z^2} + \frac{C_1A_0^4(x^2-y^2)}{768Z^2} + \left(C_3+\frac{C_1}{32}\right)\frac{A_0^2}{8Z^2}\right)}{1 + (C_1A_0^2/8)^2} -
$$

$$
- \frac{C_1A_0^6}{8\cdot(48)^2Z^2(1 + (C_1A_0^2/8)^2)^2} +  O\left(\frac{1}{Z^3}\right).  \eqno(2.22)
$$

\ni However, strictly speaking, since we started with the first perturbation in $A_0^2$, we should keep only the terms up to order $\sim A_0^2$ in the last formulas. Then they simplify to

$$
\rho^2 = 1 - \frac{x^2+y^2}{8Z^2} - \frac{A_0^2(x^2-y^2)}{48Z^2} - \frac{1}{16Z^2} +  O\left(A_0^4\right) + O\left(\frac{1}{Z^3}\right),  \eqno(2.23)
$$

$$
\sigma = \frac{C_1A_0^2}{8} - \frac{A_0^2}{48Z} + \frac{(C_1-3C_2)A_0^2(x^2+y^2)}{128Z^2} + \frac{(C_1+32C_3)A_0^2}{256Z^2} + O\left(A_0^4\right) +  O\left(\frac{1}{Z^3}\right).  \eqno(2.24)
$$

\ni Thus, we finally obtain

$$
\Phi=Ae^{i\theta}
$$

\ni with amplitude and phase given by

$$
A \approx  \frac{A_0}{4Z}\left(1 - \frac{x^2+y^2}{16Z^2} - \frac{A_0^2(x^2-y^2)}{96Z^2} - \frac{1}{32Z^2}  \right),  \eqno(2.25)
$$

$$
\theta \approx \frac{C_1A_0^2}{8} + \frac{x^2-y^2}{4Z} - \frac{A_0^2}{48Z} + \frac{(C_1-3C_2)A_0^2(x^2+y^2)}{128Z^2}  + \frac{(C_1+32C_3)A_0^2}{256Z^2}.  \eqno(2.26)
$$

\ni The main focusing/defocusing effect of the nonlinearity depends on the term $-\frac{A_0^2(x^2-y^2)}{96Z^2}$ in the amplitude eq.~(2.25). Its sign shows compression in the focusing $x$-direction and decompression in the defocusing $y$-direction as expected. (We note that in Fig.~3  of~\cite{AbBu14} the axes $Y$ and $T$ should be relabelled since there $Y$ is the focusing and $T$ is the defocusing direction.)
\par The previous formulae imply that, for the Gaussian lump initial condition of moderate amplitude $A_0$, we have the following theoretical parameters in the asymptotics:

$$
\Lambda_0 = \frac{A_0}{4},   \qquad  \theta_0 = \frac{C_1A_0^2}{8} \approx 0.2107A_0^2,   \qquad   \eta_0 = \Lambda_0^2 - \frac{A_0^2}{48} = \frac{A_0^2}{24},   \eqno(2.27)
$$

\ni see eq.~(1.2A), which we will compare with numerics in subsequent sections.

\section{General large $Z$ asymptotics -- linear and nonlinear}

\subsection{Linear case}

The linear problem has a similarity solution which describes the central  $x-y$ region for long time. Indeed, the linear problem

$$
i\prt_Z\Phi + \prt_{xx}\Phi - \prt_{yy}\Phi = 0  \eqno(3.1)
$$

\ni has a Fourier solution for general initial conditions (IC). Denoting by $\hat\Phi(k,l,Z)$ the Fourier component of $\Phi$ in $xy$-space, one gets $\hat\Phi(k,l,Z) = \hat\Phi_0(k,l)e^{-i(k^2-l^2)Z}$ where $\hat\Phi_0(k,l) = \hat\Phi(k,l,0)$ is the Fourier transform of the IC. Then, after rescaling $k\to k/\sqrt Z$ and $l\to l/\sqrt Z$, one can write the inverse Fourier transform restoring $\Phi(x,y,Z)$ as

$$
\Phi(x,y,Z) = \frac{1}{4\pi^2Z}\int\int\hat\Phi_0(k/\sqrt Z,l/\sqrt Z)e^{-i(k^2-l^2)}e^{i(kx+ly)/\sqrt Z}dkdl,  \eqno(3.2)
$$

\ni which is convenient for expansion in inverse powers of $Z$. In eq.~(3.2), we expand $\hat\Phi_0(k/\sqrt Z,l/\sqrt Z)$ in a Taylor series around the origin,

$$
\hat\Phi_0(k/\sqrt Z,l/\sqrt Z) = \hat\Phi_0(0,0) +
$$

$$
+ \prt_k\hat\Phi_0(0,0)\frac{k}{\sqrt Z} + \prt_l\hat\Phi_0(0,0)\frac{l}{\sqrt Z} + \frac{\prt_{kk}\hat\Phi_0(0,0)k^2+2\prt_{kl}\hat\Phi_0(0,0)kl+\prt_{ll}\hat\Phi_0(0,0)l^2}{2Z} + \dots   \eqno(3.3)
$$

\ni Similarly we expand the exponent $e^{i(kx+ly)/\sqrt Z}$ and get

$$
\Phi(x,y,Z) = \frac{1}{4\pi^2Z}\int\int e^{-i(k^2-l^2)}dkdl\left(\hat\Phi_0(0,0) + \prt_k\hat\Phi_0(0,0)\frac{k}{\sqrt Z} + \prt_l\hat\Phi_0(0,0)\frac{l}{\sqrt Z} + \right.
$$

$$
\left. + \frac{\prt_{kk}\hat\Phi_0(0,0)k^2+2\prt_{kl}\hat\Phi_0(0,0)+\prt_{ll}\hat\Phi_0(0,0)l^2}{2Z} + \dots\right)\left(1 + \frac{i(kx+ly)}{\sqrt Z} - \frac{(kx+ly)^2}{2Z} + \dots\right)
$$

$$
=  \frac{1}{4\pi^2Z}\int\int e^{-i(k^2-l^2)}dkdl\left(\hat\Phi_0(0,0)\left(1 - \frac{k^2x^2+l^2y^2}{2Z}\right) + \right.
$$

$$
\left. + \prt_k\hat\Phi_0(0,0)\frac{ik^2x}{Z} + \prt_l\hat\Phi_0(0,0)\frac{il^2y}{Z} +  \frac{\prt_{kk}\hat\Phi_0(0,0)k^2+\prt_{ll}\hat\Phi_0(0,0)l^2}{2Z} + O\left(\frac{1}{Z^2}\right) \right),
$$

\ni the last equality being true due to the survival of only even powers of $k$ and $l$ under the integration. This implies asymptotics of the form

$$
\Phi(x,y,Z) = \frac{C_0}{Z}\left(1 + \frac{C_1}{Z} + \frac{(C_2x+C_3y)}{Z} + \frac{i(x^2-y^2)}{4Z} + O\left(\frac{1}{Z^2}\right) \right),   \eqno(3.4)
$$

\ni where $C_0$, $C_1$, $C_2$ and $C_3$ are constants depending on the IC $\Phi_0(x,y)$. Exponentiating the expression in the parentheses one finally obtains

$$
\Phi(x,y,Z) \approx \frac{C_0}{Z}e^{\frac{i(x^2-y^2)}{4Z} + \frac{C_1}{Z} + \frac{(C_2x+C_3y)}{Z} },   \qquad C_0 = \frac{\hat\Phi_0(0,0)}{4\pi},
$$

$$
C_1 = -\frac{i(\prt_{kk}\hat\Phi_0(0,0)-\prt_{ll}\hat\Phi_0(0,0))}{4\hat\Phi_0(0,0)},   \quad C_2 = \frac{\prt_k\hat\Phi_0(0,0)}{2\hat\Phi_0(0,0)},  \quad C_3 = -\frac{\prt_l\hat\Phi_0(0,0)}{2\hat\Phi_0(0,0)},   \eqno(3.5)
$$

\ni as the approximate general asymptotic solution with $\hat\Phi_0(0,0) \neq 0$. For symmetric ICs $C_2=C_3=0$. It should be noted that the above derivation requires the initial data in Fourier space to be sufficiently smooth. This is not always the case and later we make a further comment about this, see the remark about noise in Fourier space in the discussion of the numerics.
\par We see that this solution to the linear problem, which is valid for {\it all lump type initial conditions with $\hat\Phi_0(0,0) \neq 0$}, is approximately the same as the nonlinear similarity solution eq.~(1.2). However we will see that the additional contribution in the phase in eq.~(1.2) can make a significant difference. Without this term the error in the phase of the solution can be quite substantial.

\subsection{Nonlinear case}

\ni If we assume that the solution falls like $1/Z$ at large $Z$ as we observe in all cases numerically, then it is convenient to express $\Phi=\phi/Z$ in the HNLS (1.1). Then HNLS takes form

$$
i\prt_Z\phi - \frac{i\phi}{Z} + \prt_{xx}\phi-\prt_{yy}\phi + \frac{|\phi|^2}{Z^2}\phi = 0.   \eqno(3.6)
$$

\ni Next we assume that, at large $Z$, the solution of eq.~(3.6) has the series expansion

$$
\phi(x,y,Z) = \sum_{n=0}^\infty\frac{\phi_n(x,y)}{Z^n}   \eqno(3.7)
$$

\ni in inverse powers of $Z$. Substituting eq.~(3.7) into eq.~(3.6) we find the linear wave equation

$$
\prt_{xx}\phi_0-\prt_{yy}\phi_0 = 0  \eqno(3.8)
$$

\ni at zeroth order in $1/Z$ which implies that in general $\phi_0(x,y) = \phi_+(x+y) + \phi_-(x-y)$, where functions $\phi_+$ and $\phi_-$ are arbitrary. The next order gives

$$
\prt_{xx}\phi_1-\prt_{yy}\phi_1 = i\phi_0,  \eqno(3.9)
$$

\ni so that, denoting $x_\pm = x \pm y$,

$$
\phi_1 = \frac{i}{4}\left(x_-\int_0^{x_+}\phi_+(u)du + x_+\int_0^{x_-}\phi_-(u)du\right) + g_+(x_+) + g_-(x_-),  \eqno(3.10)
$$

\ni with another two arbitrary functions $g_+$ and $g_-$. The first term coming from original nonlinearity appears only at second order in $1/Z$ which reads

$$
4\prt_{x_+x_-}\phi_2 = 2i\phi_1 - |\phi_0|^2\phi_0.   \eqno(3.11)
$$

\ni Proceeding, we obtain {\it linear} PDEs of the form $\prt_{x_+x_-}\phi_n = F(\{\phi_j, j<n\})$ allowing one to find in principle each $\phi_n$ in terms of the previous coefficients of the series (3.7). This shows the consistency of expansion (3.7) at large $Z$ and its generality since we have a sufficient number of arbitrary functions in the solution. However, the universal regime implies that a wide range of initial conditions leads to

$$
\phi_0 = \Lambda_0e^{i\theta_0} = \text{const.}   \eqno(3.12)
$$

\ni rather than the general solution of eq.~(3.8). Then the first two terms of eq.~(3.10) give the term $\phi_0\cdot i(x^2-y^2)/4Z$ which indeed universally appears in the asymptotics. As we also observe numerically, one should take $g_+ + g_- = i(\eta_0-\Lambda_0^2)\phi_0$ for a large variety of ICs, where $\eta_0$ is a real constant. Then, exponentiating the correction $\phi_1$, we obtain

$$
\Phi \approx \frac{\Lambda_0}{Z}\exp{\left[i\left(\theta_0 + \frac{x^2-y^2}{4Z} + \frac{\eta_0-\Lambda_0^2}{Z}\right)\right]}.  \eqno(3.13)
$$

\ni The last formula is the corrected self-similar solution (1.2A) which is now seen to be consistent with all our analytical estimates. In what follows with extensive numerical calculations we verify this solution and determine the values of the constants. We will see that this solution is valid in the central zone $s\sim O(1)\ll Z$. This solution is observed for a large class of ICs -- virtually all that contain a lump of energy around the center $x=y=0$.

\section{Numerical results}

The numerical simulations in this paper employed the ETD2 scheme (exponential time differencing, spectral in space and second-order in time) proposed in~\cite{CoxMatt02}. The computation domain is taken to be a square of size $L$ and the number of gridpoints  $N$ was chosen such that $L/N = 400/1024 = 600/1536 = 800/2048 = 0.390625$ depending on size requirements. The time ($Z$) step was $0.01$ in these simulations.
\par Corresponding to each of the initial conditions in the table below there are five figures. In the top row, in the leftmost figure the real part of the numerical solution of eq.~(1.1) is plotted in the $x,y$ plane at the final value of time $Z=Z_{max}$ in the simulation. In the top row, center figure, the real part of the exact solution eq.~(1.2) is plotted for the same $Z$ with fitted amplitude and phase constants $\Lambda_0$ and $\theta_0$ based on the numerical solution, and the absolute value of their difference is shown next in the top right figure; there is very good agreement in the central spot, for all these initial conditions. The chosen initial conditions include various Gaussian lumps which cover a wide range of parameters -- amplitude and the widths along $x$ and $y$ axes, as well as some other functions. In some cases to the initial conditions a small amount of randomness was added (10\%). Apart from an expected spread  in the slope of the numerical line in the right figure of the bottom row the results are largely the same. This is discussed more fully below.

\par In the left figure of the bottom row of two figures, the maximum amplitudes of the above two solutions are plotted together versus $\log Z$; in each case they approach each other rather fast as $Z$ grows. The agreement is already excellent when $Z\sim 10$, for the properly chosen amplitude $\Lambda_0$ parameter of eq.~(1.2) specified to agree with numerics asymptotically at large $Z$. Thus, the amplitude is well described by the exact similarity solution. The parameter $\theta_0$ in eq.~(1.2) was also specified to fit the numerics. In the right figure of the bottom two figures, the differences $\Delta\Theta=\theta-\theta_0-s$ taken at the center $x=y=0$, and therefore also $s=0$, were plotted versus $1/Z$ together for the above two solutions. There one observes gradual approach to a straight line in almost all cases; however, in most cases the {\it slopes} are seen to be different for the numerical and the exact solution. The slope of the line corresponding to the numeric solution on the center phase plot is equal to $\eta_0-\Lambda_0^2$, from which, with already known (determined by amplitude fitting) $\Lambda_0$, the parameter $\eta_0$ is found.
\par Thus, the phase exhibits a significant error. This discrepancy is the motivation to consider the corrected approximate analytic solution presented in section 3, with the additional parameter $\eta_0$. It is in turn determined from the numerics.

\par The parameters computed from the numerical data are presented in the following table (there are more cases presented here than in figures - due to restrictions on space):  \\

\begin{tabular}{|c|c|c|c|c|c|}
\hline
Initial condition & $\Lambda_0$ & $\theta_0$ & $Z_{max}$ & $\eta_0-\Lambda_0^2$ & $\eta_0$ \\
\hline
$e^{-x^2-y^2}$ & 0.25 & 0.21 & 16 & -0.0153 &  0.0472 \\
\hline
$2e^{-x^2-y^2}$ & 0.5 & 0.841 & 16 & -0.0751 &  0.1749 \\
\hline
$3e^{-x^2-y^2}$ & 0.7296 & 1.84 & 16 & -0.1736 & 0.3587 \\
\hline
$3.5e^{-x^2-y^2}$ & 0.6986 & 2.36 & 16 & -0.2521 & 0.2359  \\
\hline
$4e^{-x^2-y^2}$ & 0.5216 & 2.71 & 16 & -0.3283 & -0.0562  \\
\hline
$0.25\cdot4e^{-4x^2-4y^2}$ & 0.25 & 0.053 & 16 & -0.0085 & 0.054  \\
\hline
$0.5\cdot4e^{-4x^2-4y^2}$ & 0.5 & 0.2115 & 16 & -0.0087 & 0.2413  \\
\hline
$2\cdot0.5e^{-0.5x^2-0.5y^2}$ & 0.25 & 0.43 & 16 & -0.2203 & -0.1578  \\
\hline
$5\cdot0.5e^{-0.5x^2-0.5y^2}$ & 0.505 & 2.41 & 16 & -0.091 & 0.164  \\
\hline
$5\cdot0.2e^{-0.2x^2-0.2y^2}$ & 0.2439 & 1.043 & 16 & -0.286 & -0.2265  \\
\hline
$10\cdot0.1e^{-0.1x^2-0.1y^2}$ & 0.2449 & 2.01 & 32 & -1.221 & -1.161  \\
\hline
$e^{-x^2-2y^2}$ & 0.192 & 0.145 & 16 & 0.0431 & 0.08 \\
\hline
$e^{-2x^2-y^2}$ & 0.1768 & 0.145 & 16 & -0.0685 & -0.0372 \\
\hline
$0.5(e^{-(x+1)^2-y^2}+e^{-(x-1)^2-y^2})$ & 0.25 & 0.105 & 16 & 0.2464 & 0.3089 \\
\hline
$0.5(e^{-x^2-(y+1)^2}+e^{-x^2-(y-1)^2})$ & 0.25 & 0.105 & 16 & -0.232 & -0.1695 \\
\hline
$\text{sech}(x^2+y^2)$ & 0.395 & 0.42 & 24 &  -0.0498 &  0.1062 \\
\hline
$2e^{-x^2-y^2}\cos(2(x+y))$ & 0.0746 & 0.607 & 16 & -0.0345 & -0.0289 \\
\hline
$2.5\tanh(|x^2-y^2|)e^{-x^2-y^2}$ & 0.258 & 0.132 & 16 & -0.0198 & 0.0468 \\
\hline
$(x+iy)^4e^{-x^2-y^2}$ & 0.0021 & -1.45 & 16 & -0.7144 & -0.7144  \\
\hline
\end{tabular}

\section{Discussion of the results}

As one might expect, corresponding to initial conditions with larger energy, whether due to larger amplitude or width,  larger $Z$ are required to achieve the same degree of agreement between the numerical and the asymptotic solution described by eq.~(1.2) or eq.~(1.2A). The actual solution is closer to the exact similarity solution eq.~(1.2) (i.e.~the correction parameter $\eta_0$ is smaller by absolute value) when initially one has a moderate lump of energy with the maximum density at the center. When the initial amplitude is much bigger or the lump is much more narrow than the ones presented in the table/figures, more accurate numerics are required. Previous numerical investigations, e.g.~\cite{BerRasMultispl96, BerEtAl01, LitEtAl00}, also found that for larger initial amplitudes of HNLS high resolutions are required to obtain reliable results. In this work we do not investigate large or rapidly varying functions. This is in the spirit of the asymptotic derivation of the HNLS equation. 

\par From the numerical values of the parameters presented in the table, one can see that for most initial conditions featuring a localized lump of energy at the center, the parameter $\eta_0$ turns out to be of the same order as $\Lambda_0^2$. The absolute value of their difference is  smaller for narrower initial beams while it becomes of the same order as the parameters themselves for initial widths $\sim 1$ or greater. For input beams of amplitude $1$, narrow in one direction and width $1$ in the other, we find that $|\eta_0|$ is less than $\Lambda_0^2$.
\par We see in Figs.~15--20 that an initial lump with additional noise (which was taken to be of moderate amplitude ten times smaller than that of the deterministic part) leads to similar pictures as the corresponding lump without noise. But the amplitude and phase undergo random shifts so that after many (100) realizations we obtain thick curves for the numerical amplitude and especially numerical phase. The amplitude shifts due to the randomness are smaller (cf. the numerical curves  in figs.~15--20). The average amplitude and phase are consistent with the corresponding values without noise. When the noise is added to every spatial grid point, it creates relatively large effective gradients which lead to the wide spread of the phase curves around the average. The amplitude curves, however, remain close to the mean curve even in this case. As different realizations of the random noise present the various possible initial conditions in a neighborhood of their average, these results are especially indicative of the main point we emphasize: essentially all initial conditions without large gradients lead to the same universal asymptotic regime that we exposed here.
\par {\it Remark.} The picture turns out to be very different if one adds a similar type of noise in the spectral (Fourier) space instead. Then the asymptotics become drastically modified and we observe oscillations of significant amplitude. This occurs for both the HNLS eq.~and its linearized version. Therefore the discrepancy with the asymptotics eq.~(1.2A) can be understood looking at the derivation of the asymptotic formula for the linear case in section 3.1. There it was necessary for the Fourier transform of the IC to be smooth enough in order for its Taylor expansion at the origin in spectral space to be valid. The spectral noise, however, makes this function rough. In contrast, for the noise in physical space considered above, the IC in Fourier space turns out to be smooth which explains the discrepancy in the large $Z$ behavior.

\par If there is a hole rather than a lump at the center of the $xy$-plane in the initial condition, then one sees the hole spreading at later times and the values of the solution become tiny around the origin. Still, even in such cases where the most energy is well away from the center, the hyperbolic structure of the solution (1.2) can be observed (see fig.~14, the intitial condition $(x+iy)^4e^{-x^2-y^2}$). This phenomenon is a feature of both the full HNLS equation and its linearized version. 

\par For relatively small initial amplitudes, the reshaping of the wave packet can be well described by considering nonlinearity as a perturbation to the linearized equation. This way one can quantitatively understand the dumbbell shapes often observed forming from the initial round beam both in two and three spatial dimensions~\cite{LitEtAl83, NewellEtAl94, LitEtAl00, BerEtAl01, AbBu14}. We analyzed this situation in section 2 for an initial Gaussian beam and showed, as expected, that the beam is compressed in the focusing $x$-direction and decompressed in the defocusing $y$-direction. For the Gaussian beam of small or moderate amplitude, we have theoretical values, in the first perturbative approximation in the initial amplitude $A_0$, for the parameters $\Lambda_0$, $\theta_0$ and $\eta_0$, see eq.~(2.27). Their numerical values are presented in the table below:

\bigskip

\begin{tabular}{|c|c|c|c|}
\hline
$A_0$ & $\Lambda_0$ & $\theta_0$ & $\eta_0$ \\
\hline
1 & 0.25 & 0.2107 &  0.0417 \\
\hline
2 & 0.5 & 0.8429 &  0.1667 \\
\hline
3 & 0.75 & 1.8965 & 0.375 \\
\hline
4 & 1 & 3.3715 & 0.6667  \\
\hline
\end{tabular}

\bigskip

One sees that there is a relatively good agreement with the numerical results up to $A_0\sim3$, and for larger $A_0$ it rapidly worsens. Still the qualitative agreement with the universal regime eq.~(1.2A) often exists even for larger amplitudes.
\par An interesting observation is that in the HNLS with initial lump of energy that is large/wide, rings of low amplitude are observed to develop and move away from the center. Later they can disconnect, reconnect with parts of other structures in various ways forming intricate patterns. These complex processes or sometimes a simpler initial deformation of the wave beam (energy lump) are followed later by outbursts of energy from the central region to both directions of the defocusing coordinate axis ($y$-axis in our case) corresponding to beam splitting.
\par Thus, if the initial condition has relatively large energy or is substantially different from a single packet of small energy, much more complicated pictures than we show in this paper appear at intermediate times $Z\sim0.5-5$ and often persist to larger $Z$. Still on the edges of these sometimes exotic patterns one clearly sees the development of the same familiar hyperbolic structure described above. Based on our numerical findings the universal regime with a central hyperbolic structure eventually develops even for initial lumps of relatively large amplitude. We believe that the phenomena discussed in literature like spiky hyperbolic structures numerically observed in~\cite{LitEtAl00, BerEtAl01, BerEtAl02, LitEtAl02, LitEtAl03} as well as observed X-waves~\cite{ContiEtAl03, Conti04, CouaironEtAl06, MoshEtAl09, KeNaZe11} correspond to intermediate regimes just at the onset of the hyperbolic long-time asymptotic structure, at least in the (2+1)-D case considered here. We expect these waves to eventually develop into the universal regime, perhaps with many centers as we also observed in our simulations. We also note that~X-waves are known to appear in both linear and nonlinear situations; this can be also be said about our universal hyperbolic structure.

\section{Conclusion}
\par The main conclusion is that the similarity solution eq.~(1.2) with the corrections described by eq.~(1.2A) appears universally in the central zone of the HNLS at long times/large propagation distances. As long as there are no large or rapidly varying initial data
the universal regime outlined here is expected to be observed in the long-time limit. This universal behavior also may help select among many existing large energy solutions of the HNLS equation at intermediate times; this also might be relevant to the transient behavior observed in different types of beam propagation.

\par Our results are supported by analytical estimates and numerical computations. Analytically we consider the nonlinear term as a perturbation of gaussian initial conditions and consider linear and nonlinear problems via their long time limits. Numerically we consider a wide range of initial conditions including random initial data. We also investigate the averaged variational method in the context of a Gaussian ansatz in Appendix B. We find that while the method reproduces the similarity solution (1.2) it does not reproduce the important modification of the phase in (1.2A).

\bigskip
{\bf\large Acknowledgments} \\
This research was partially supported by the the NSF under grant CHE 1125935 and the U.S. Air Force Office of Scientific Research, under grant FA9550-16-1-0041.

\section*{Appendix A: Some exact reductions of (2+1)-D HNLS}

A larger class of reductions of the HNLS eq.~(1.1) is obtained if we consider the following ansatz. Letting $\Phi=Ae^{i\theta}$ with

$$
A = \frac{\Lambda(\xi, \eta)}{R(Z)},  \qquad  \theta = \sigma(\xi, \eta) + \frac{(\alpha_1Z+\alpha_0)\xi^2}{4} + \frac{(\beta_1Z+\beta_0)\eta^2}{4} + \frac{(\gamma_1Z+\gamma_0)\xi\eta}{2} +
$$

$$
+ (\mu_1Z+\mu_0+\mu_*R)\xi + (\nu_1Z+\nu_0+\nu_*R)\eta + h(z),   \eqno(A1)
$$

\ni where

$$
\xi = \frac{C_{11}x+C_{12}y}{R(Z)} + \xi_0(Z),  \qquad \eta = \frac{C_{21}x+C_{22}y}{R(Z)} + \eta_0(Z),   \eqno(A2)
$$

$$
R^2(Z) = HZ^2 + H_1Z + R_0^2,   \qquad  \alpha_1 = -\frac{C_2H}{\Delta^2}, \quad \beta_1 = -\frac{C_1H}{\Delta^2}, \quad \gamma_1 = \frac{C_3H}{\Delta^2},   \eqno(A3)
$$

$$
C_1 = C_{11}^2-C_{12}^2, \qquad C_2 = C_{21}^2-C_{22}^2, \qquad C_3 = C_{11}C_{21}-C_{12}C_{22},
$$

$$
\Delta^2 = C_3^2-C_1C_2 = (C_{11}C_{22}-C_{12}C_{21})^2\neq 0.  \eqno(A4)
$$

\ni All coefficients in the above formulae are constant except those with explicitly given $Z$ dependence.
When eq.~(A1) is substituted into eq.~(1.1), its imaginary part eq.~(B1) multiplied by $\Lambda(\xi, \eta)$ becomes a conservation law,

$$
\prt_{\xi}[\Lambda^2(C_1\prt_{\xi}\sigma + C_3\prt_{\eta}\sigma + C_4\xi + C_6\eta + C_8)] + \prt_{\eta}[\Lambda^2(C_3\prt_{\xi}\sigma + C_2\prt_{\eta}\sigma + C_5\xi + C_7\eta + C_9)] = 0,   \eqno(A5)
$$

\ni where the newly introduced constants are

$$
C_4 = \frac{C_1\alpha_0+C_3\gamma_0-H_1/2}{2}, \qquad C_5 = \frac{C_3\alpha_0+C_2\gamma_0}{2},
$$

$$
C_6 = \frac{C_1\gamma_0+C_3\beta_0}{2}, \qquad C_7 = \frac{C_2\beta_0+C_3\gamma_0-H_1/2}{2}. \eqno(A6)
$$

\ni The constants $C_8$ and $C_9$ are at this point arbitrary and functions $\xi_0(Z)$, $\eta_0(Z)$ are determined by equations whose solutions are written out later. The real part of eq.~(1.1),
after the substitution of ansatz (A1) then yields

$$
C_1[\prt_{\xi\xi}\Lambda-\Lambda(\prt_{\xi}\sigma)^2] + C_2[\prt_{\eta\eta}\Lambda-\Lambda(\prt_{\eta}\sigma)^2] + 2C_3[\prt_{\xi\eta}\Lambda-\Lambda\prt_{\xi}\sigma\prt_{\eta}\sigma] +
$$

$$
+ \Lambda[\Lambda^2 - 2(C_4\xi+C_6\eta+C_8)\prt_{\xi}\sigma - 2(C_5\xi+C_7\eta+C_9)\prt_{\eta}\sigma - K_1\xi^2 - K_2\eta^2 - K_3\xi\eta - K_4\xi - K_5\eta - K_6] = 0,  \eqno(A7)
$$

\ni where now $C_8$ and $C_9$ must satisfy

$$
HC_8 = H(C_1\mu_0-C_3\nu_0) + H_1(C_3\nu_1-C_1\mu_1)/2,  \qquad  HC_9 = H(C_2\nu_0-C_3\mu_0) + H_1(C_3\mu_1-C_2\nu_1)/2,   \eqno(A8)
$$

\ni and constants $K_i$, $i=1,\dots,5$ are given by

$$
4K_1 = C_1\alpha_0^2+C_2\gamma_0^2+2C_3\alpha_0\gamma_0-\alpha_0H_1+\alpha_1R_0^2,   \eqno(A9)
$$

$$
4K_2 = C_1\gamma_0^2+C_2\beta_0^2+2C_3\beta_0\gamma_0-\beta_0H_1+\beta_1R_0^2,   \eqno(A10)
$$

$$
2K_3 = C_1\alpha_0\gamma_0+C_2\beta_0\gamma_0+C_3(\alpha_0\beta_0+\gamma_0^2)-\gamma_0H_1+\gamma_1R_0^2,   \eqno(A11)
$$

$$
2K_4 = 2(\alpha_0C_8+\gamma_0C_9)-\mu_0H_1+2\mu_1R_0^2,  \qquad  2K_5 = 2(\gamma_0C_8+\beta_0C_9)-\nu_0H_1+2\nu_1R_0^2,  \eqno(A12)
$$

\ni and constant $K_6$ is arbitrary. Finally, after using eq.~(A8), the functions $\xi_0$ and $\eta_0$ take form

$$
\xi_0(Z) = \frac{2C_3(\nu_1H_1-2\nu_0H)}{HR(Z)}\int^Z\frac{dZ}{R(Z)} - \frac{2(C_1\mu_*+C_3\nu_*)Z}{R(Z)} - \frac{2(C_1\mu_1+C_3\nu_1)}{H},
$$

$$
\eta_0(Z) = \frac{2C_3(\mu_1H_1-2\mu_0H)}{HR(Z)}\int^Z\frac{dZ}{R(Z)} - \frac{2(C_3\mu_*+C_2\nu_*)Z}{R(Z)} - \frac{2(C_3\mu_1+C_2\nu_1)}{H},  \eqno(A13)
$$

\ni and function $h(Z)$ can be found from

$$
R^2h'(Z) = C_1(\mu_1Z+\mu_0+\mu_*R)^2 + C_2(\nu_1Z+\nu_0+\nu_*R)^2 + 2C_3(\mu_1Z+\mu_0+\mu_*R)(\nu_1Z+\nu_0+\nu_*R) -
$$

$$
- 2C_8(\mu_1Z+\mu_0+\mu_*R) - 2C_9(\nu_1Z+\nu_0+\nu_*R) + K_6.   \eqno(A14)
$$

\ni Further reduction of eqs.~(A5) and (A7) to ODEs is achieved if e.g.~one takes $\Lambda=\Lambda(\xi)$ and $\sigma=\sigma(\xi)$ (without loss of generality, taking functions of $\eta$ only gives identical results up to relabeling of constant parameters). Then eq.~(A5) reduces to

$$
[\Lambda^2(C_1\sigma' + C_4\xi + C_8)]' + C_7\Lambda^2 = 0,  \eqno(A15)
$$

\ni where prime now means $d/d\xi$, together with the further restriction $C_6=0$, i.e.~$C_1\gamma_0+C_3\beta_0=0$. Eq.~(A7) then implies that $K_2=K_3=K_5=0$ and reduces to

$$
C_1[\Lambda'' - \Lambda(\sigma')^2] + \Lambda[\Lambda^2 - 2(C_4\xi+C_8)\sigma' - K_1\xi^2 - K_4\xi - K_6] = 0.   \eqno(A16)
$$

\ni Introducing function $U(\xi)$ such that $U' = \Lambda^2$ allows one to integrate eq.~(A15) once and get

$$
U'(C_1\sigma' + C_4\xi+C_8) + C_7U = C_I = \text{const.},   \eqno(A17)
$$

\ni where $C_I$ is an arbitrary constant. Plugging $\sigma'$ from eq.~(A17) into eq.~(A16) multiplied by $2\Lambda$ finally yields a third-order ODE for $U$,

$$
C_1\left(U''' - \frac{(U'')^2}{2U'}\right) + 2(U')^2 +
$$

$$
+ 2\left((C_4^2-K_1)\xi^2 + (2C_4C_8-K_4)\xi + C_8^2-K_6\right)U' - \frac{2(C_I-C_7U)^2}{U'} = 0.  \eqno(A18)
$$

\ni For special values of the parameters this ODE can be reduced to the Painlev\'e IV, II or I equations as well as to ODEs for elliptic functions and their elementary degenerations. The above reductions cover the majority of the cases considered in~\cite{Taj83, GagWin89, GagEtAl89, Gag90, Clark92, OzGun06, GuoLin10}. Such reductions might be valuable in studying various intermediate regimes of the HNLS equation such as those related to descriptions of rogue waves~\cite{KharPel03, Rub15}.

\section*{Appendix B: the approximate solution of HNLS corresponding to the Gaussian variational ansatz}
Expressing $\Phi=Ae^{i\theta}$ in the HNLS eq.~(1.1), we can rewrite the HNLS equation as two real equations for the amplitude $A$ and the phase $\theta$,

$$
\prt_Z A + A(\prt_{xx}\theta-\prt_{yy}\theta) + 2(\prt_xA\prt_x\theta - \prt_yA\prt_y\theta) = 0,    \eqno(B1)
$$

$$
A\prt_Z\theta = \prt_{xx}A - \prt_{yy}A - A((\prt_x\theta)^2 - (\prt_y\theta)^2) + A^3.   \eqno(B2)
$$

\ni If, in accordance with the Gaussian variational ansatz of~\cite{Ber94, Rub15}, we substitute the expressions

$$
A = \frac{\Lambda_0}{\sqrt{L(Z)R(Z)}}e^{-\frac{x^2}{L^2(Z)}-\frac{y^2}{R^2(Z)}},  \qquad   \Lambda_0=\text{const.}, \quad  L(Z)>0, \quad R(Z)>0,  \eqno(B3)
$$

$$
\theta = U(Z)x^2 + V(Z)y^2 + \sigma(Z),   \eqno(B4)
$$

\ni we find that eq.~(B1) is satisfied exactly if

$$
U(Z) = \frac{1}{4L}\frac{dL}{dZ},  \qquad   V(Z) = -\frac{1}{4R}\frac{dR}{dZ}.   \eqno(B5)
$$

\ni As for eq.~(B2), it cannot be satisfied exactly this way. However, it can be satisfied approximately, in two different regions of $xy$-plane. First, in the region $|x|\lesssim L(Z), |y|\lesssim R(Z)$, eq.~(B2) is satisfied if the scaling functions $L(Z)$ and $R(Z)$ satisfy the system found in~\cite{Ber94, Rub15} from variational principle,

$$
\frac{d^2L}{dZ^2} = \frac{16}{L^3} - \frac{8\Lambda_0^2}{L^2R},   \qquad   \frac{d^2R}{dZ^2} = \frac{16}{R^3} + \frac{8\Lambda_0^2}{LR^2},   \eqno(B6)
$$

\ni where the function $\sigma(Z)$ in eq.~(B4) is chosen so that

$$
\frac{d\sigma}{dZ} = \frac{\Lambda_0^2}{LR} - \frac{2}{L^2} + \frac{2}{R^2},   \eqno(B7)
$$

\ni and we use the approximation

$$
e^{-\frac{x^2}{L^2(Z)}-\frac{y^2}{R^2(Z)}} \approx 1 - \frac{x^2}{L^2(Z)} - \frac{y^2}{R^2(Z)},   \eqno(B8)
$$

\ni valid in the stated region. Similarly, in the region $|x| \gg L(Z), |y| \gg R(Z)$, eq.~(B2) can be approximately satisfied if one uses there eqs.~(B6) and (B7) with $\Lambda_0=0$ in them and approximates the exponent in eq.~(B8) by zero i.e.~neglects the last term $A^3$ in eq.~(B2). Thus, the error of the approximation here is bounded above by

$$
\left| e^{-\frac{x^2}{L^2(Z)}-\frac{y^2}{R^2(Z)}} - 1 + \frac{x^2}{L^2(Z)} + \frac{y^2}{R^2(Z)} \right| \leq \frac{1}{2}\left( \frac{x^2}{L^2(Z)} + \frac{y^2}{R^2(Z)}\right)^2,   \eqno(B9)
$$

\ni which yields information about the nature of the approximation of the the variational ansatz for this class of problems~\cite{Ber94, Rub15}.
\par Considering the large $Z$ asymptotics of the approximate solution given by eqs.~(B3)--(B8) in the central region, one can see that they are compatible with the the scales $L(Z)$ and $R(Z)$ changing as $L(Z) \sim Z + O(1)$ and $R(Z) \sim Z + O(1)$, i.e.~both approaching $Z$. This corresponds to the amplitude decreasing as $1/Z$ which is consistent with all our found asymptotics. Besides, as follows from eqs.~(B4), (B5) and (B7), under such symmetric along $x$ and $y$ asymptotic scaling the phase of the solution behaves as

$$
\theta = \theta_0 + \frac{x^2-y^2}{4Z} - \frac{\Lambda_0^2}{Z} + O\left(\frac{1}{Z^2}\right).
$$

\ni The last expression shows that the important parameter $\eta_0$, see eq.~(1.2A), which we found both analytically and numerically, is missing here. It could be recovered if we consider asymmetric scaling at large $Z$ i.e.~$L(Z) \sim C_1Z$ and $R(Z) \sim C_2Z$ with constants $C_1\neq C_2$. However, in section 2 we found nonzero $\eta_0$ for symmetric scales which follow from symmetric Gaussian IC. The kind of approximate solution presented here, and thus also the Gaussian variational ansatz of~\cite{Ber94, Rub15}, misses this possibility which demonstrates its serious limitations. Exact similarity solutions from appendix A are more flexible in this respect as are their two-scale generalizations which we plan to consider elsewhere.

\newpage

\begin{figure}
\includegraphics[width=4cm]{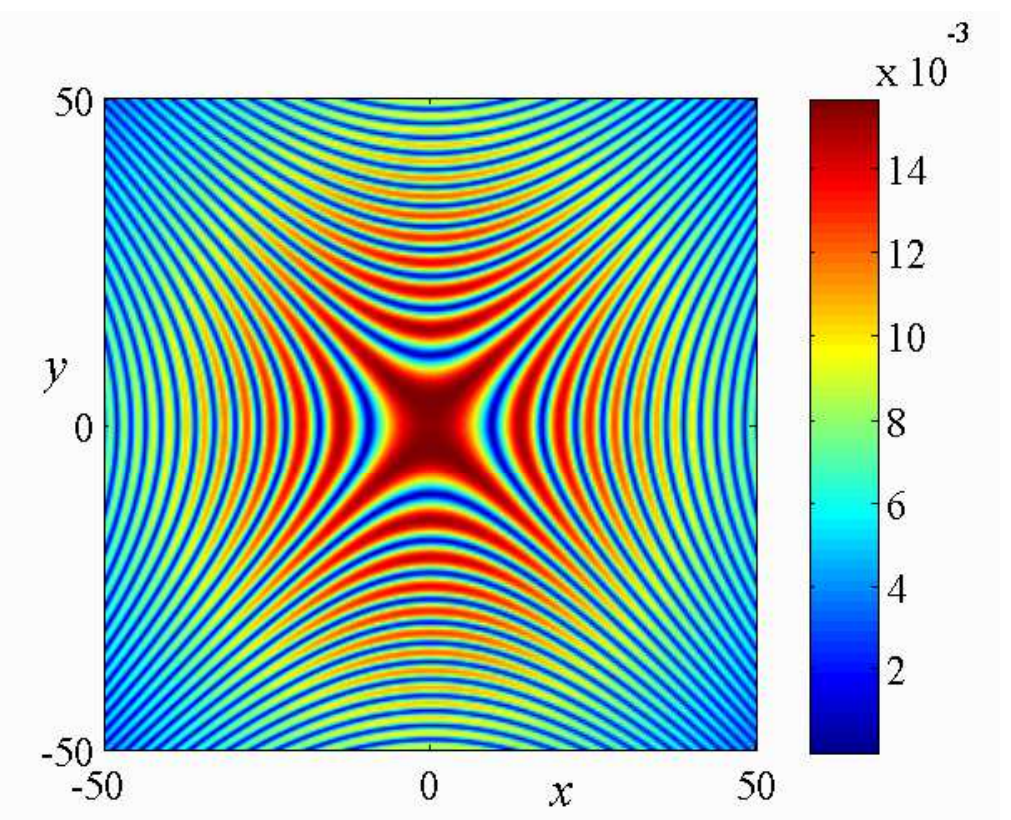}  \hspace{1cm}  \includegraphics[width=4cm]{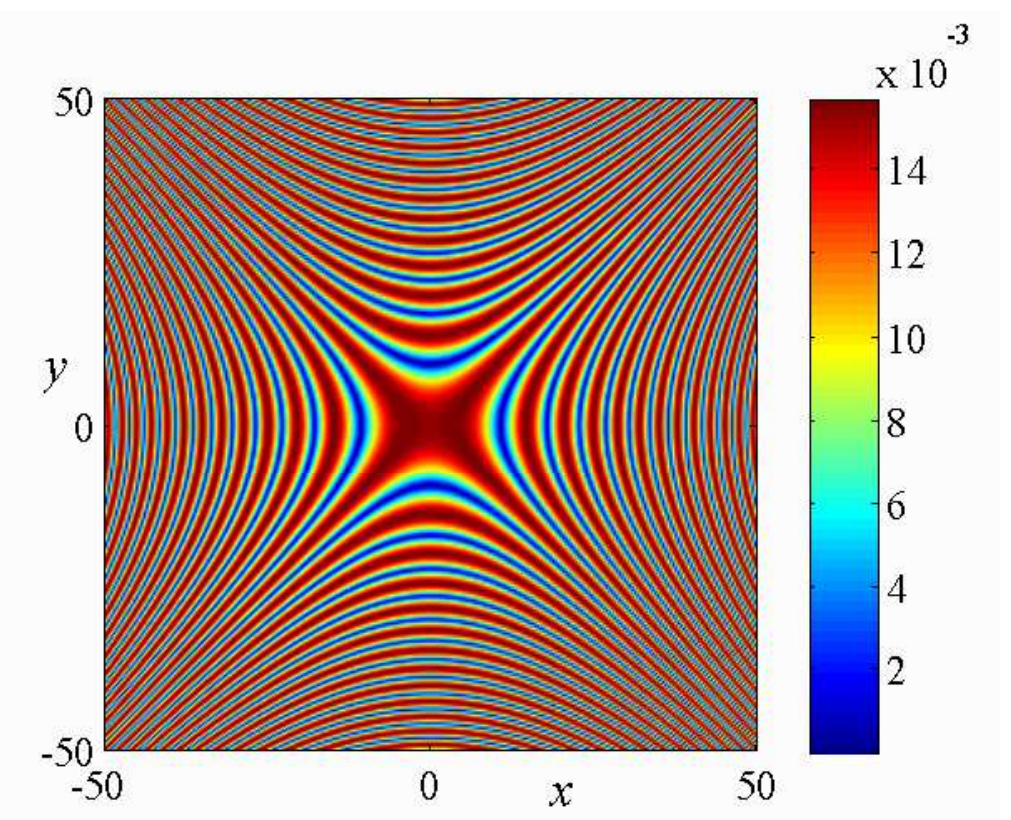} \hspace{1cm}  \includegraphics[width=4cm]{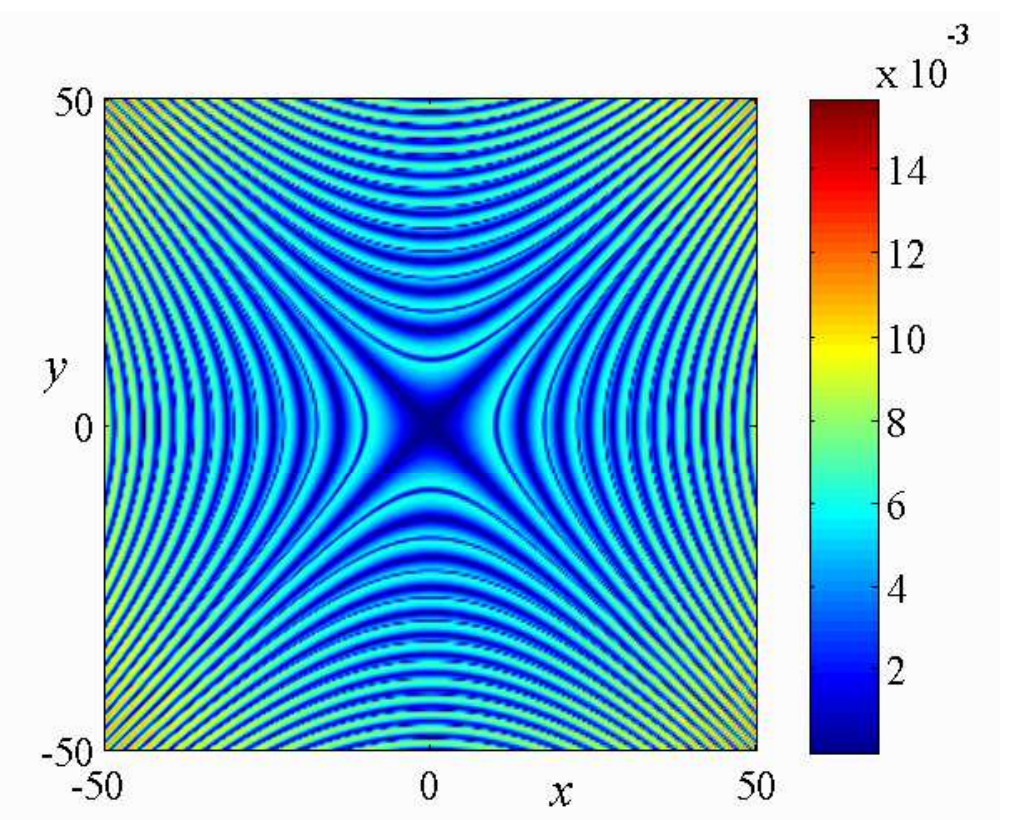}

\caption{Top: Initial condition $e^{-x^2-y^2}$: Numerical solution, Exact similarity solution and absolute value of their difference at $Z=16$.  Bottom: Log-amplitude vs. $\log Z$, $\Delta\theta = \theta-\theta_0-s$ vs. $1/Z$.}

\includegraphics[width=4cm]{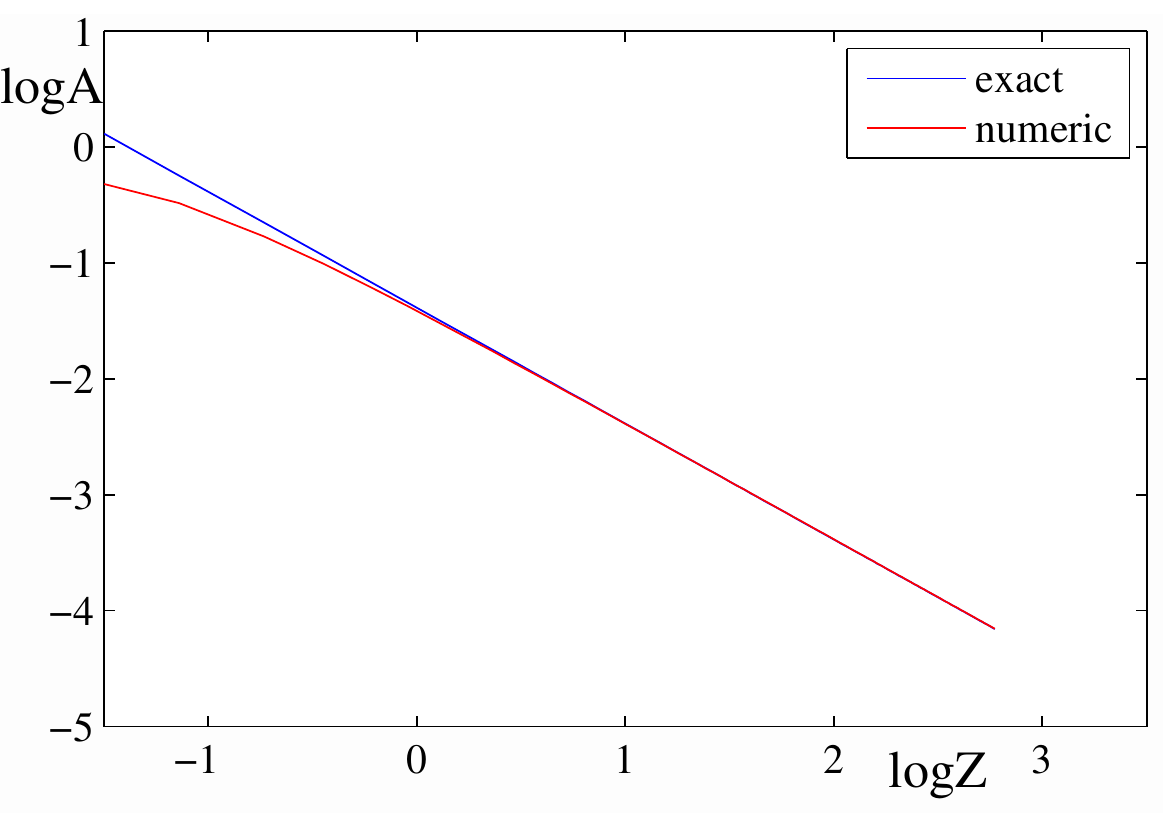}  \hspace{1cm}  \includegraphics[width=4cm]{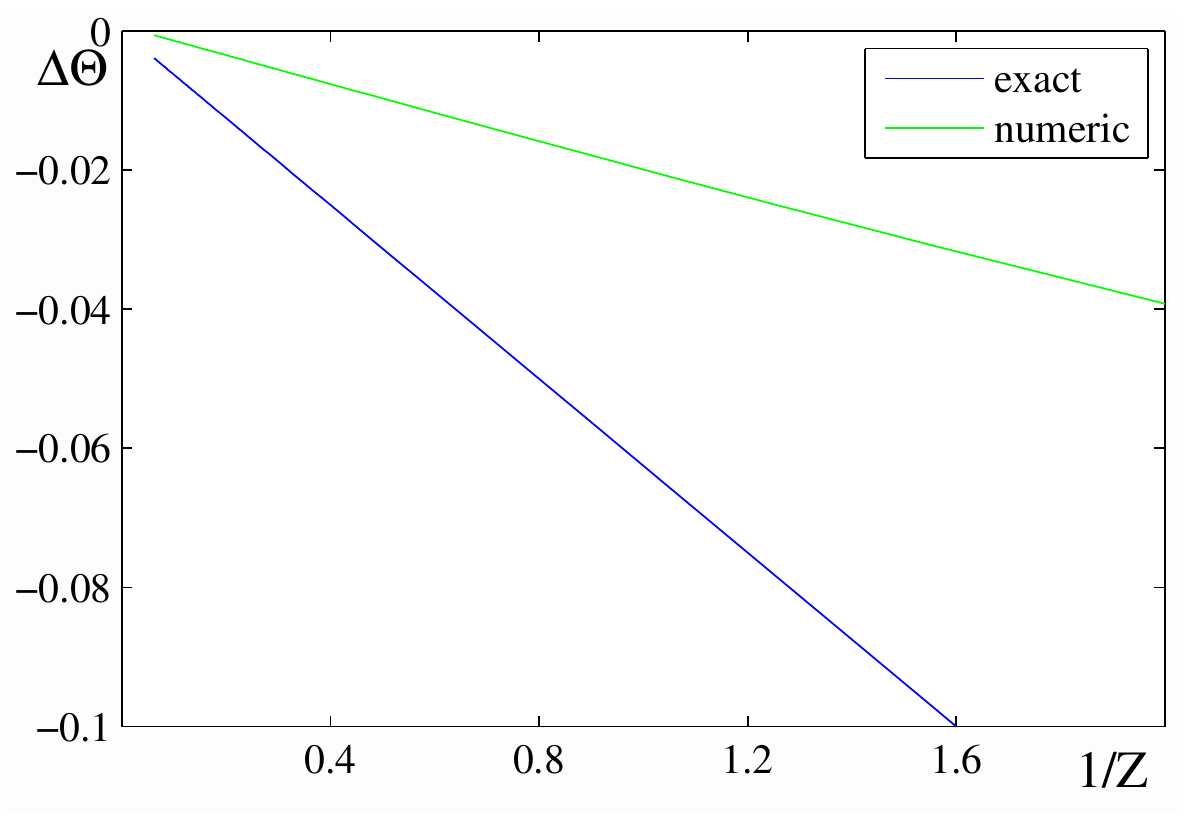}

\end{figure}

\begin{figure}
\includegraphics[width=4cm]{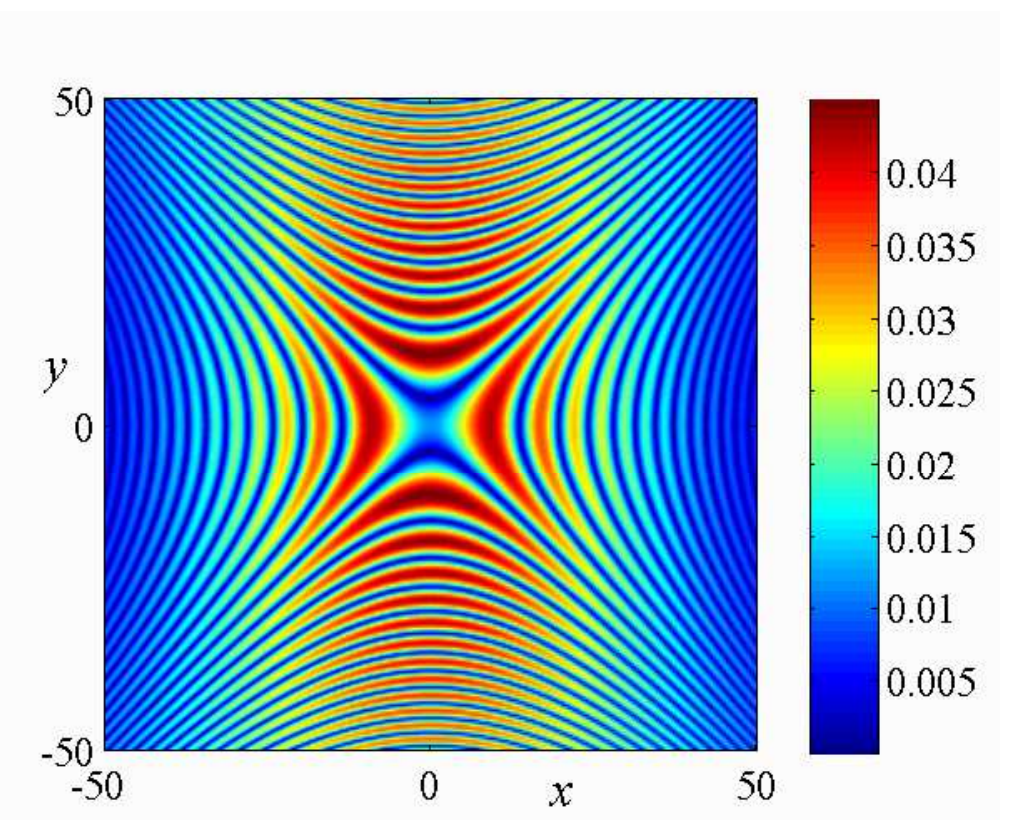}  \hspace{1cm}  \includegraphics[width=4cm]{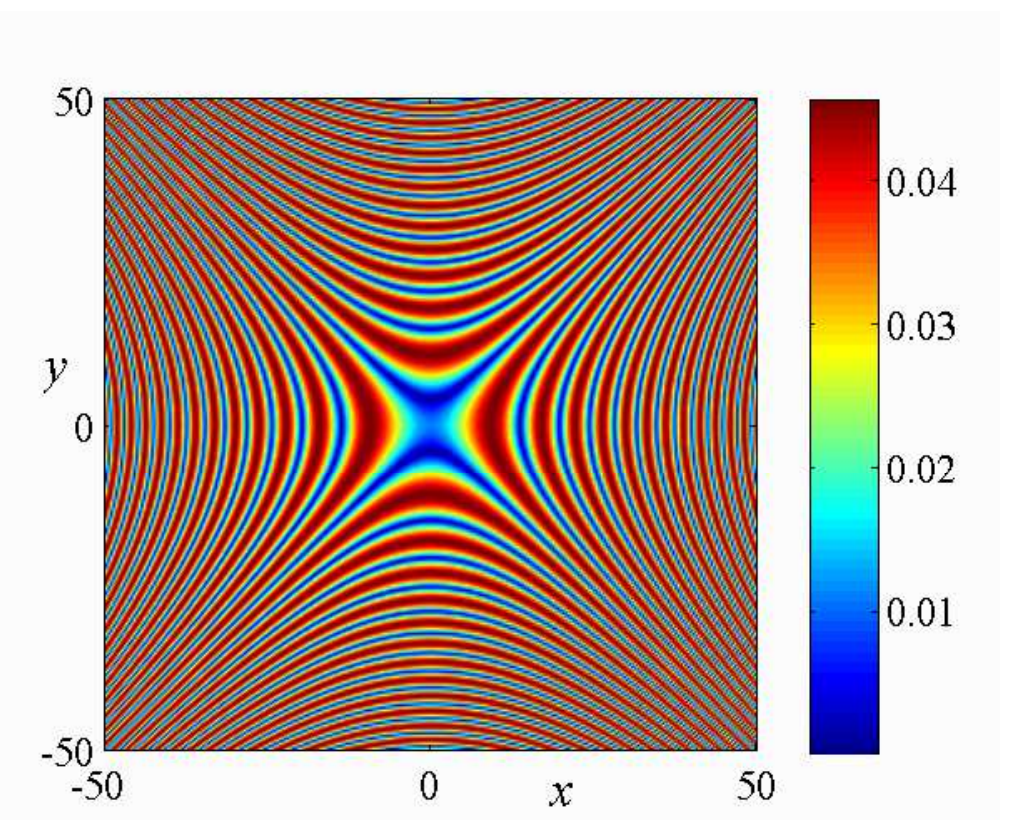} \hspace{1cm}  \includegraphics[width=4cm]{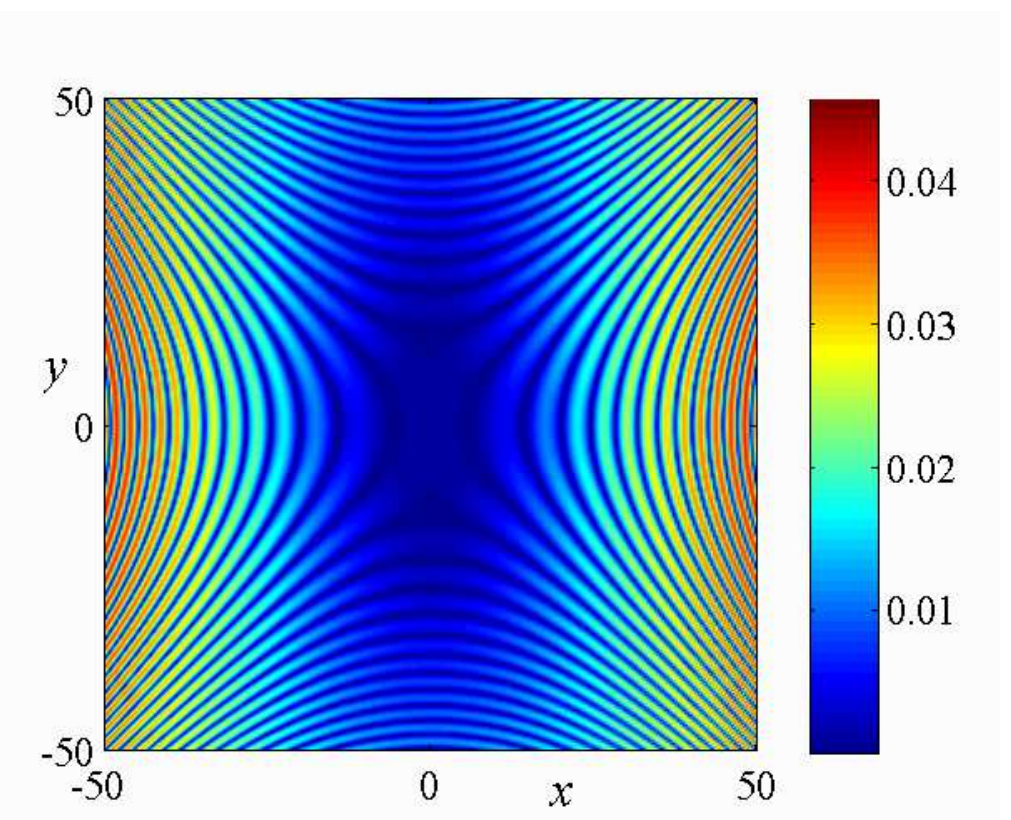}

\caption{Top:  Initial condition $3e^{-x^2-y^2}$: Numerical solution, Exact similarity solution and absolute value of their difference at $Z=16$.  Bottom: Log-amplitude vs. $\log Z$, $\Delta\theta = \theta-\theta_0-s$ vs. $1/Z$.}

\includegraphics[width=4cm]{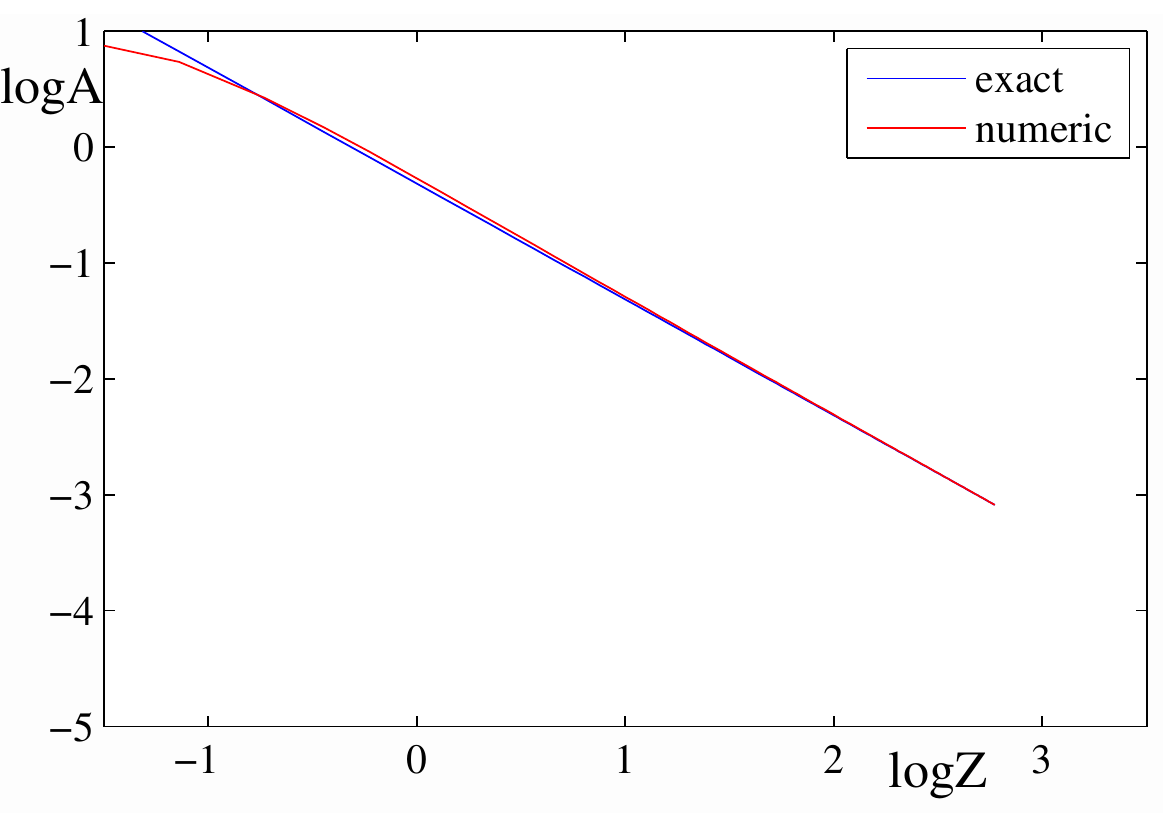} \hspace{1cm}  \includegraphics[width=4cm]{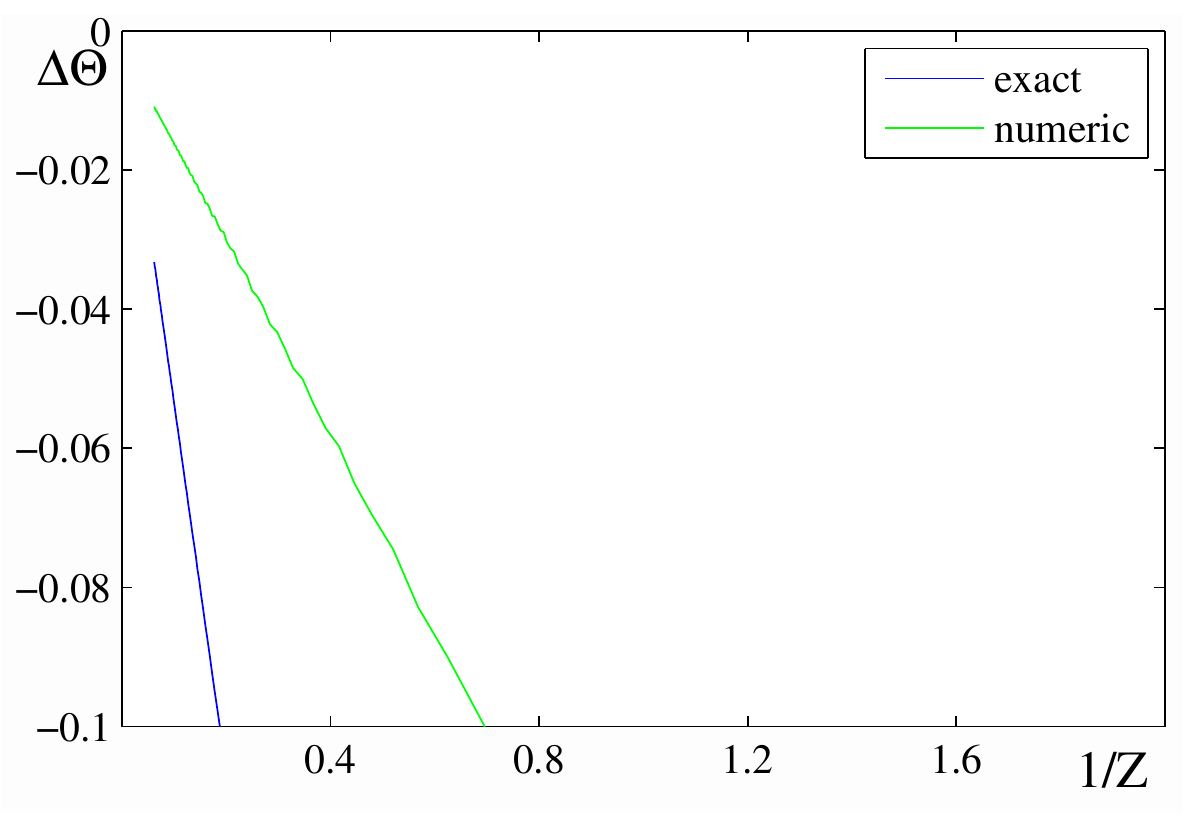}
\end{figure}

\newpage

\begin{figure}
\includegraphics[width=4cm]{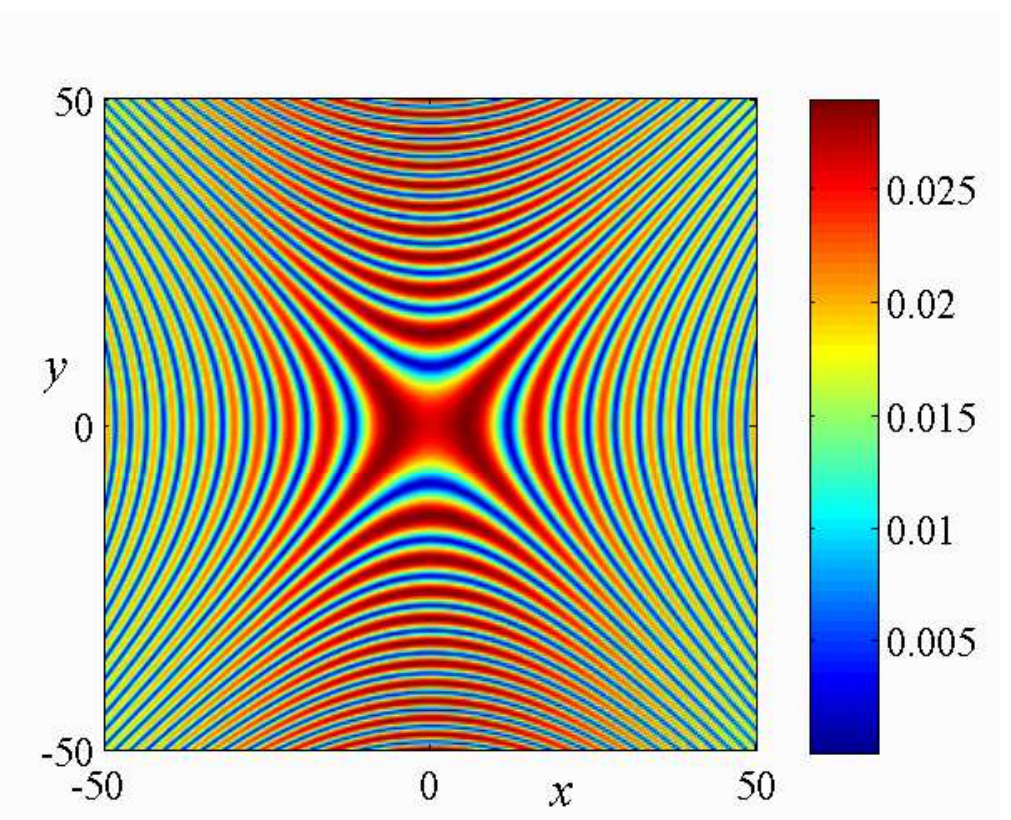}  \hspace{1cm}  \includegraphics[width=4cm]{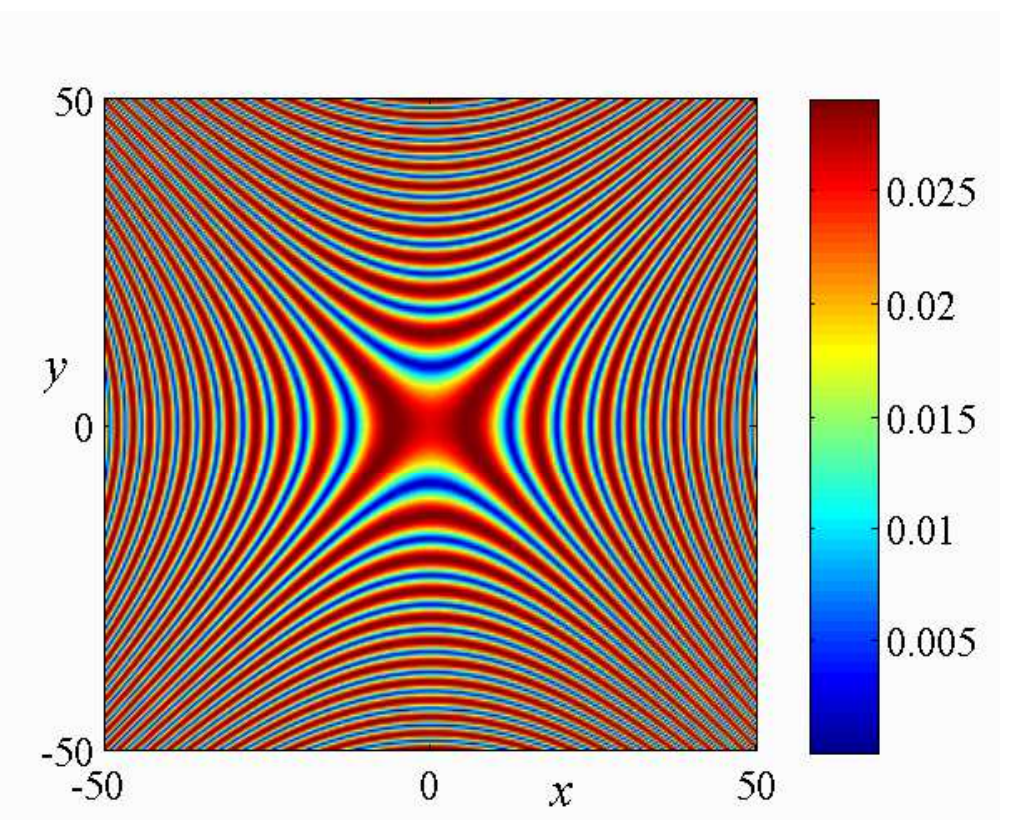} \hspace{1cm}  \includegraphics[width=4cm]{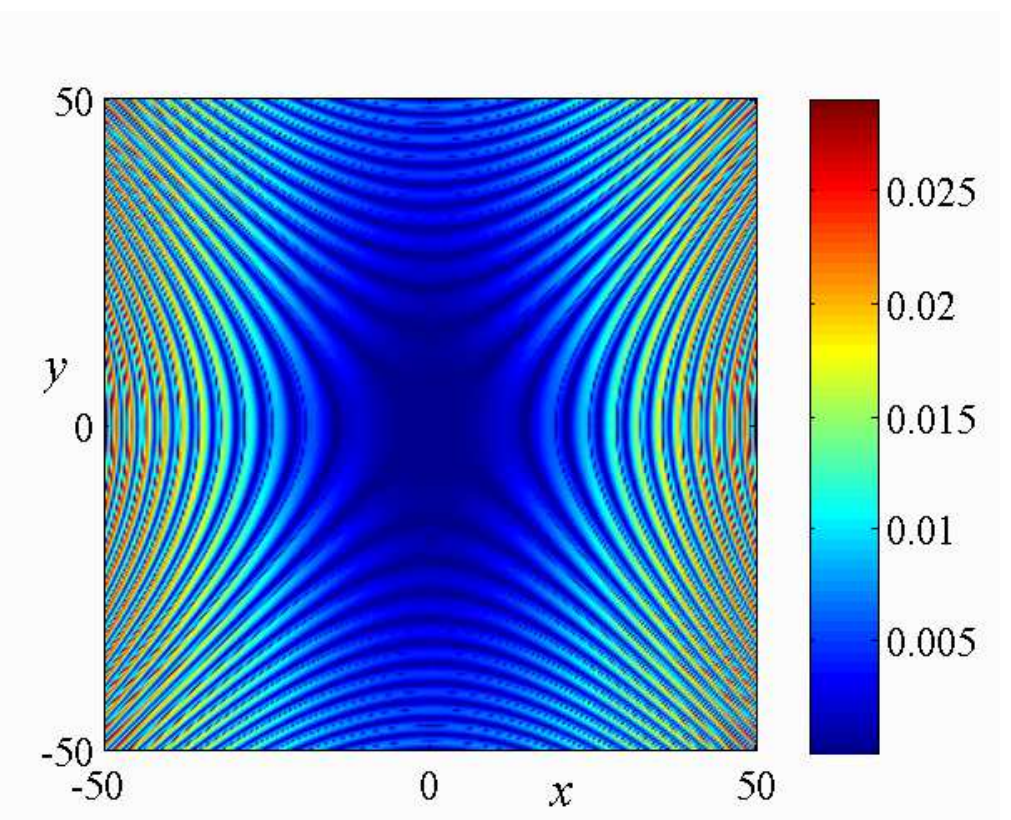}

\caption{Top:  Initial condition $4e^{-x^2-y^2}$: Numerical solution, Exact similarity solution and absolute value of their difference at $Z=16$.  Bottom: Log-amplitude vs. $\log Z$, $\Delta\theta = \theta-\theta_0-s$ vs. $1/Z$.}

\includegraphics[width=4cm]{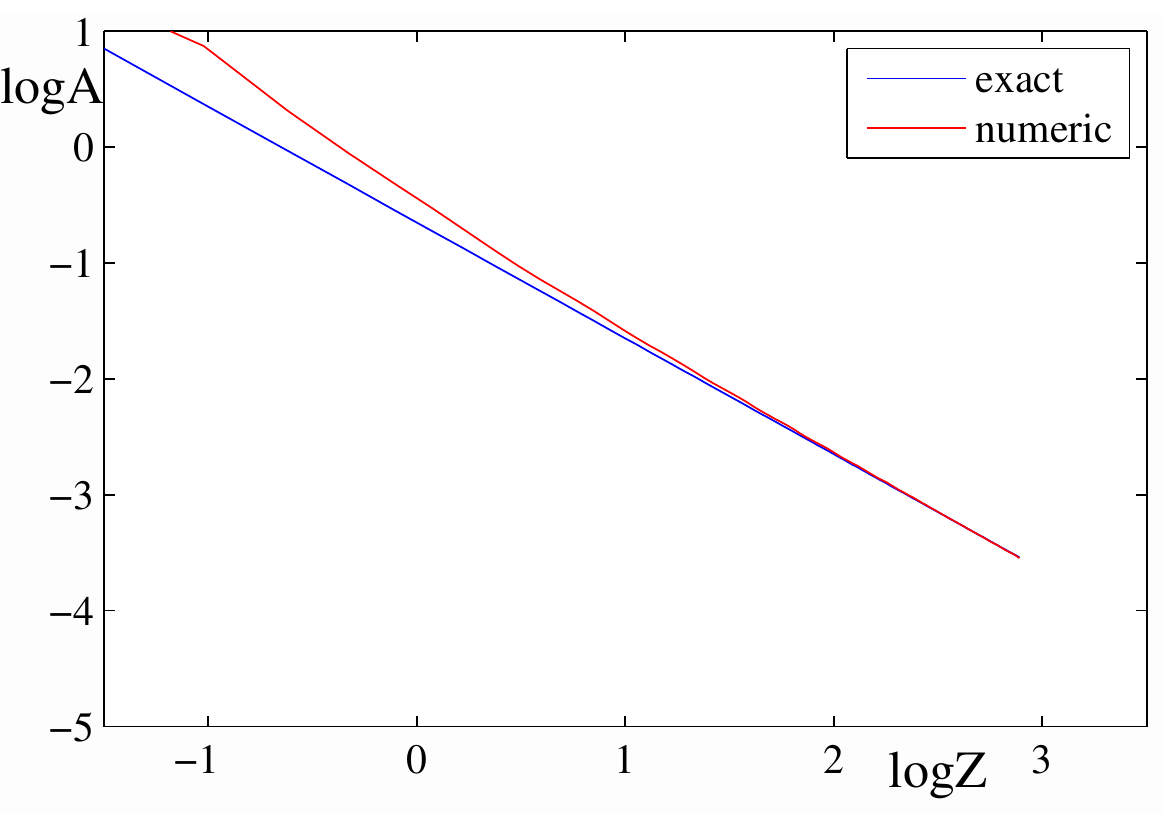} \hspace{1cm}  \includegraphics[width=4cm]{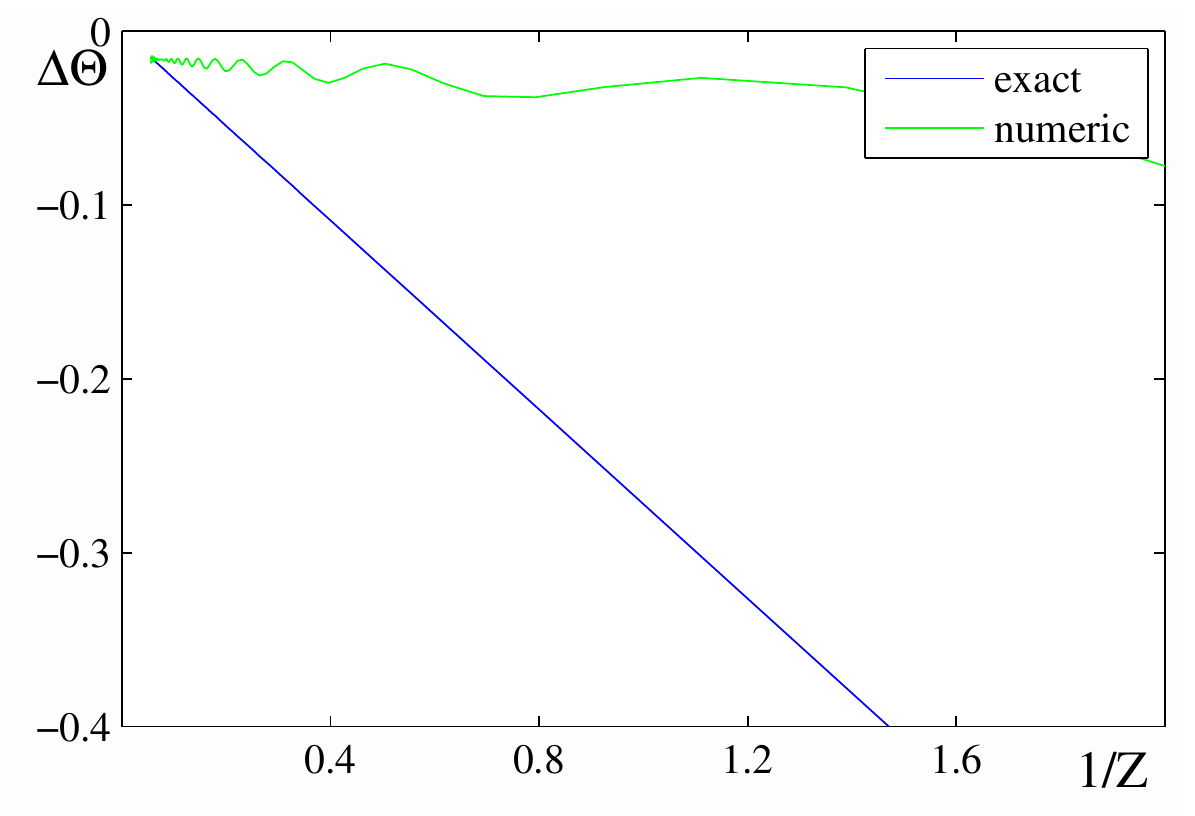}
\end{figure}

\begin{figure}
\includegraphics[width=4cm]{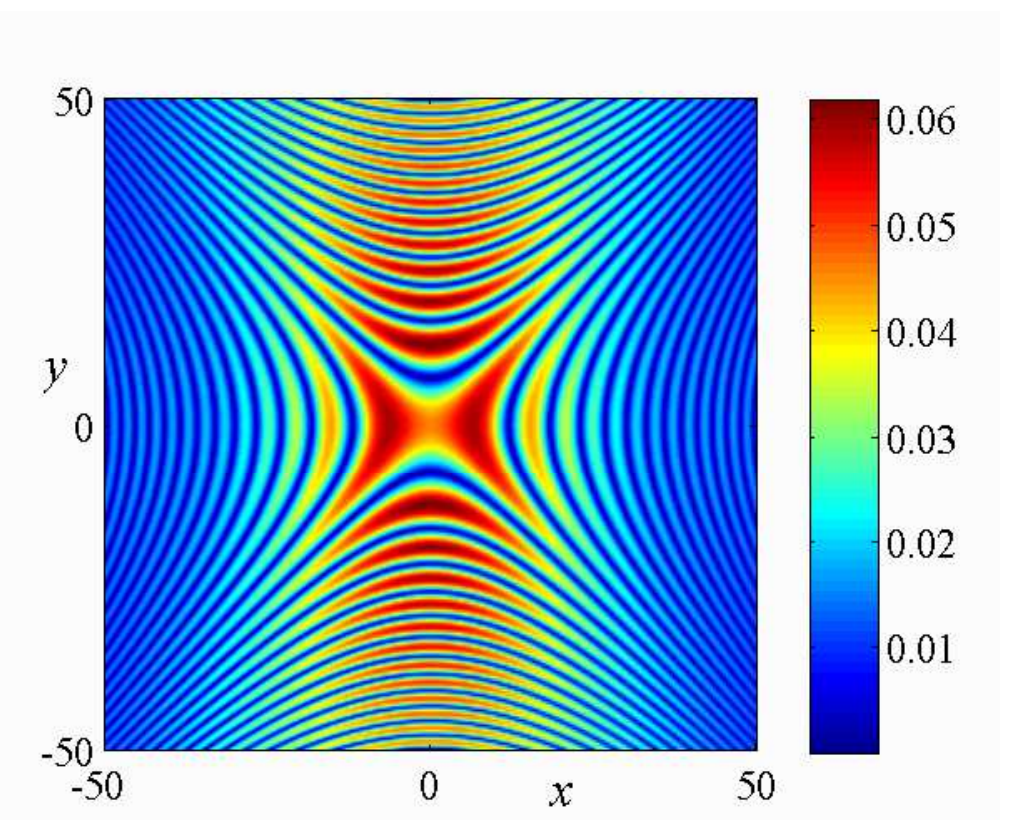}  \hspace{1cm}  \includegraphics[width=4cm]{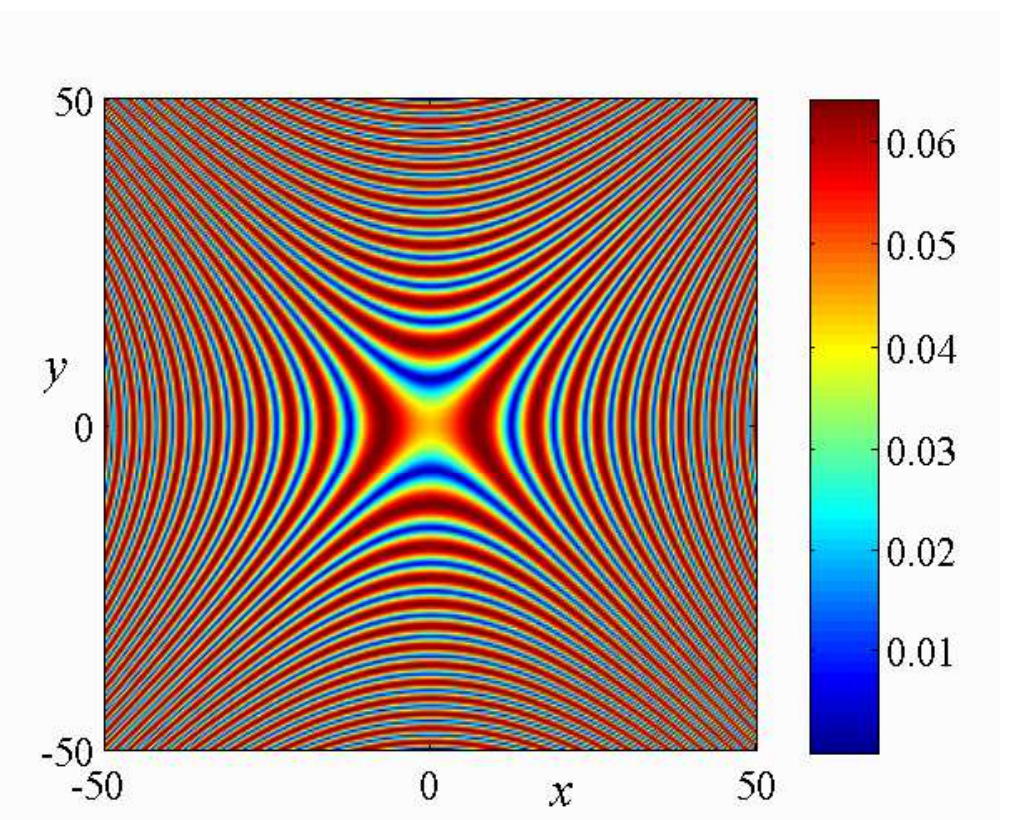} \hspace{1cm}  \includegraphics[width=4cm]{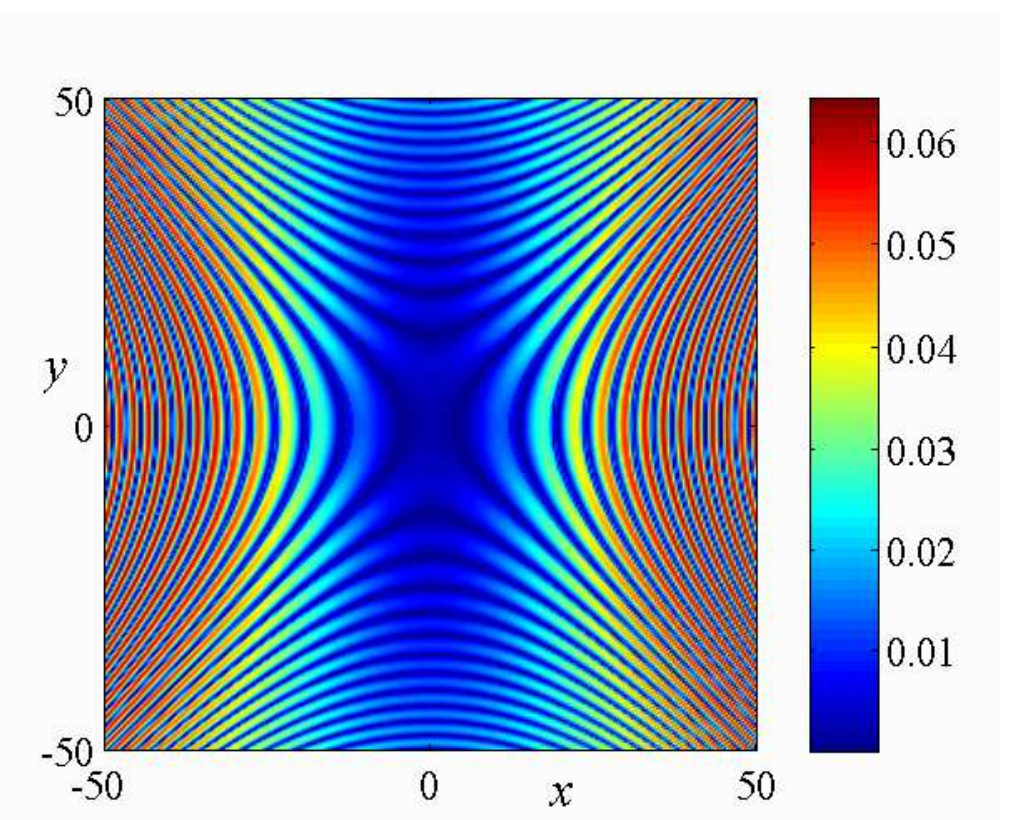}

\caption{Top:  Initial condition $2.5e^{-0.5x^2-0.5y^2}$: Numerical solution, Exact similarity solution and absolute value of their difference at $Z=16$.  Bottom: Log-amplitude vs. $\log Z$, $\Delta\theta = \theta-\theta_0-s$ vs. $1/Z$.}

\includegraphics[width=4cm]{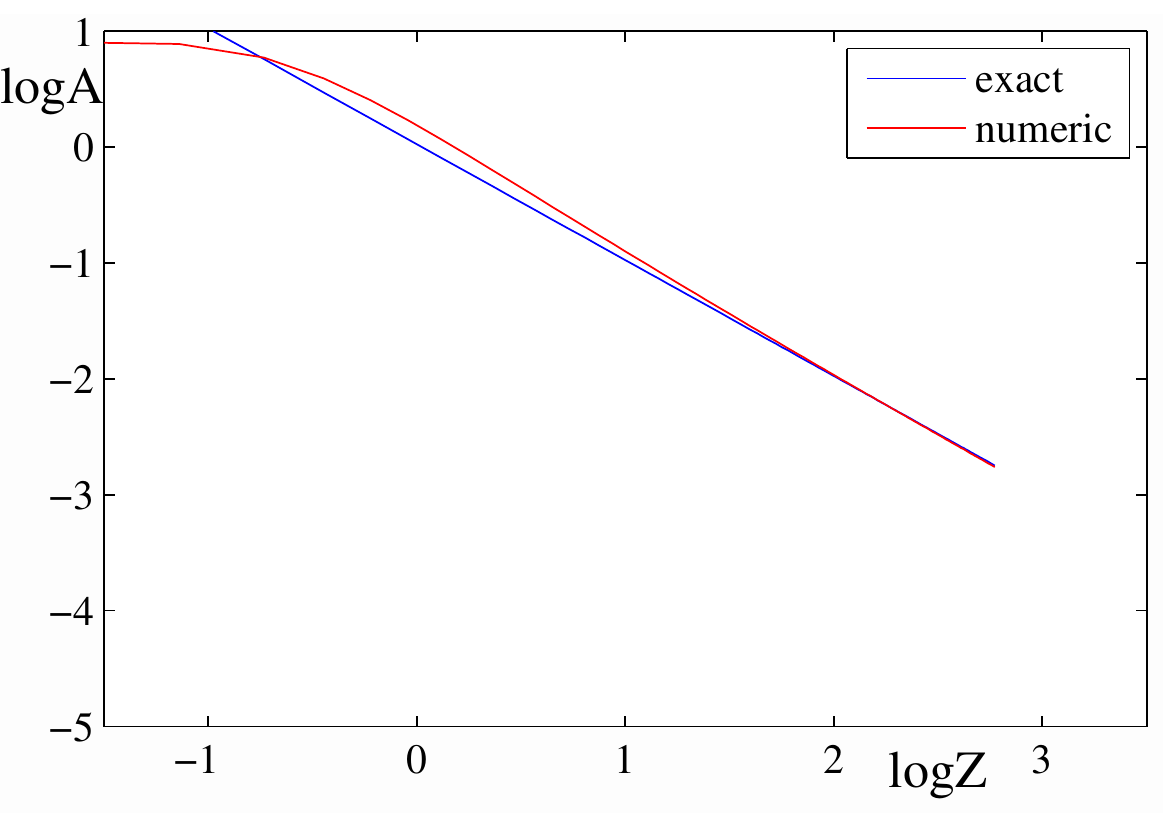}  \hspace{1cm}  \includegraphics[width=4cm]{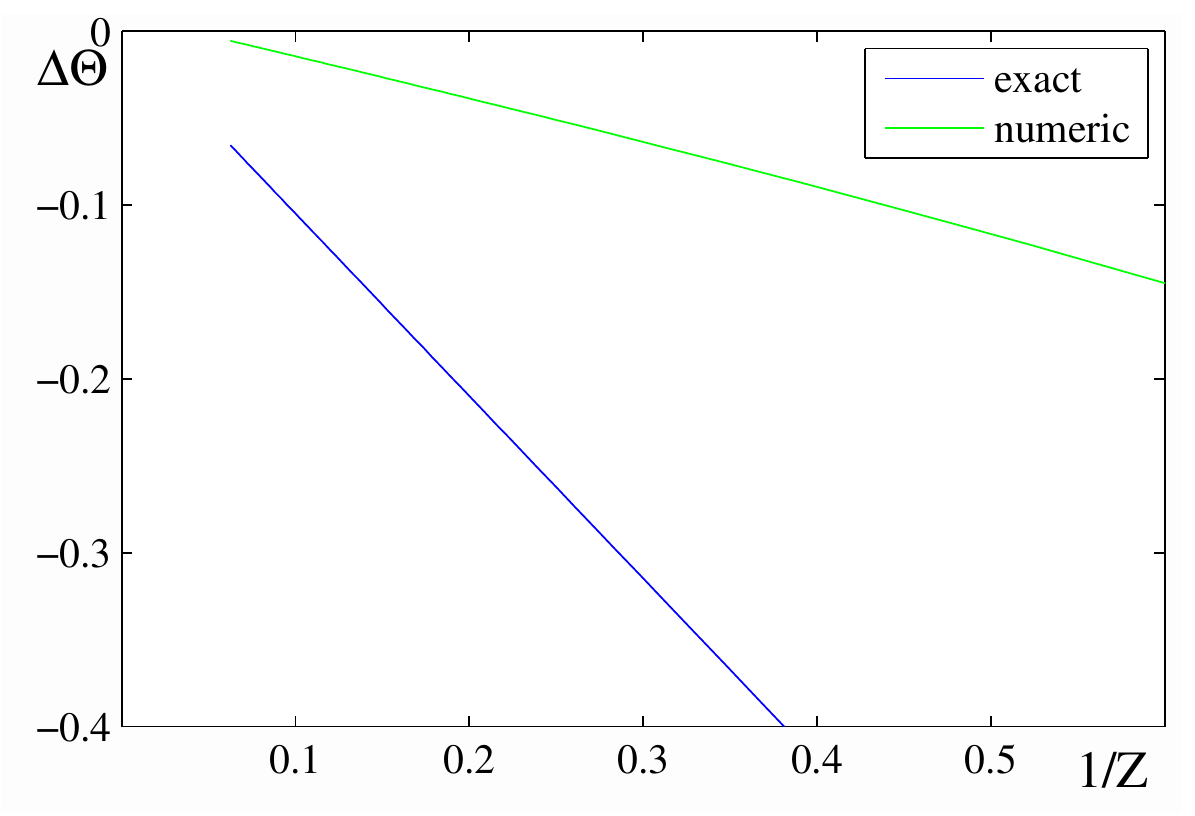}

\end{figure}

\newpage

\begin{figure}
\includegraphics[width=4cm]{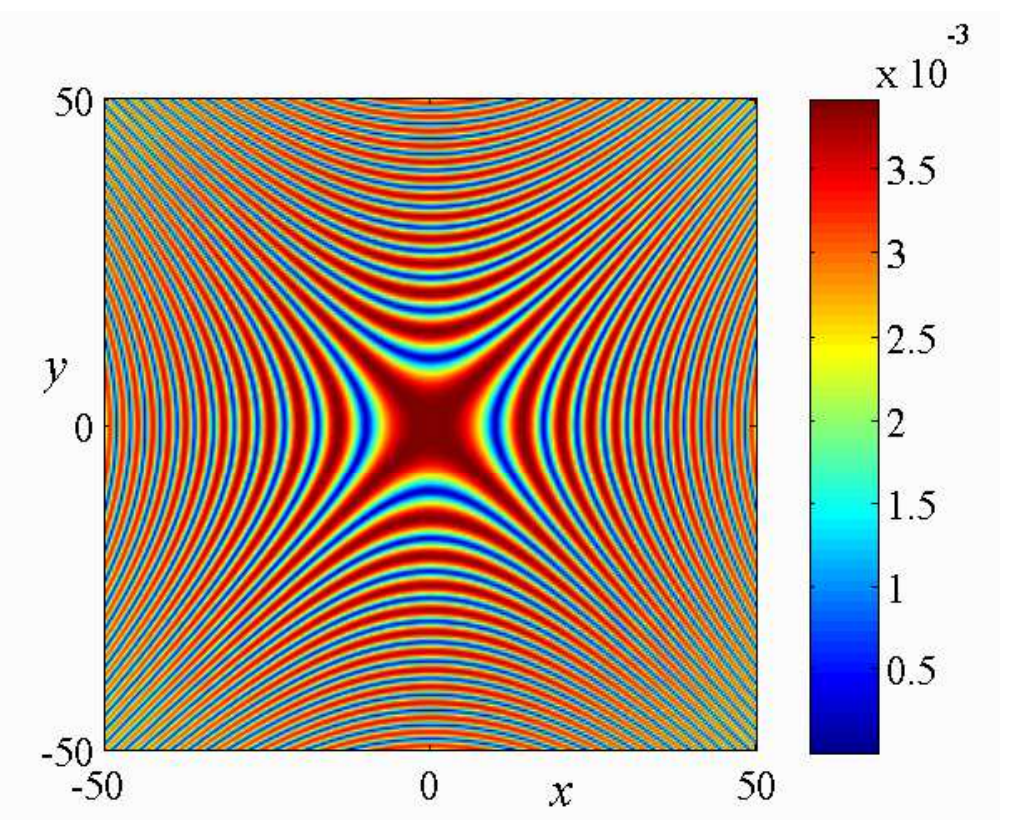}  \hspace{1cm}  \includegraphics[width=4cm]{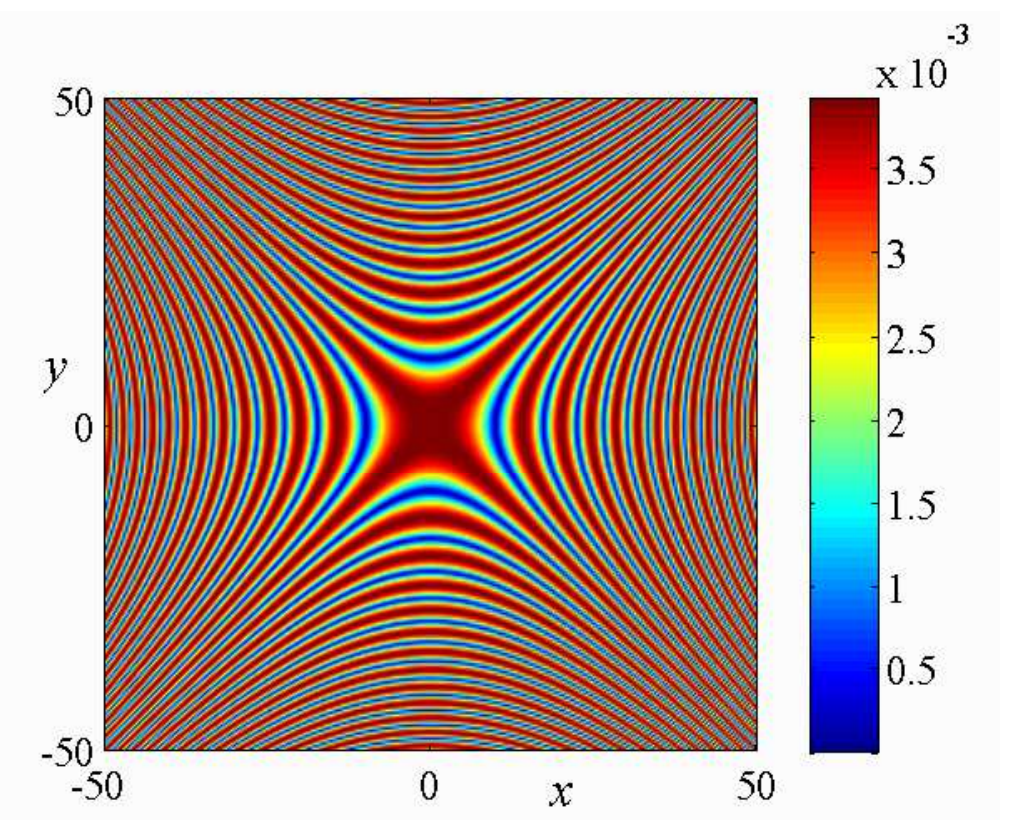} \hspace{1cm}  \includegraphics[width=4cm]{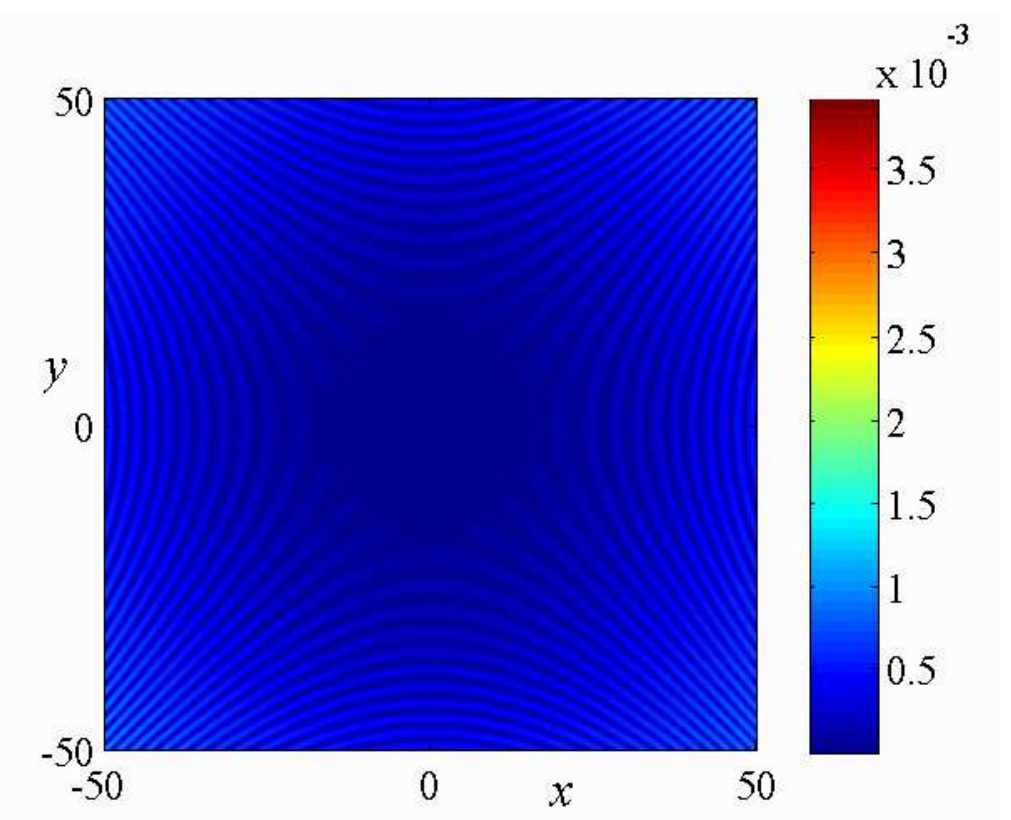}

\caption{Top:  Initial condition $e^{-4x^2-4y^2}$: Numerical solution, Exact similarity solution and absolute value of their difference at $Z=16$.  Bottom: Log-amplitude vs. $\log Z$, $\Delta\theta = \theta-\theta_0-s$ vs. $1/Z$.}

\includegraphics[width=4cm]{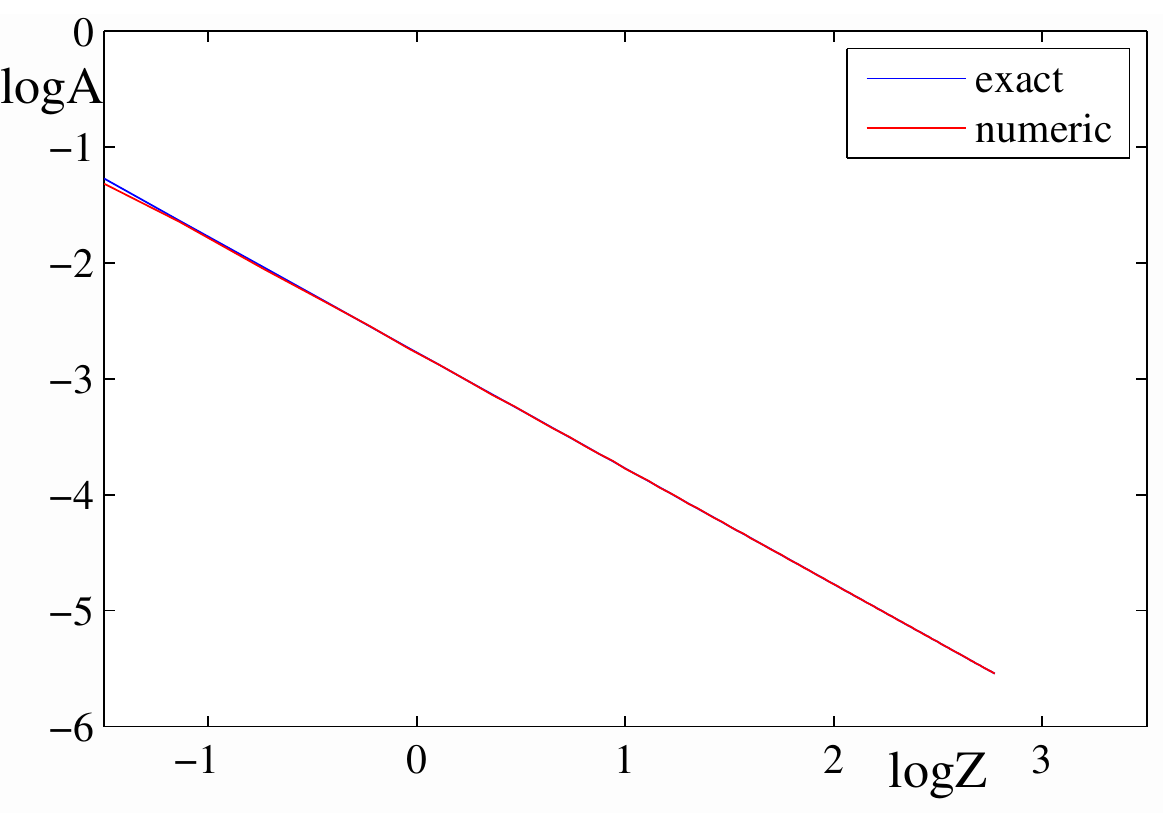}  \hspace{1cm}  \includegraphics[width=4cm]{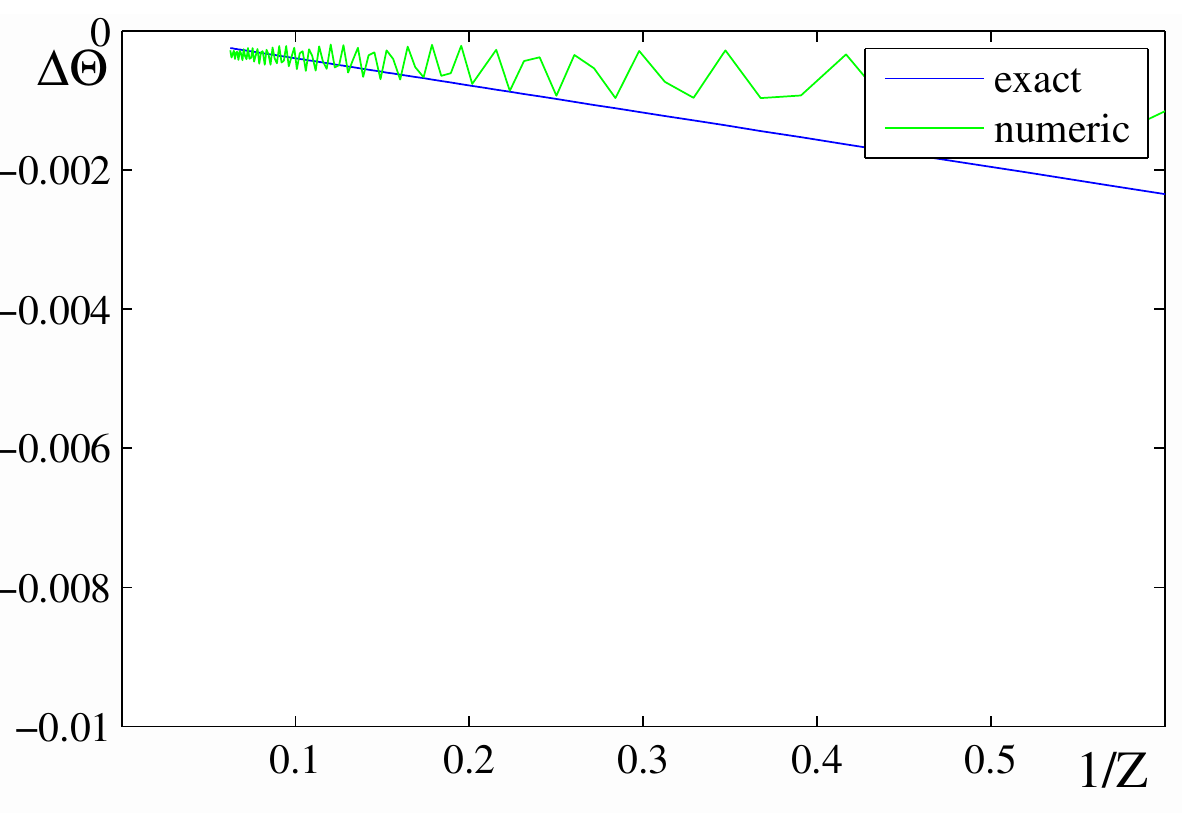}

\end{figure}

\begin{figure}
\includegraphics[width=4cm]{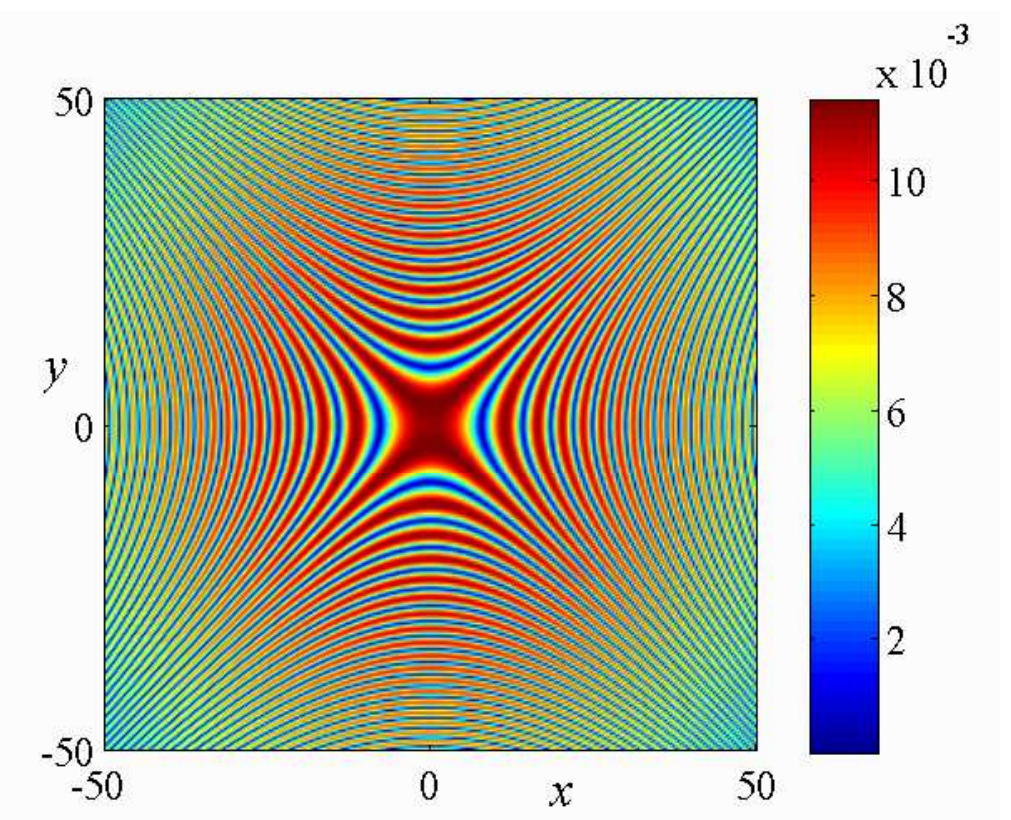}  \hspace{1cm}  \includegraphics[width=4cm]{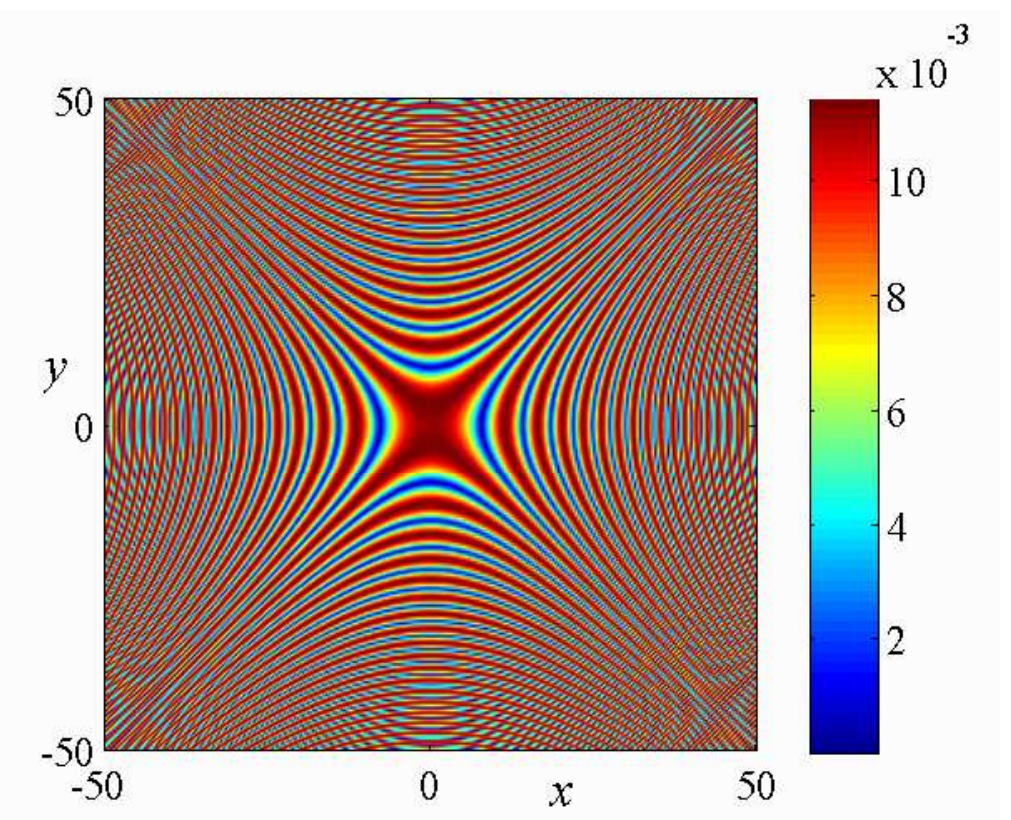} \hspace{1cm}  \includegraphics[width=4cm]{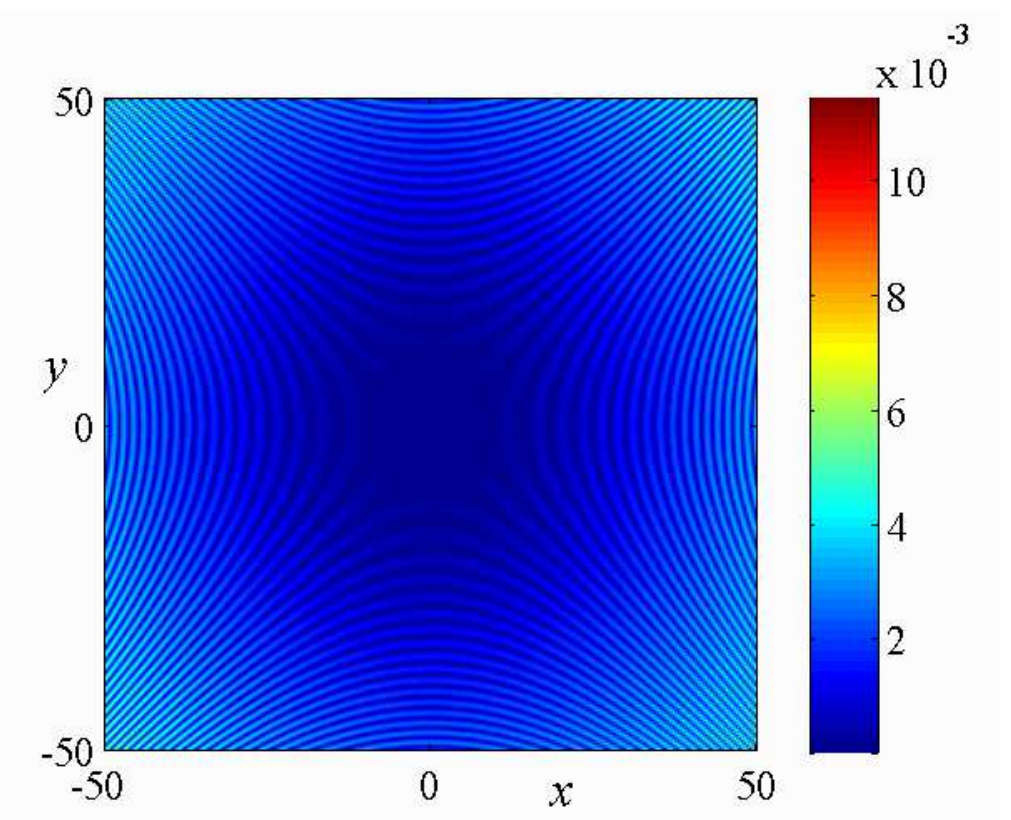}

\caption{Top:  Initial condition $2e^{-4x^2-4y^2}$: Numerical solution, Exact similarity solution and absolute value of their difference at $Z=16$.  Bottom: Log-amplitude vs. $\log Z$, $\Delta\theta = \theta-\theta_0-s$ vs. $1/Z$.}

\includegraphics[width=4cm]{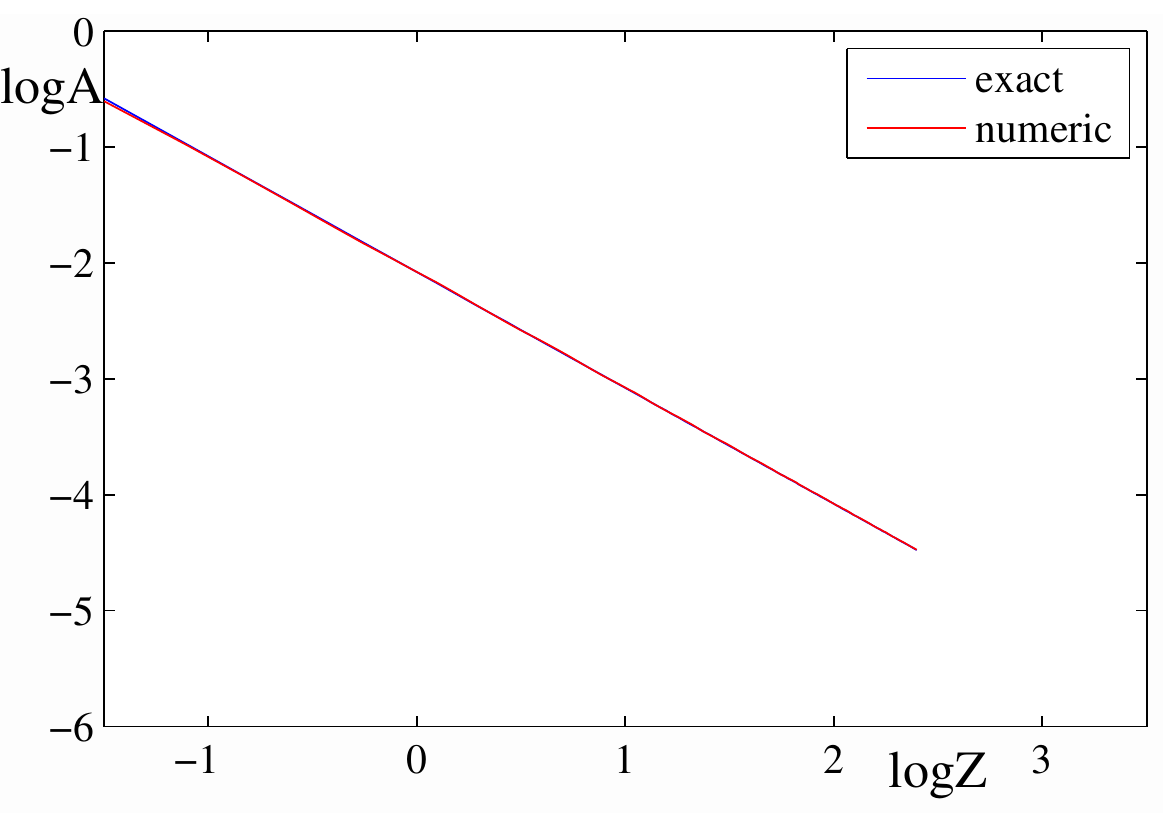}  \hspace{1cm}  \includegraphics[width=4cm]{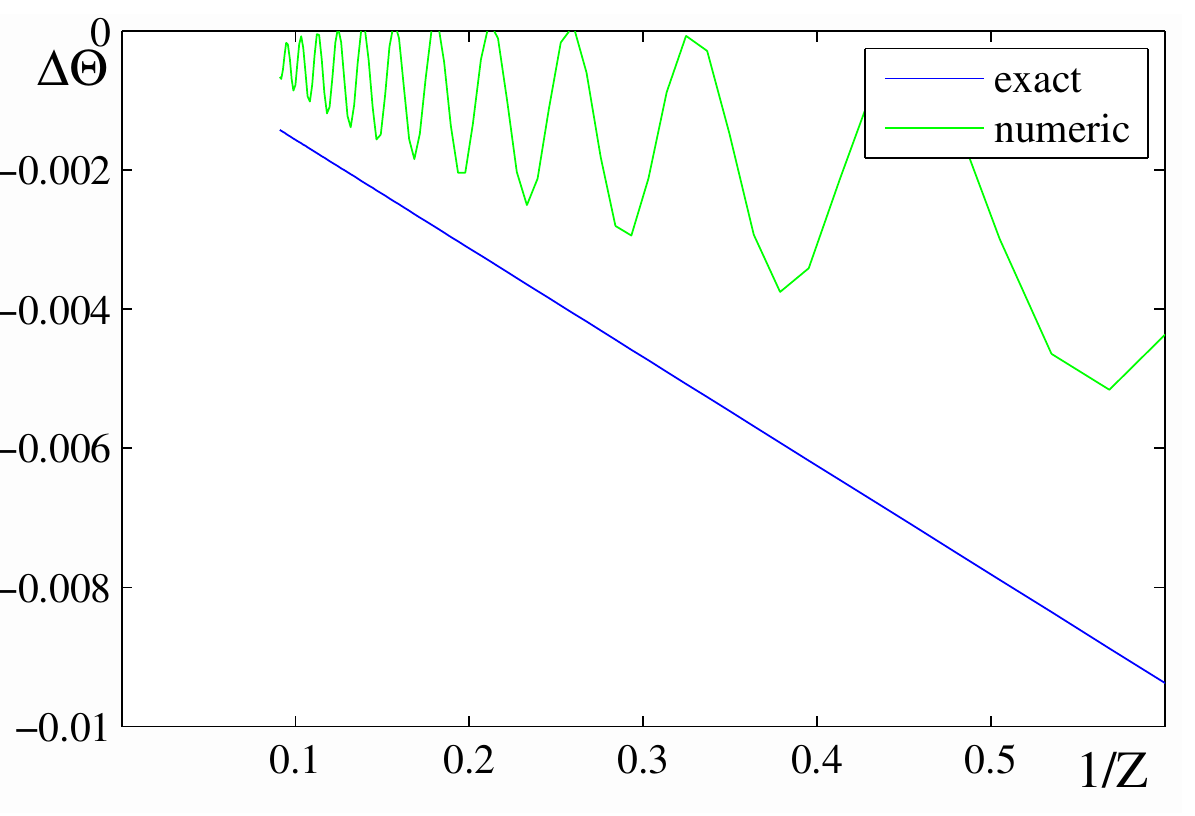}

\end{figure}

\newpage

\newpage

\begin{figure}
\includegraphics[width=4cm]{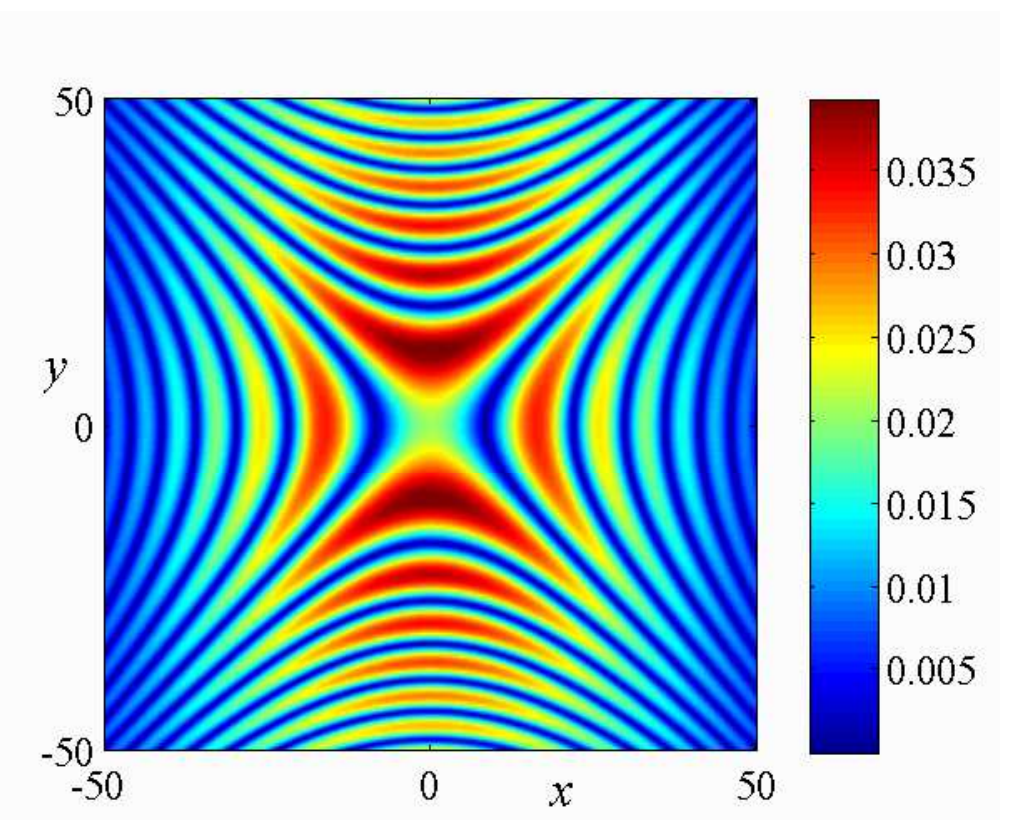}  \hspace{1cm}  \includegraphics[width=4cm]{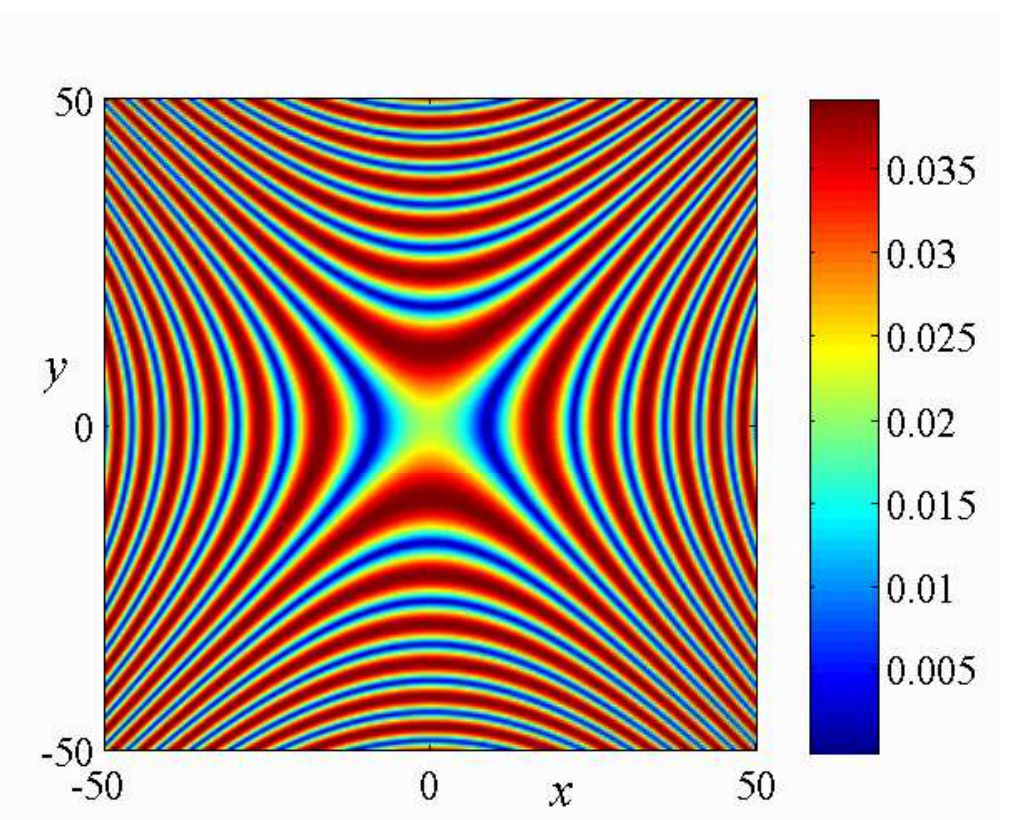} \hspace{1cm}  \includegraphics[width=4cm]{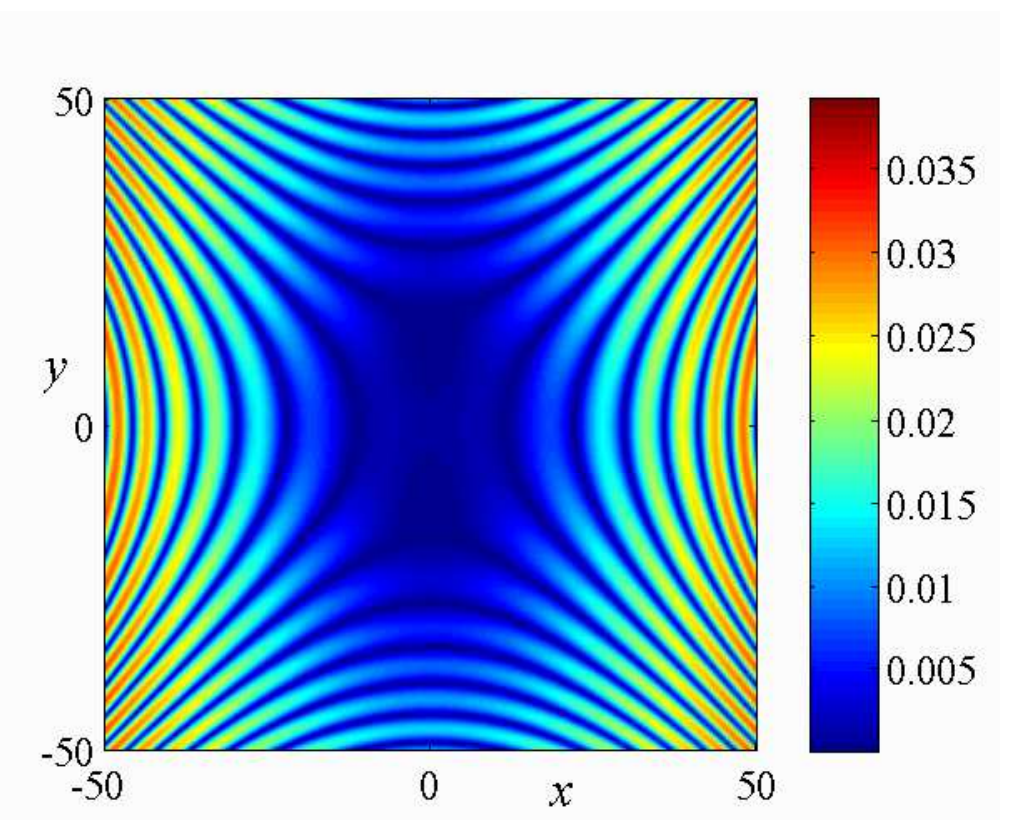}

\caption{Top:  Initial condition $e^{-0.2x^2-0.2y^2}$: Numerical solution, Exact similarity solution and absolute value of their difference at $Z=24$.  Bottom: Log-amplitude vs. $\log Z$, $\Delta\theta = \theta-\theta_0-s$ vs. $1/Z$.}

\includegraphics[width=4cm]{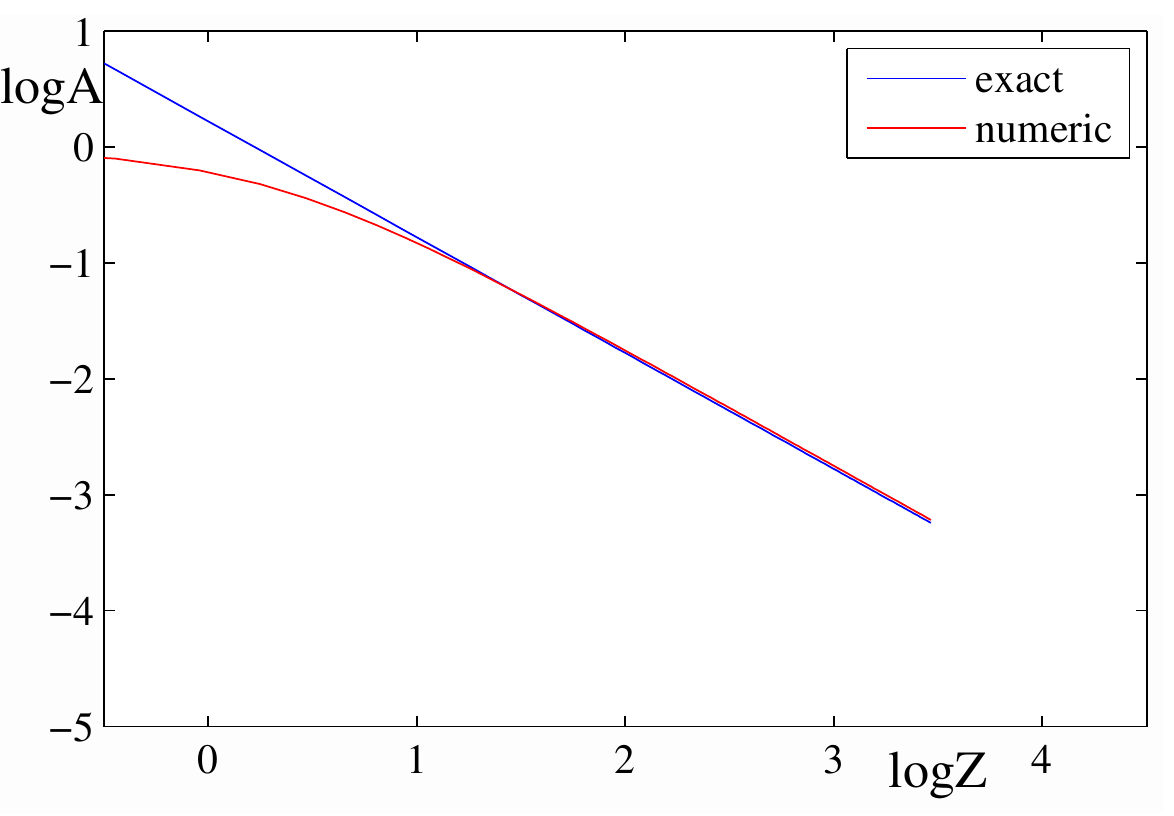}  \hspace{1cm}  \includegraphics[width=4cm]{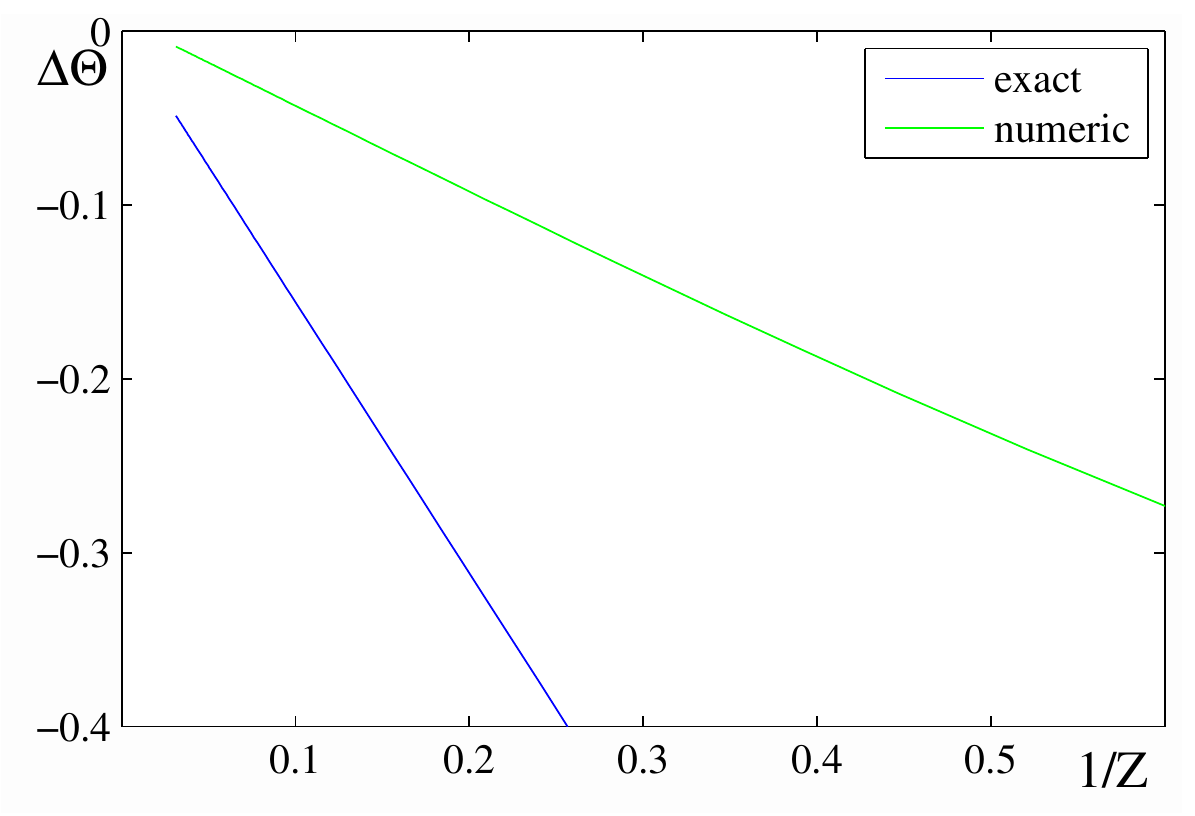}

\end{figure}

\begin{figure}
\includegraphics[width=4cm]{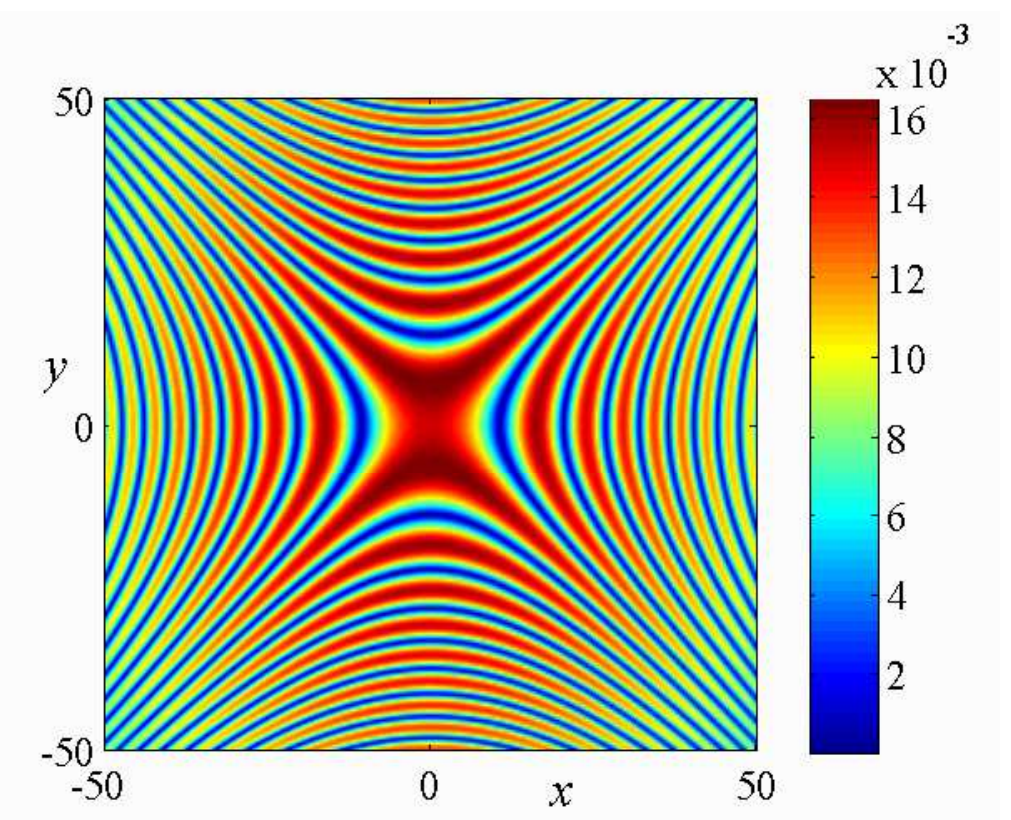}   \hspace{1cm}  \includegraphics[width=4cm]{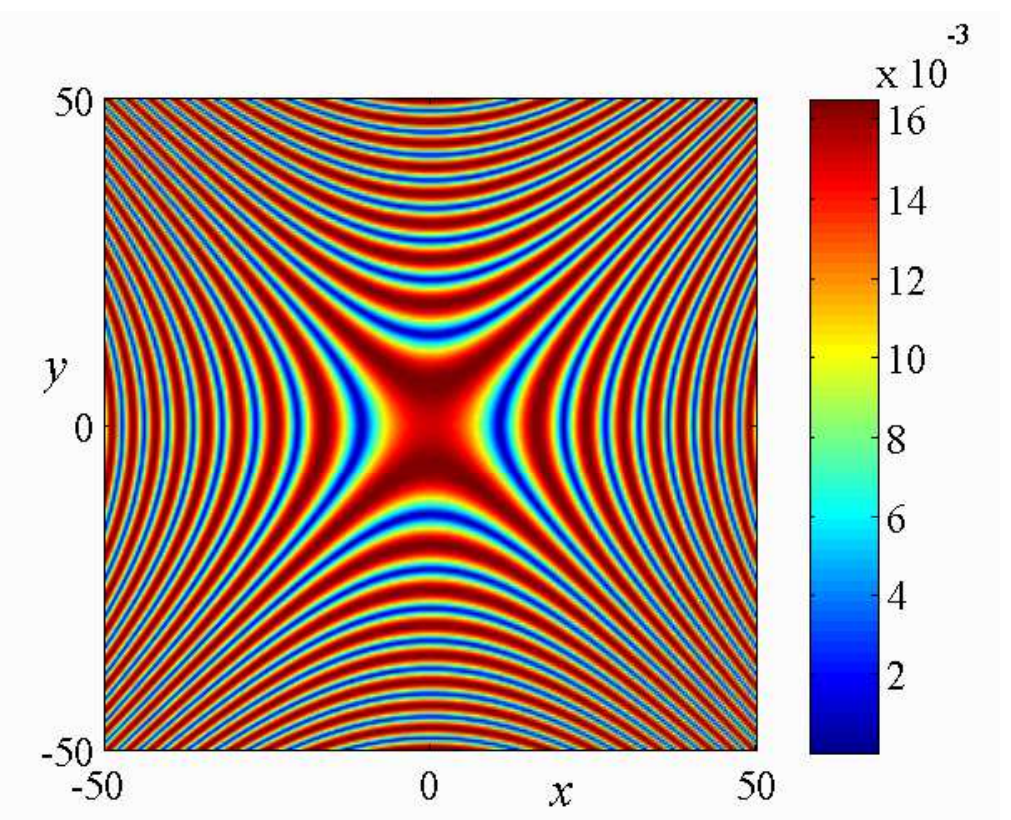}  \hspace{1cm}  \includegraphics[width=4cm]{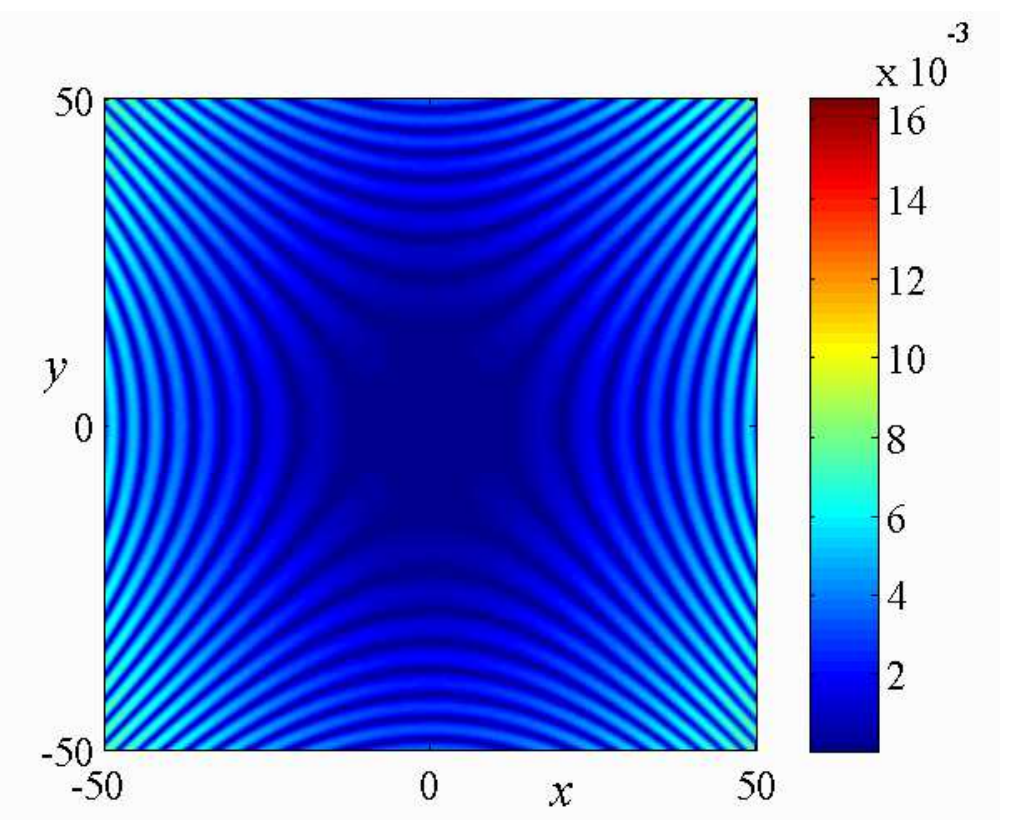}

\caption{Top:  Initial condition $\text{sech}(x^2+y^2)$: Numerical solution, Exact similarity solution and absolute value of their difference at $Z=24$.  Bottom: Log-amplitude vs. $\log Z$, $\Delta\theta = \theta-\theta_0-s$ vs. $1/Z$.}

\includegraphics[width=4cm]{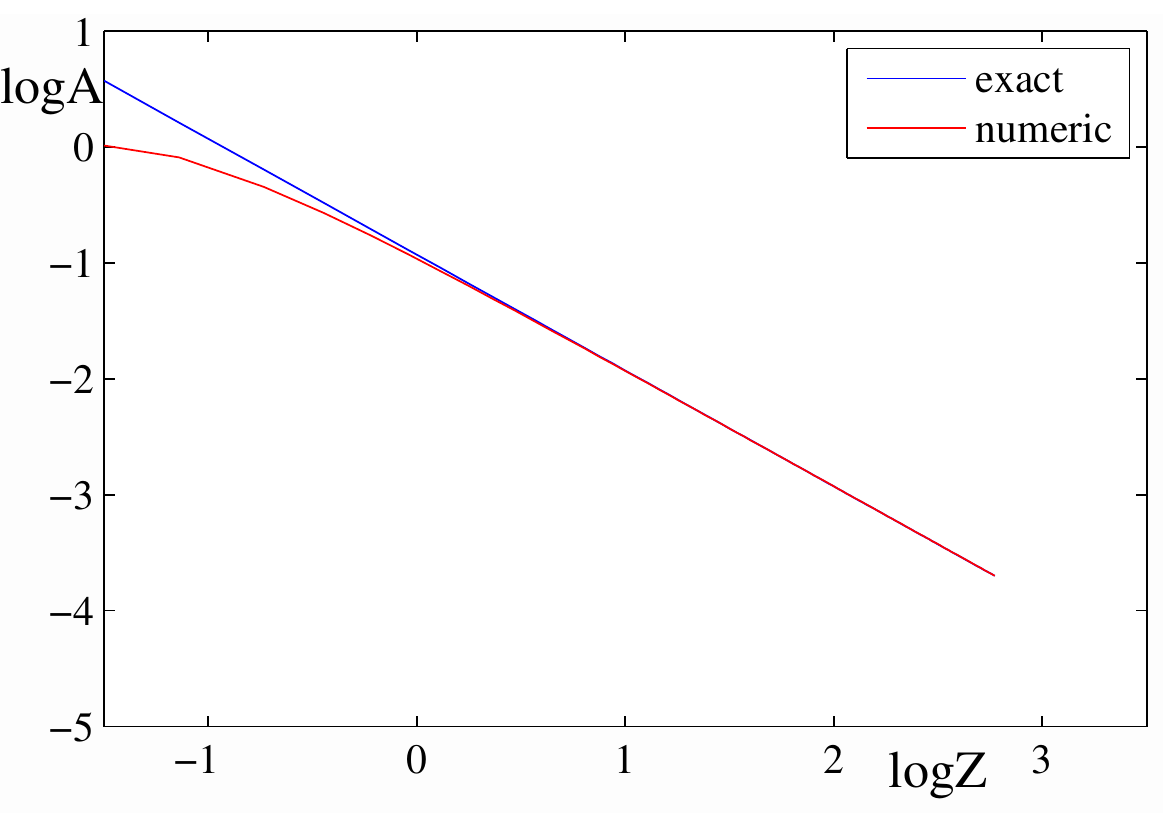}  \hspace{1cm}  \includegraphics[width=4cm]{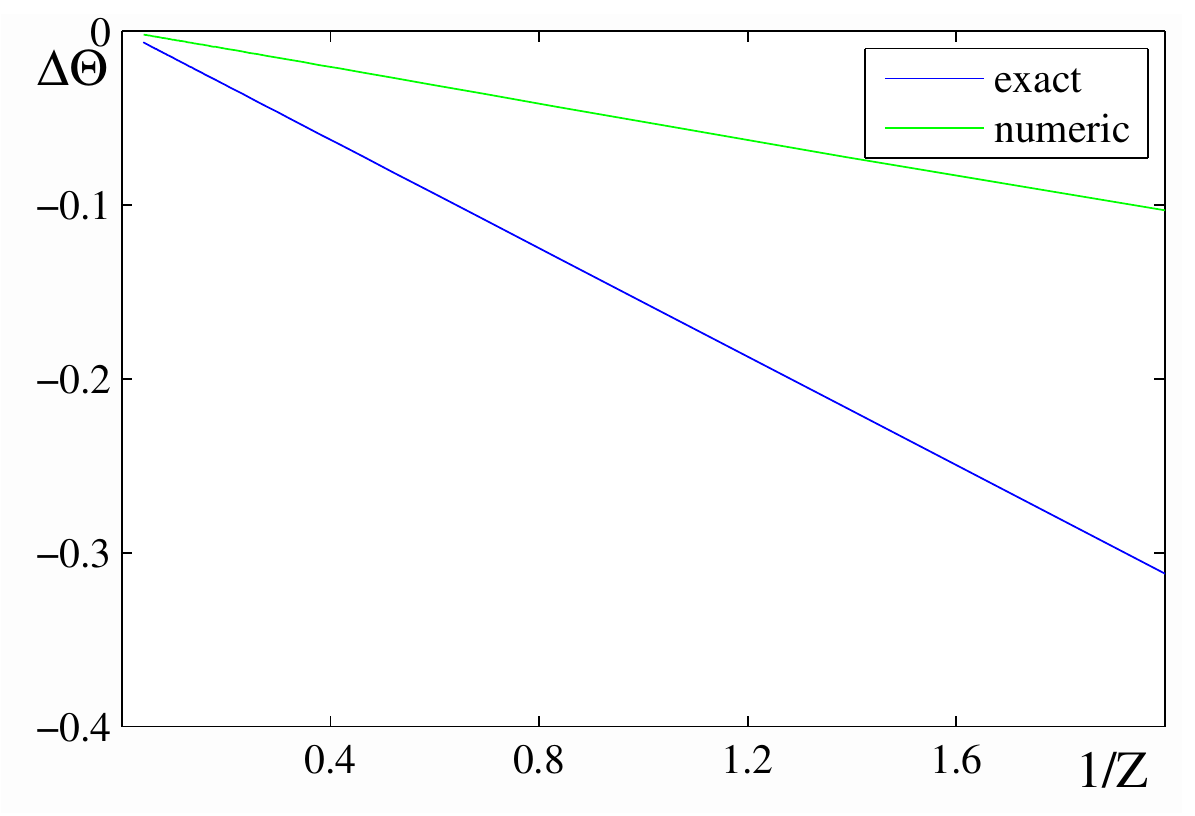}
\end{figure}

\newpage

\newpage

\begin{figure}
\includegraphics[width=4cm]{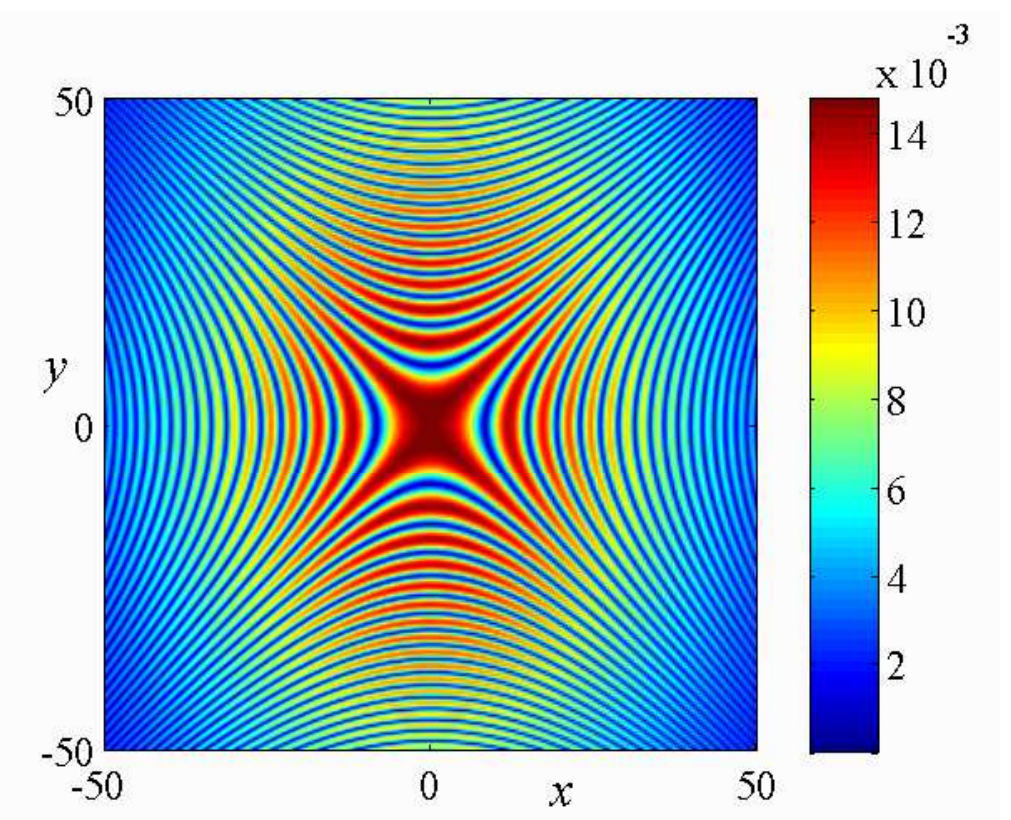}  \hspace{1cm}  \includegraphics[width=4cm]{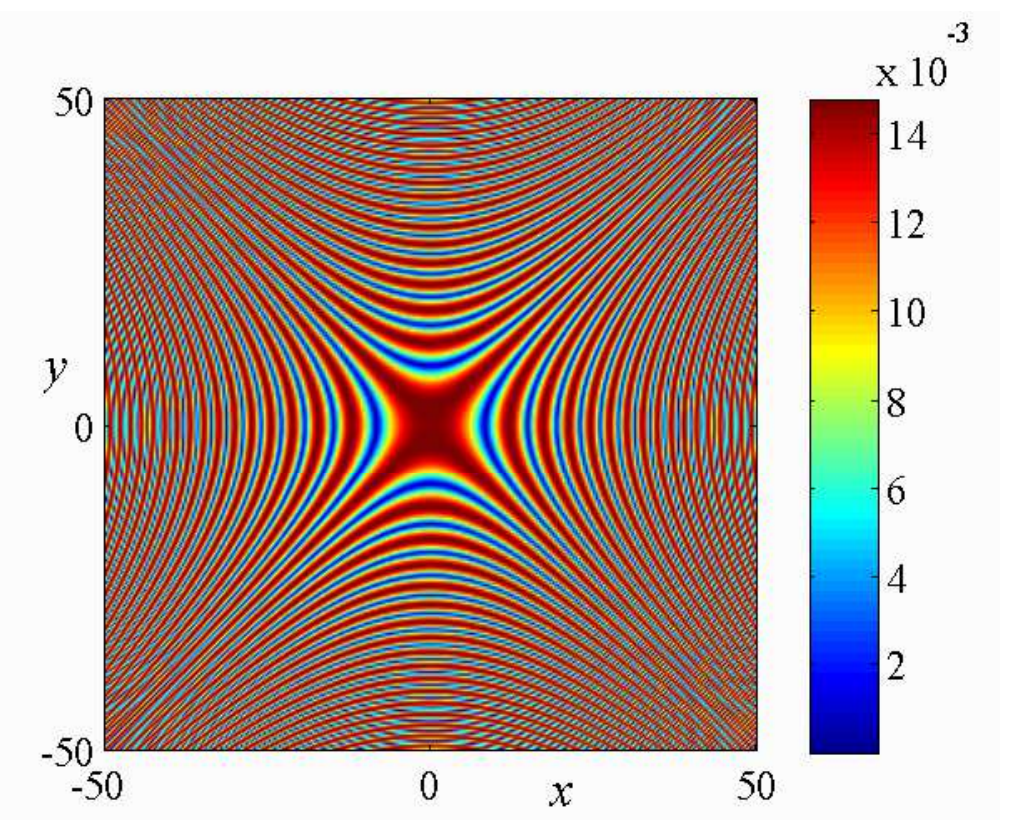} \hspace{1cm}  \includegraphics[width=4cm]{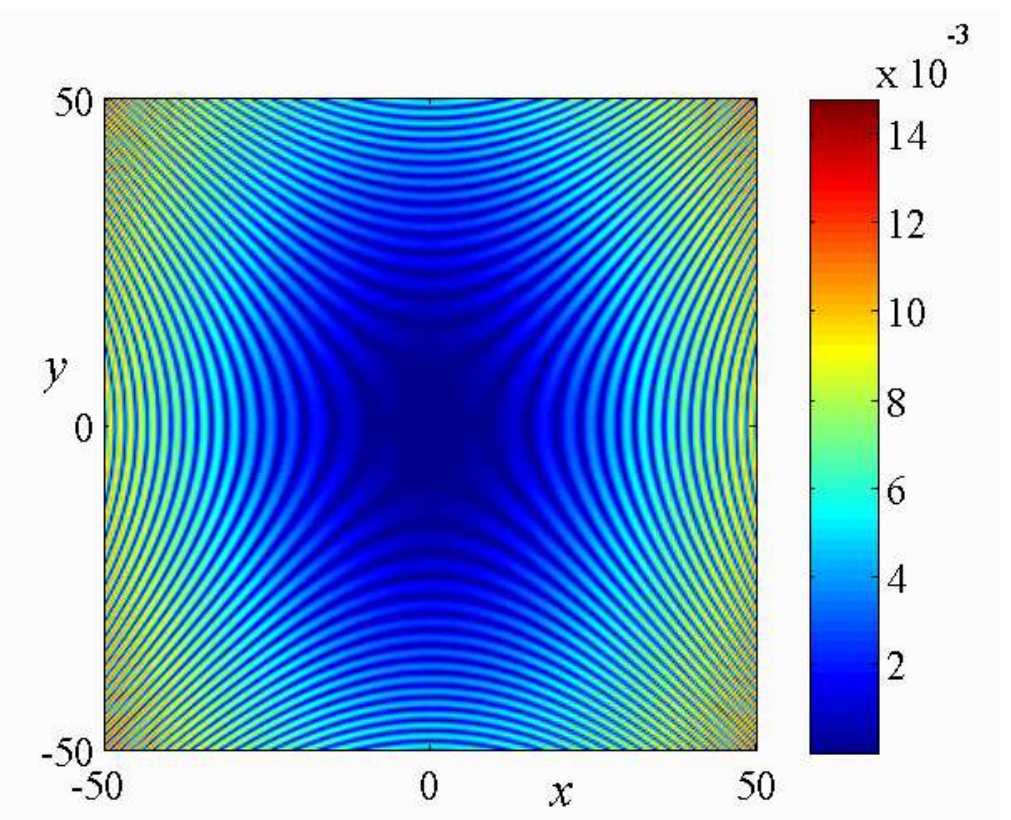}

\caption{Top: Initial condition $e^{-x^2-2y^2}$: Numerical solution, Exact similarity solution and absolute value of their difference at $Z=16$.  Bottom: Log-amplitude vs. $\log Z$, $\Delta\theta = \theta-\theta_0-s$ vs. $1/Z$.}

\includegraphics[width=4cm]{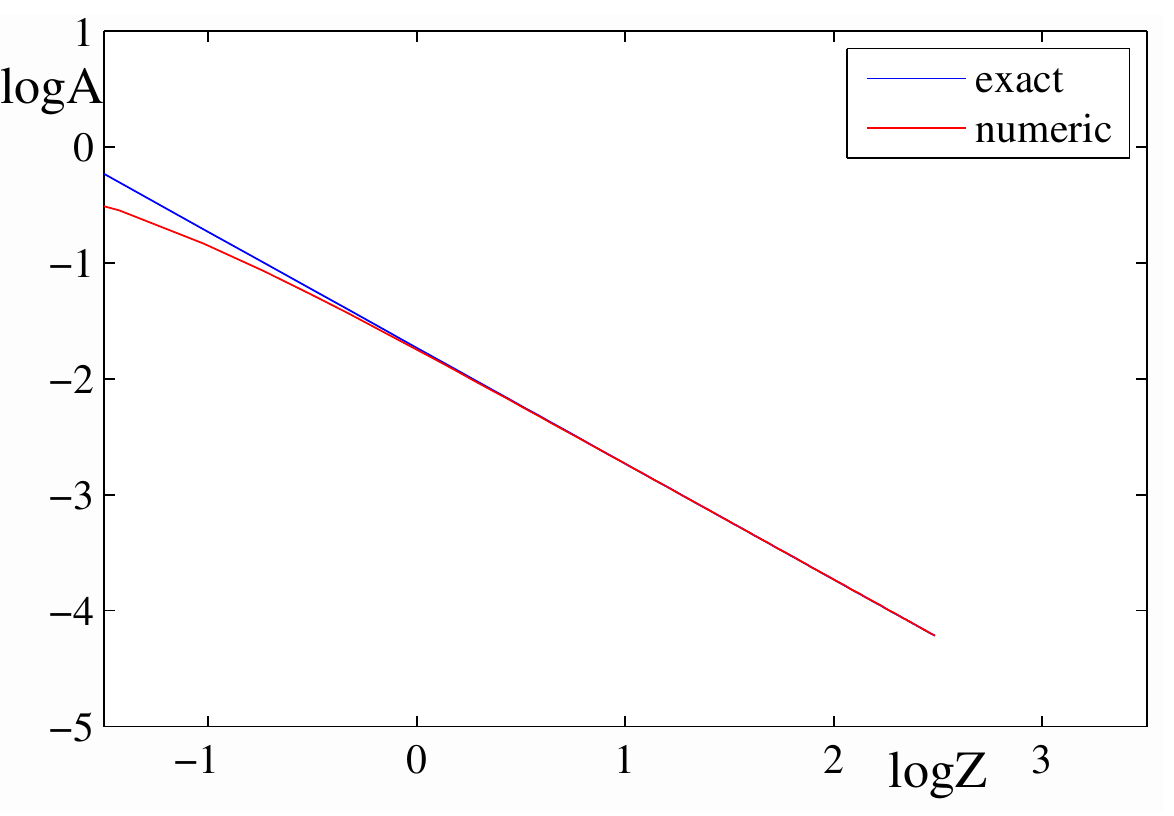}  \hspace{1cm}  \includegraphics[width=4cm]{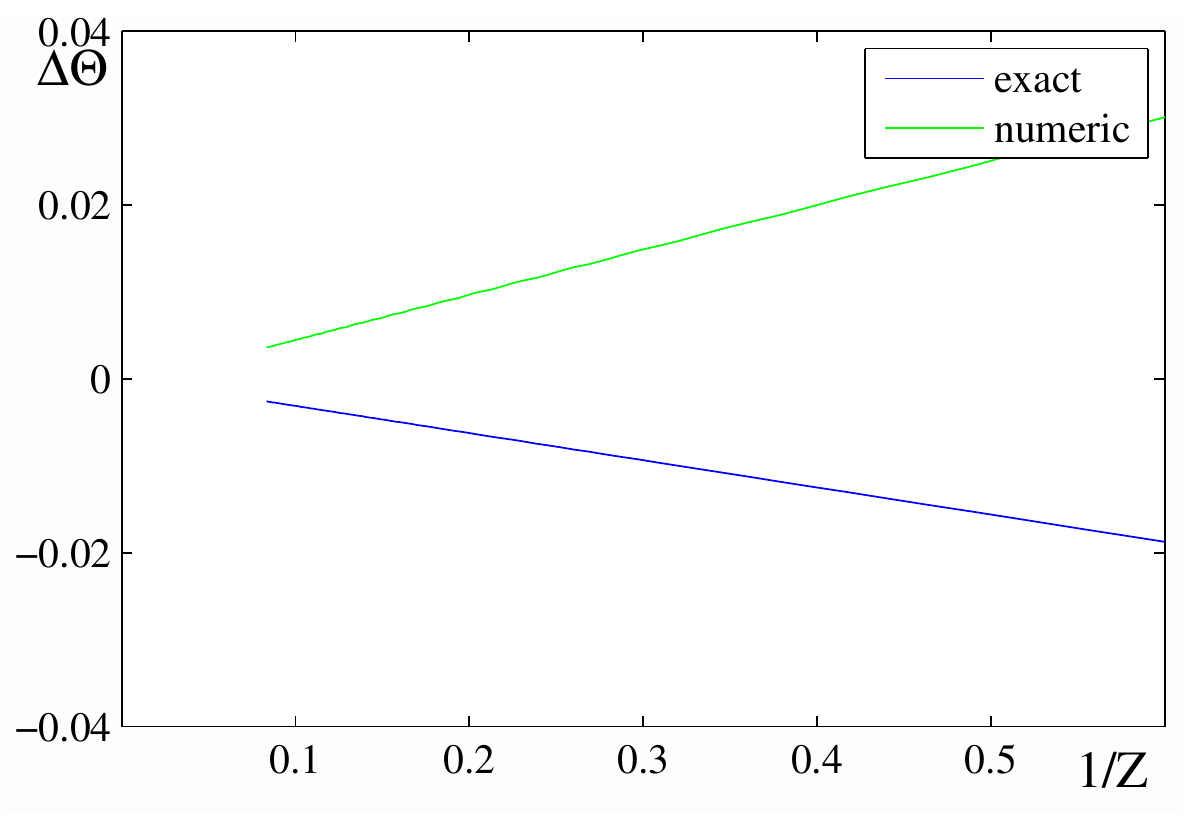}

\end{figure}

\begin{figure}
\includegraphics[width=4cm]{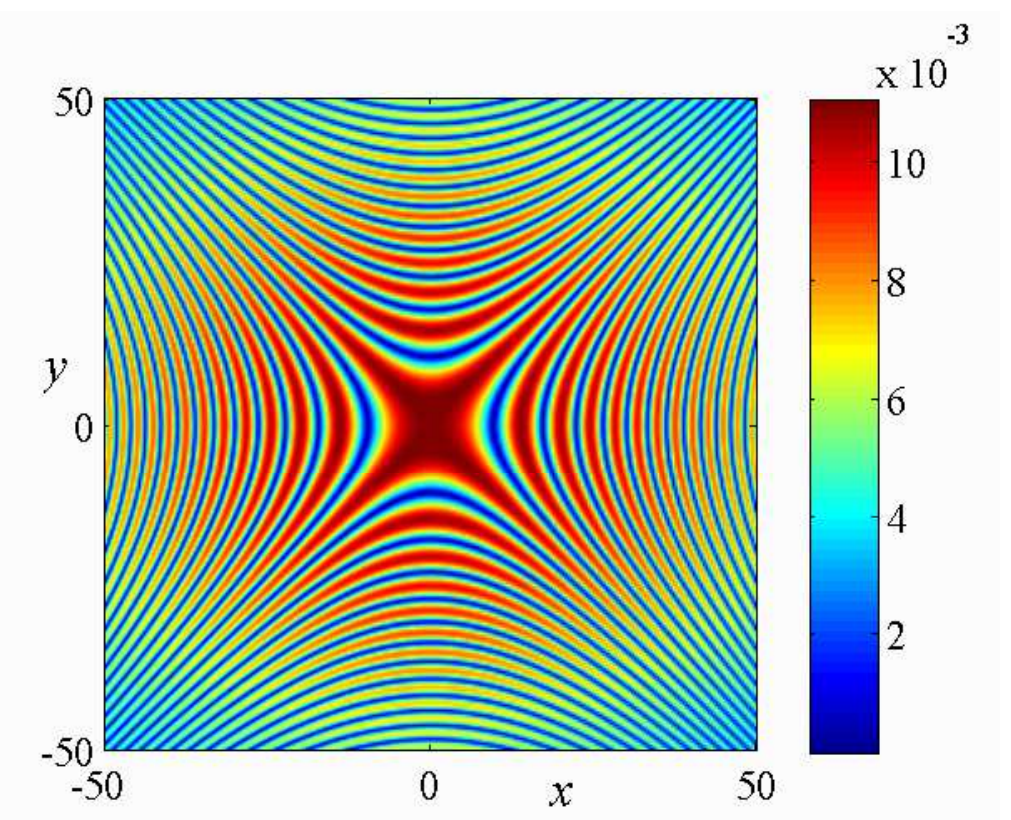}  \hspace{1cm}  \includegraphics[width=4cm]{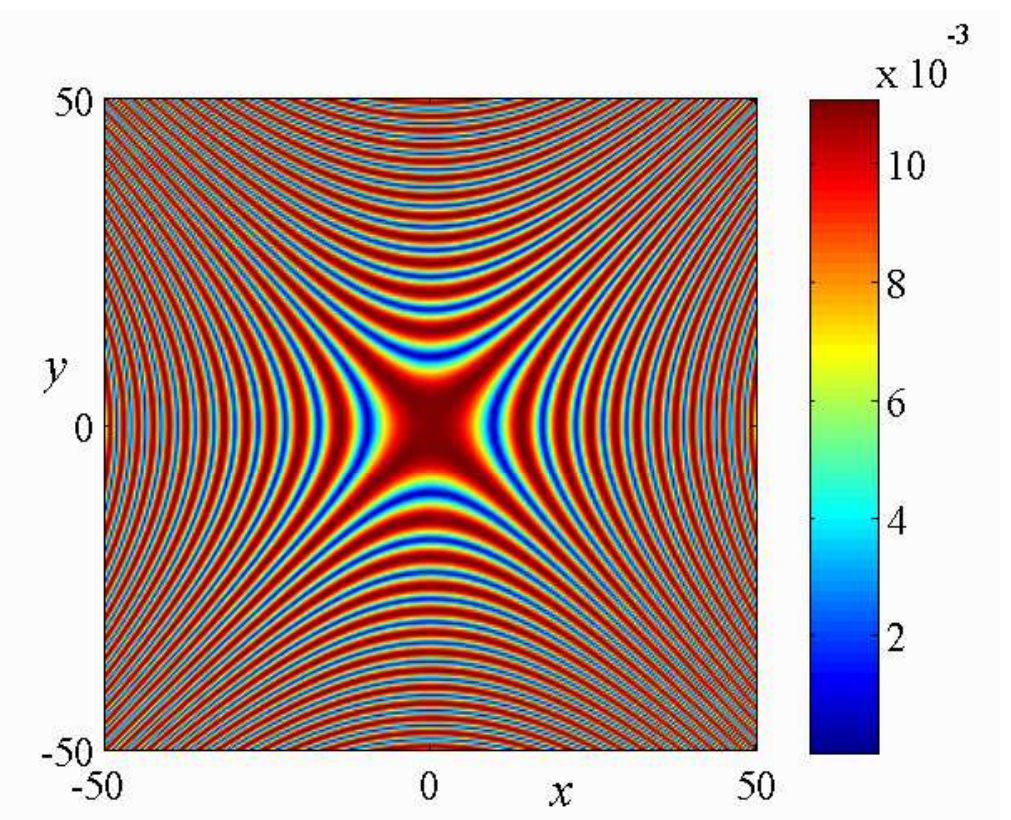} \hspace{1cm}  \includegraphics[width=4cm]{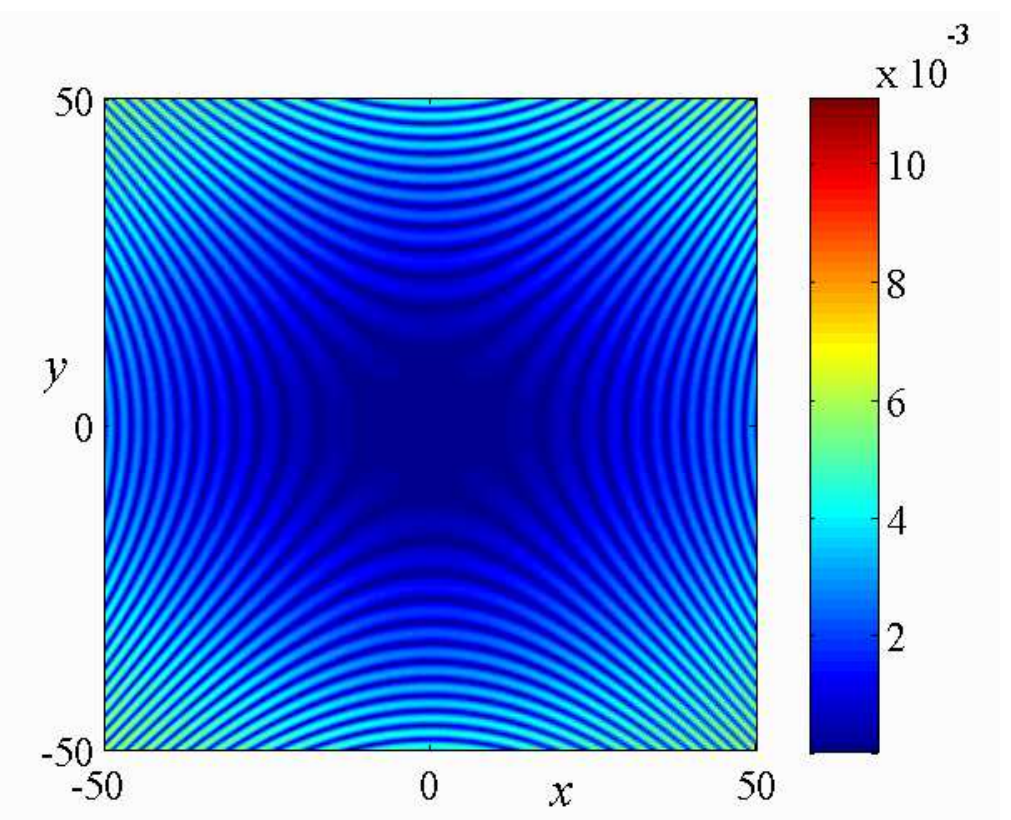}

\caption{Top: Initial condition $e^{-2x^2-y^2}$: Numerical solution, Exact similarity solution and absolute value of their difference at $Z=16$.  Bottom: Log-amplitude vs. $\log Z$, $\Delta\theta = \theta-\theta_0-s$ vs. $1/Z$.}

\includegraphics[width=4cm]{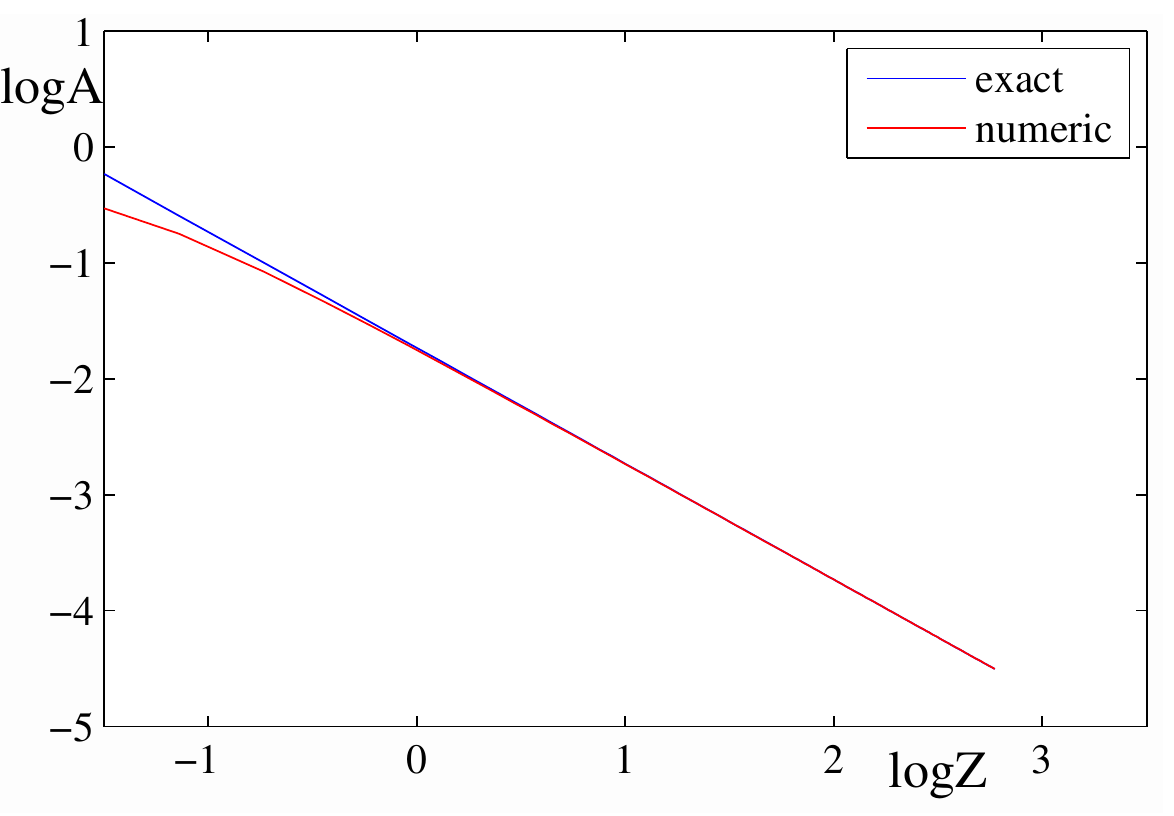}  \hspace{1cm}  \includegraphics[width=4cm]{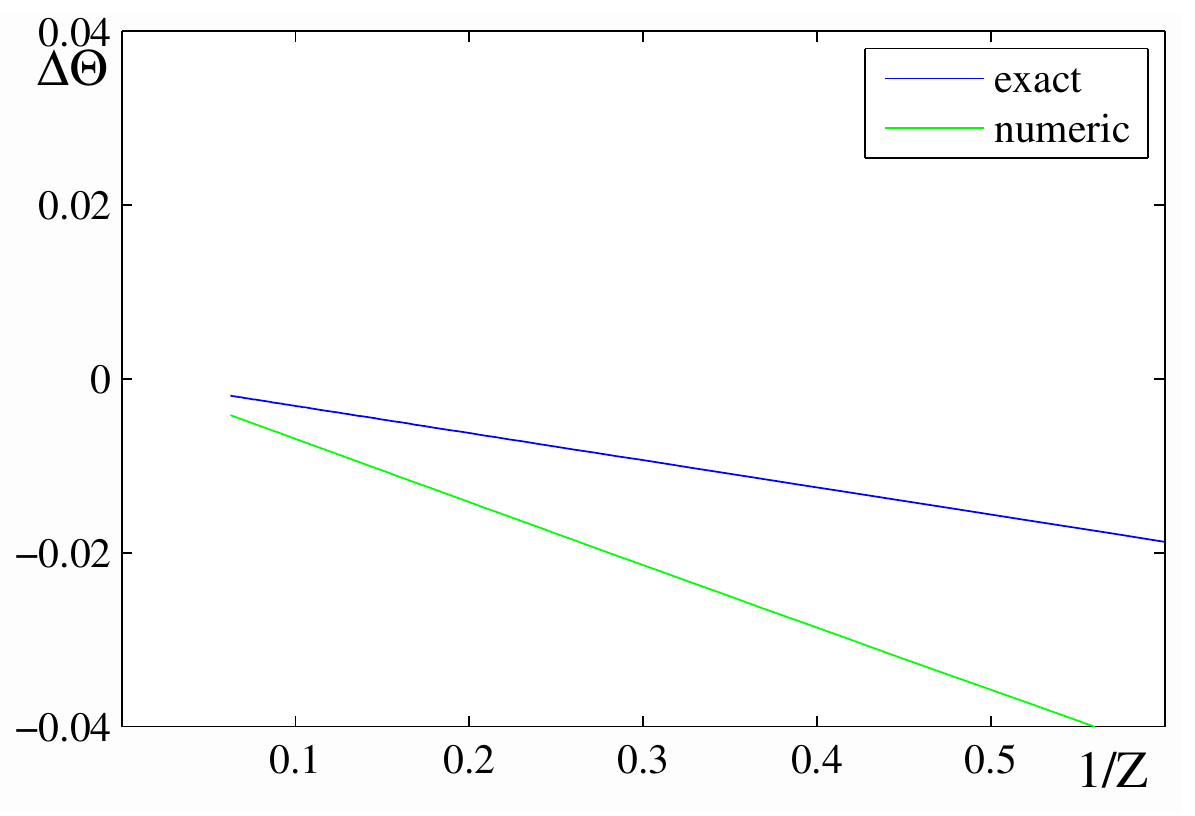}

\end{figure}

\newpage

\begin{figure}
\includegraphics[width=4cm]{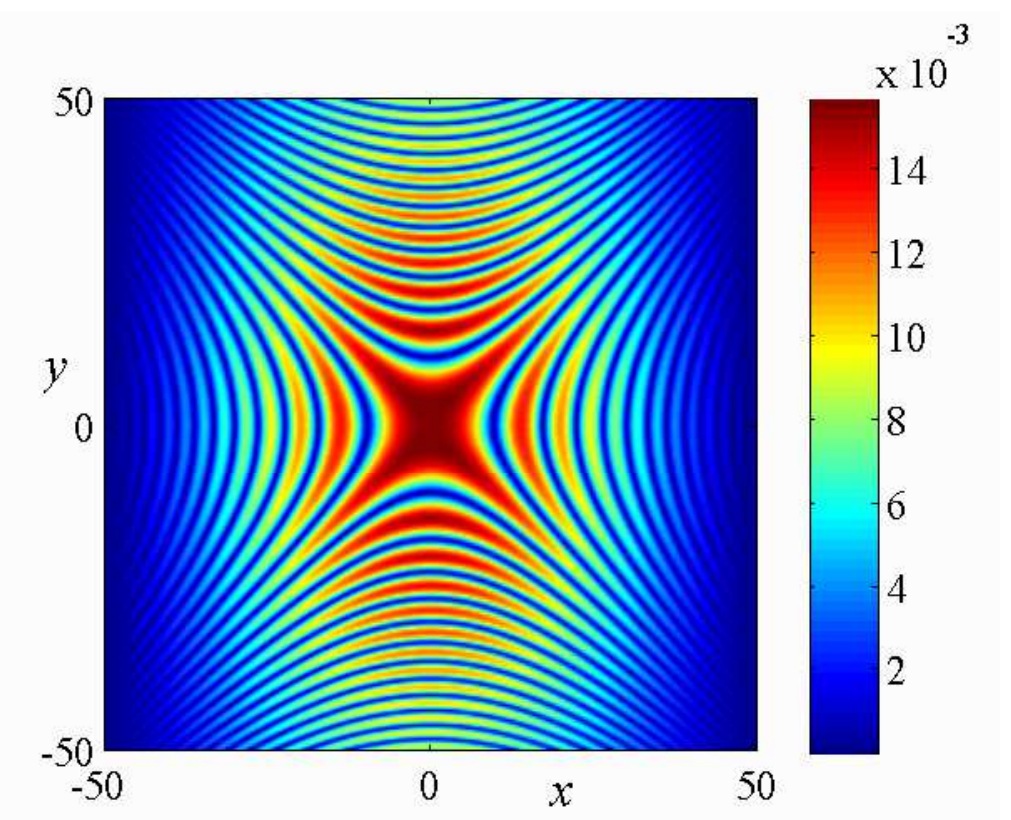}  \hspace{1cm}  \includegraphics[width=4cm]{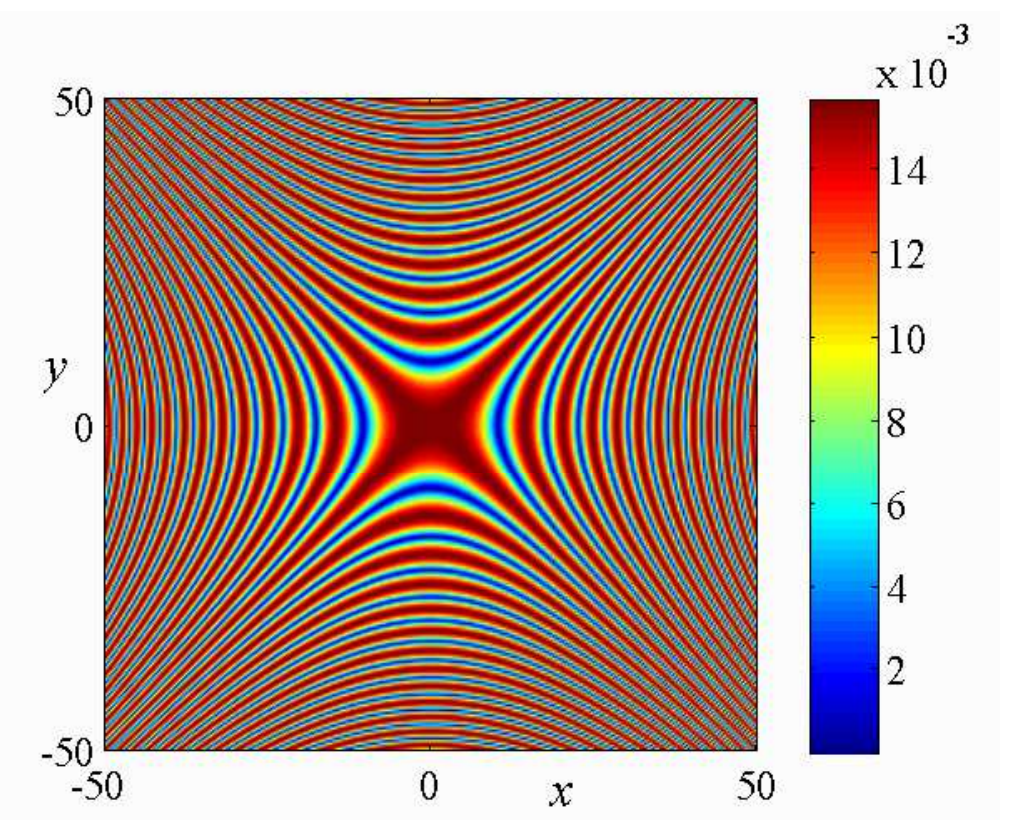} \hspace{1cm}  \includegraphics[width=4cm]{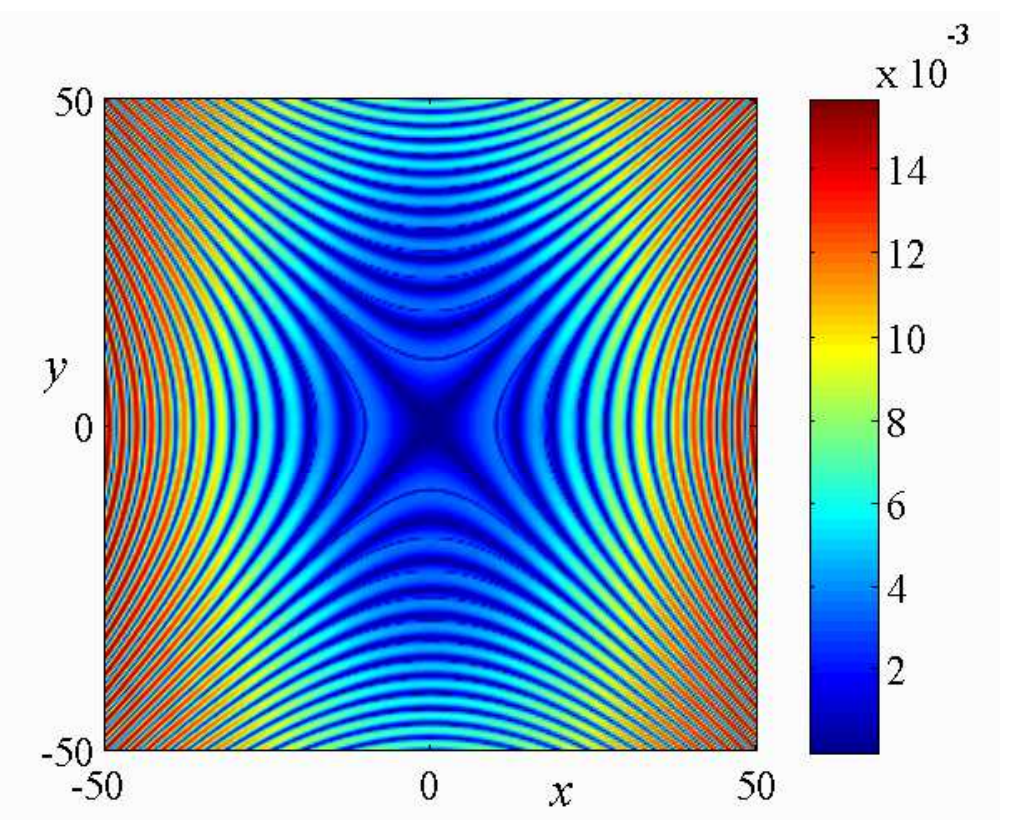}

\caption{Top: Two Gaussian peaks $0.5(e^{-(x+1)^2-y^2}+e^{-(x-1)^2-y^2})$ initial condition: Numerical solution, Exact similarity solution and absolute value of their difference at $Z=16$.  Bottom: Log-amplitude vs. $\log Z$, $\Delta\theta = \theta-\theta_0-s$ vs. $1/Z$.}

\includegraphics[width=4cm]{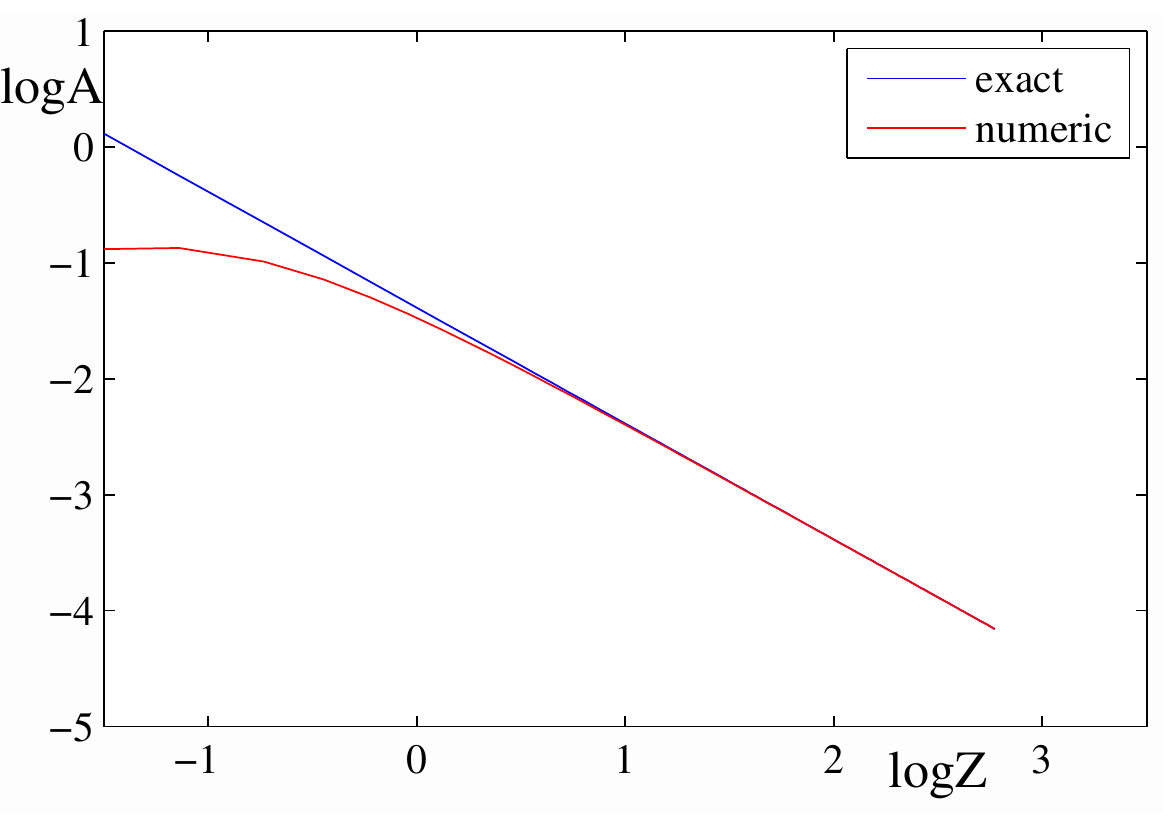}  \hspace{1cm}  \includegraphics[width=4cm]{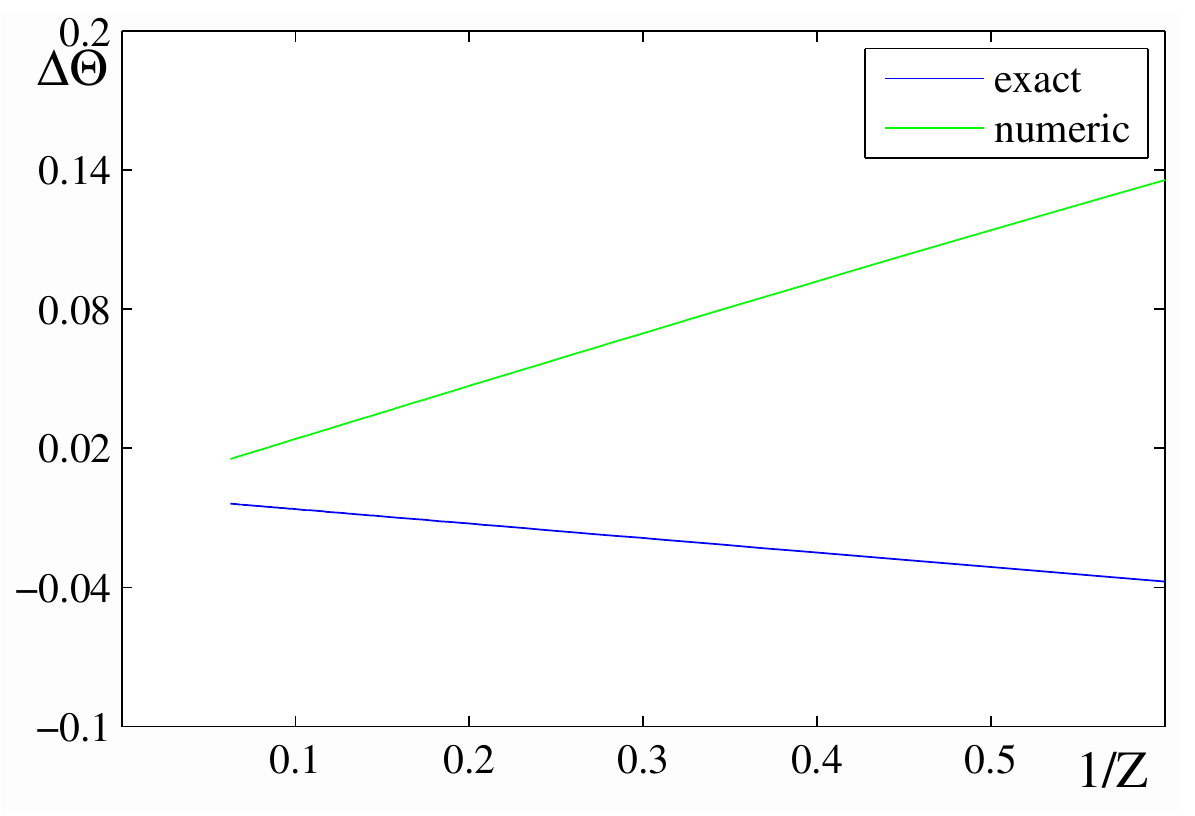}

\end{figure}

\begin{figure}
\includegraphics[width=4cm]{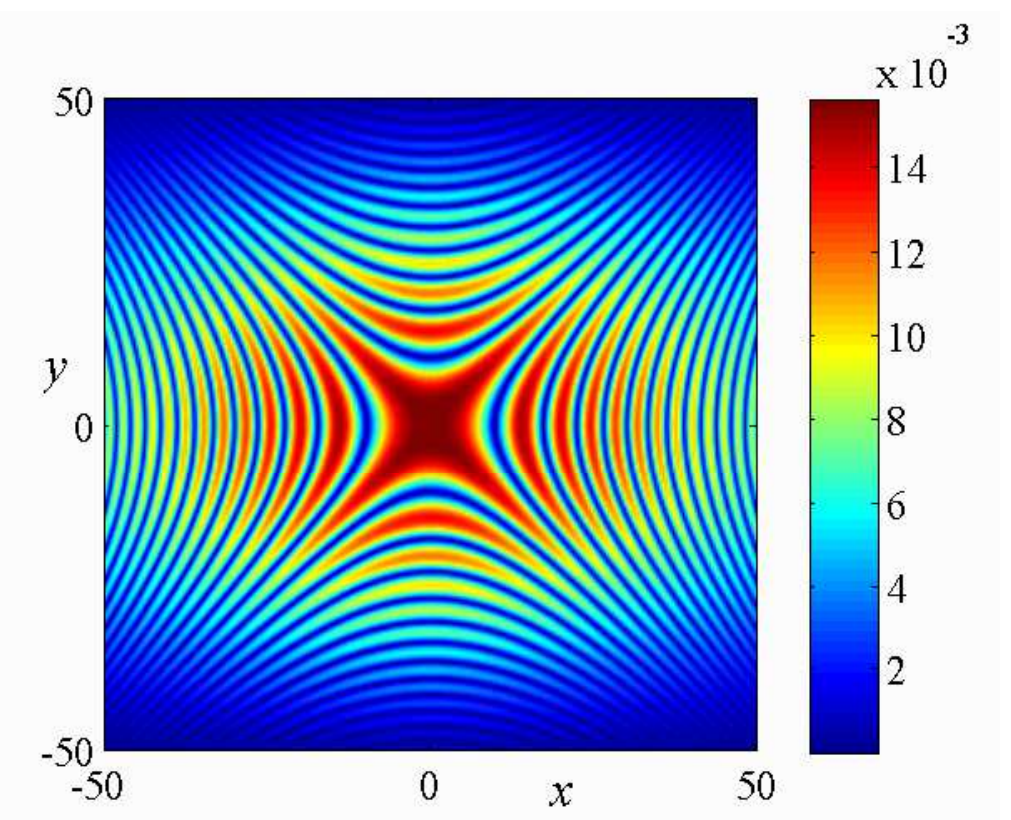}  \hspace{1cm}  \includegraphics[width=4cm]{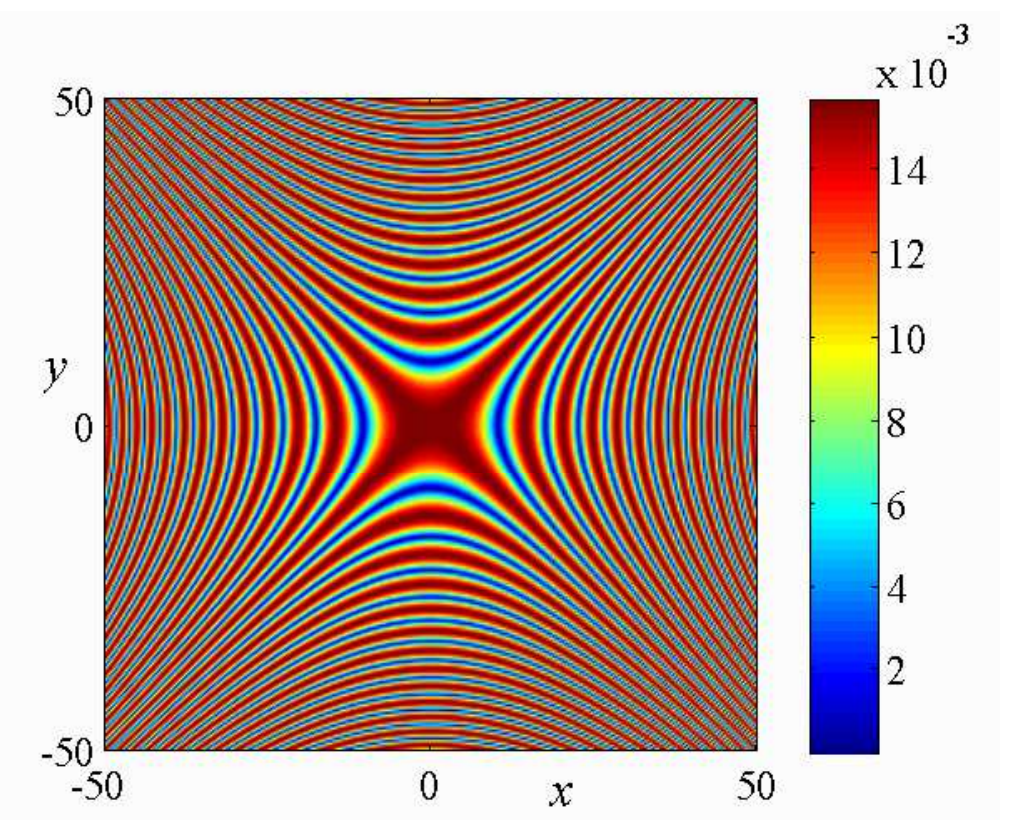} \hspace{1cm}  \includegraphics[width=4cm]{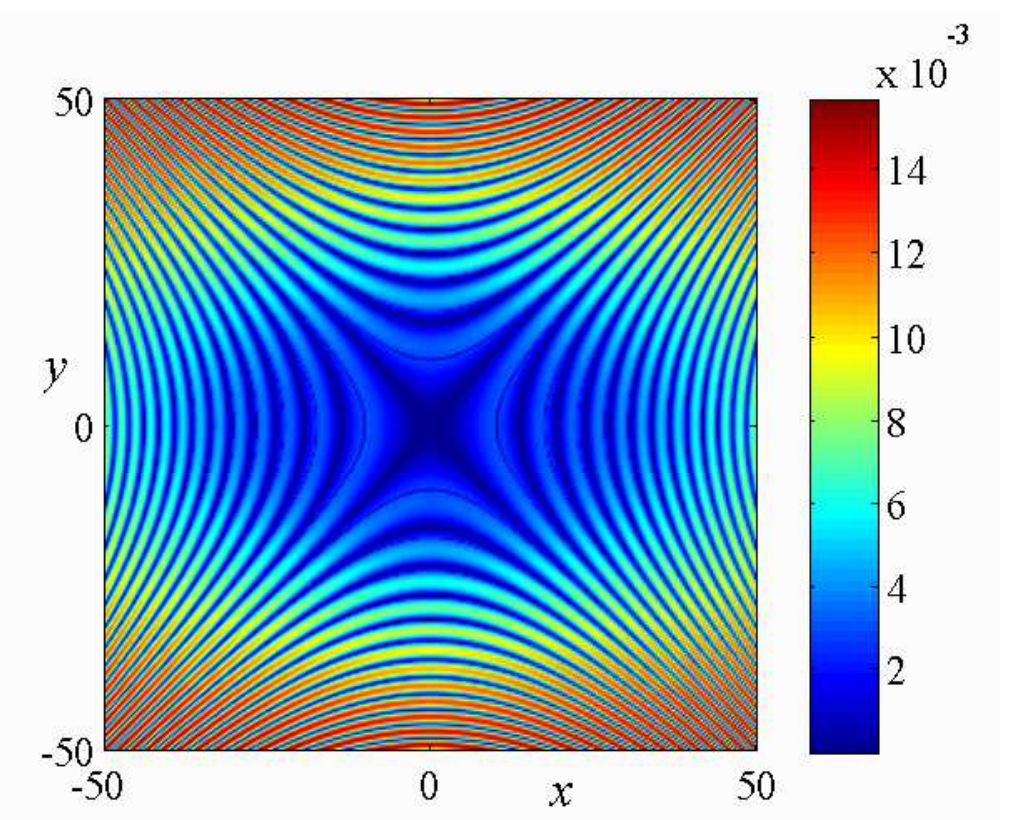}

\caption{Top:  Two Gaussian peaks $0.5(e^{-x^2-(y+1)^2}+e^{-x^2-(y-1)^2})$ initial condition: Numerical solution, Exact similarity solution and absolute value of their difference at $Z=16$.  Bottom: Log-amplitude vs. $\log Z$, $\Delta\theta = \theta-\theta_0-s$ vs. $1/Z$.}

\includegraphics[width=4cm]{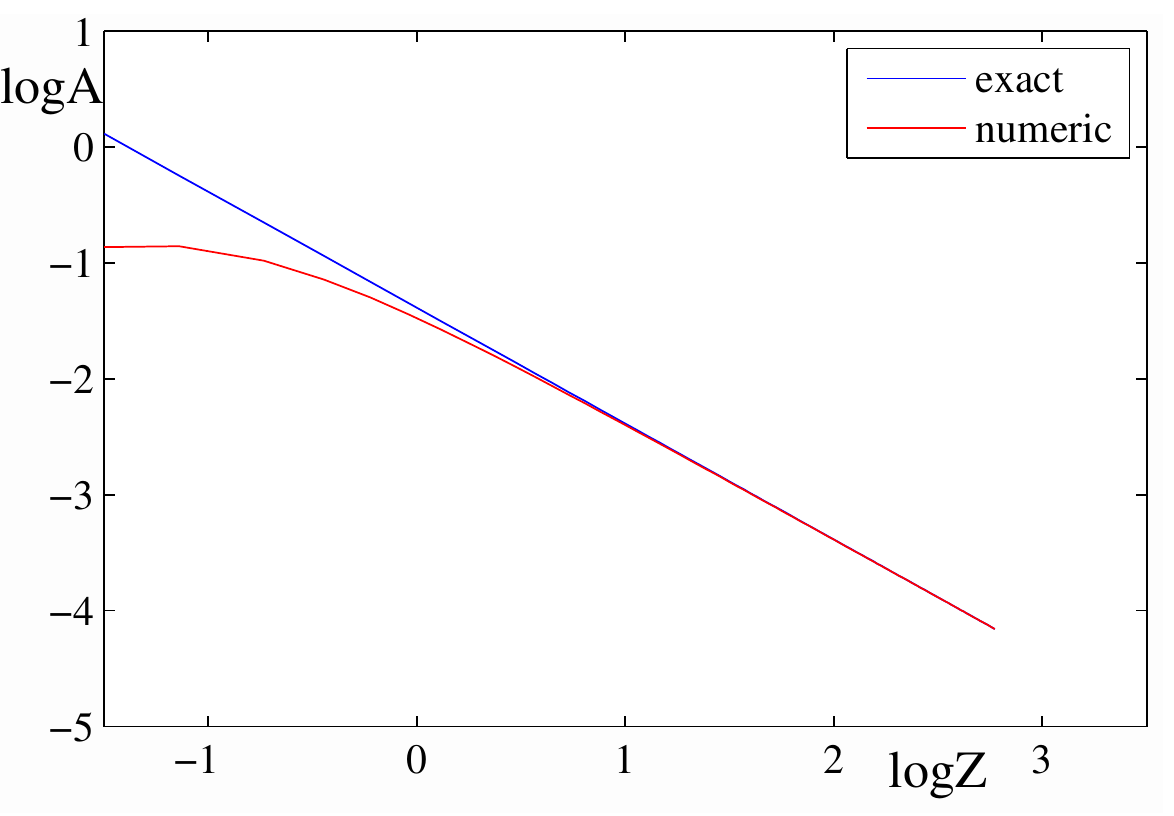}  \hspace{1cm}  \includegraphics[width=4cm]{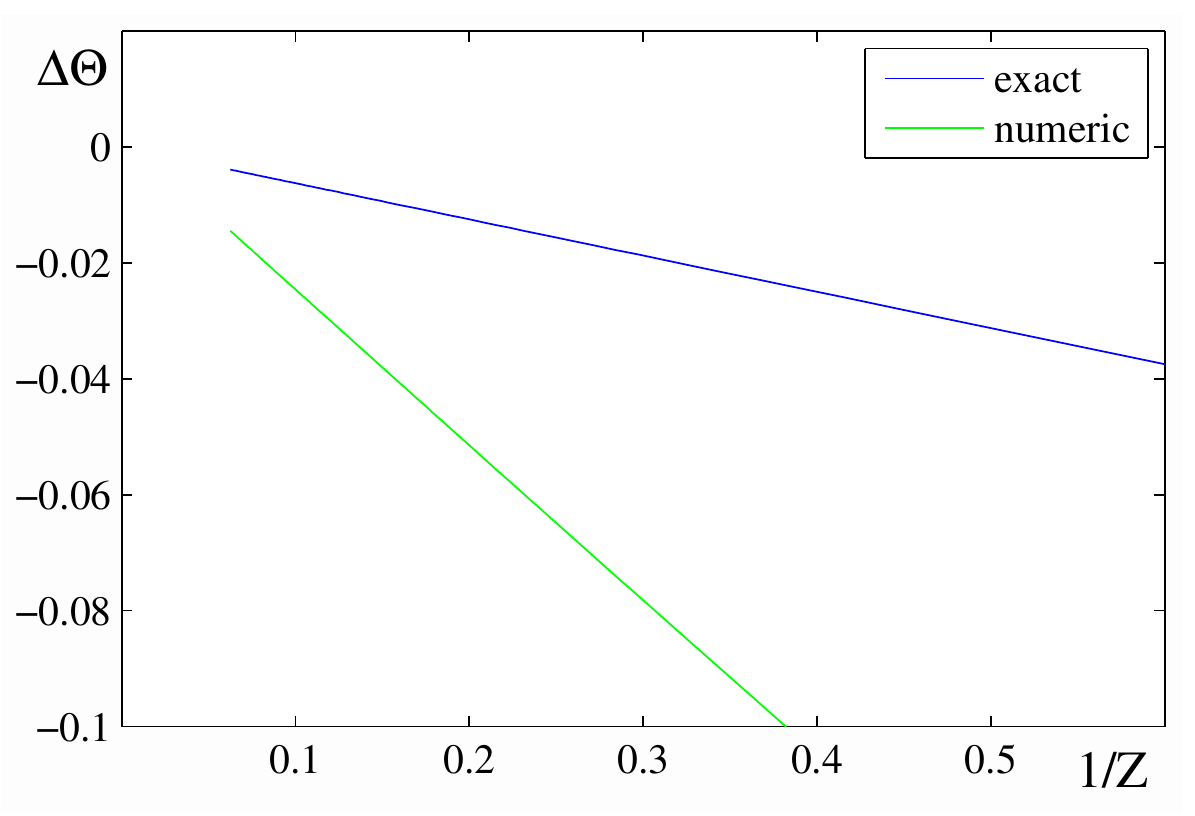}

\end{figure}

\newpage

\begin{figure}
\includegraphics[width=4cm]{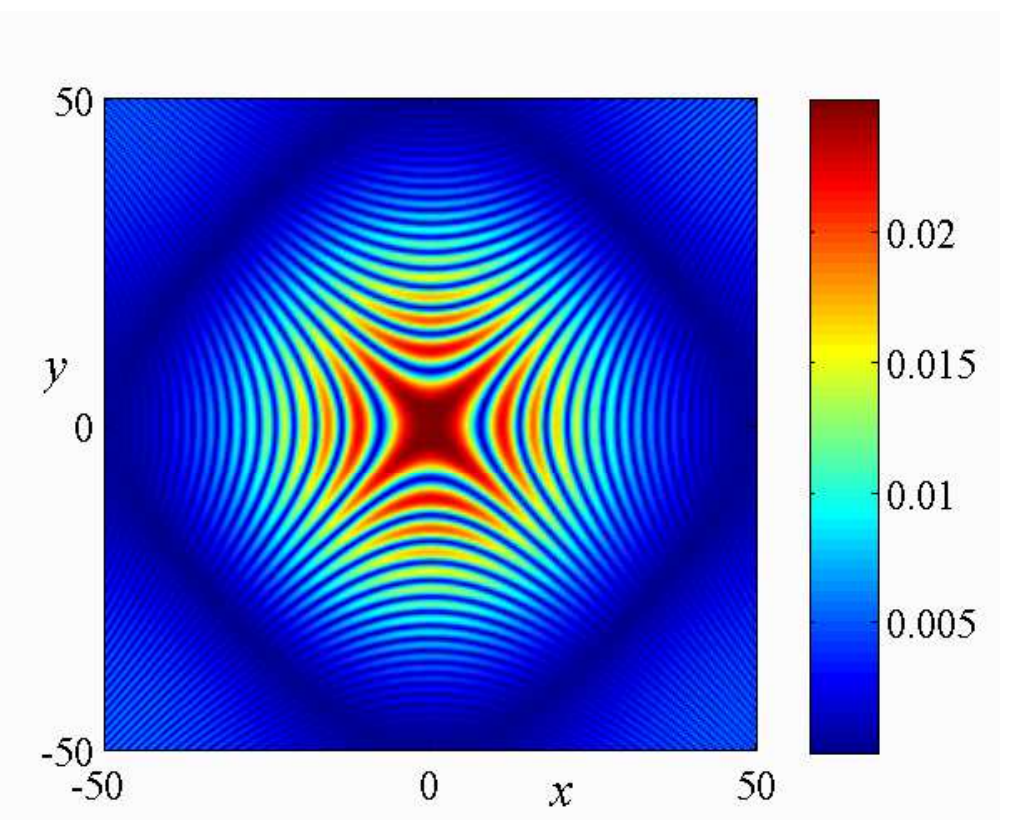}  \hspace{1cm}  \includegraphics[width=4cm]{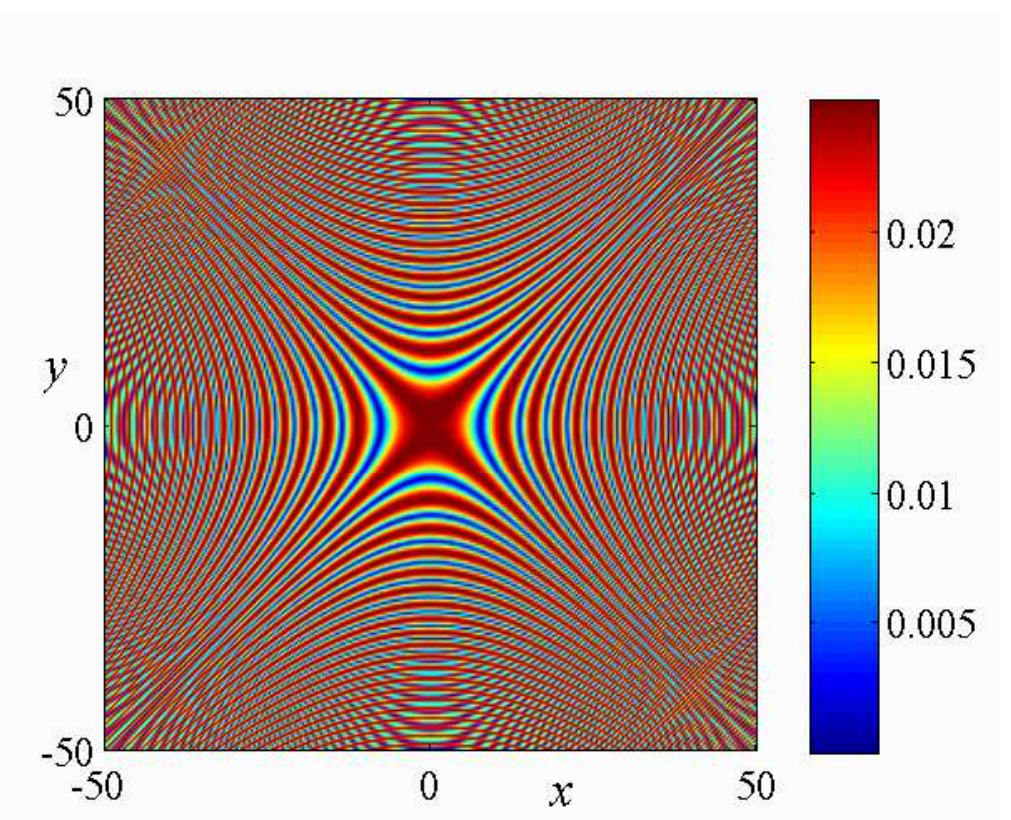} \hspace{1cm}  \includegraphics[width=4cm]{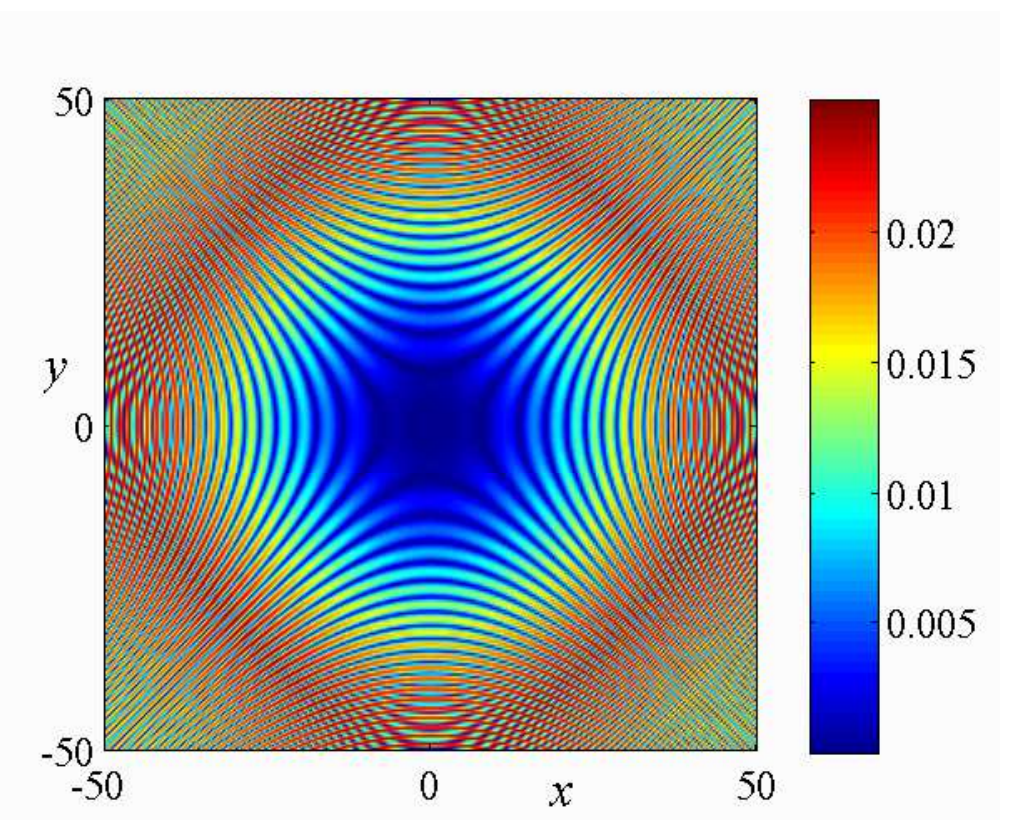}

\caption{Top: Initial condition $2.5\tanh(|x^2-y^2|)e^{-x^2-y^2}$: Numerical solution, Exact similarity solution and absolute value of their difference at $Z=16$.  Bottom: Log-amplitude vs. $\log Z$, $\Delta\theta = \theta-\theta_0-s$ vs. $1/Z$.}

\includegraphics[width=4cm]{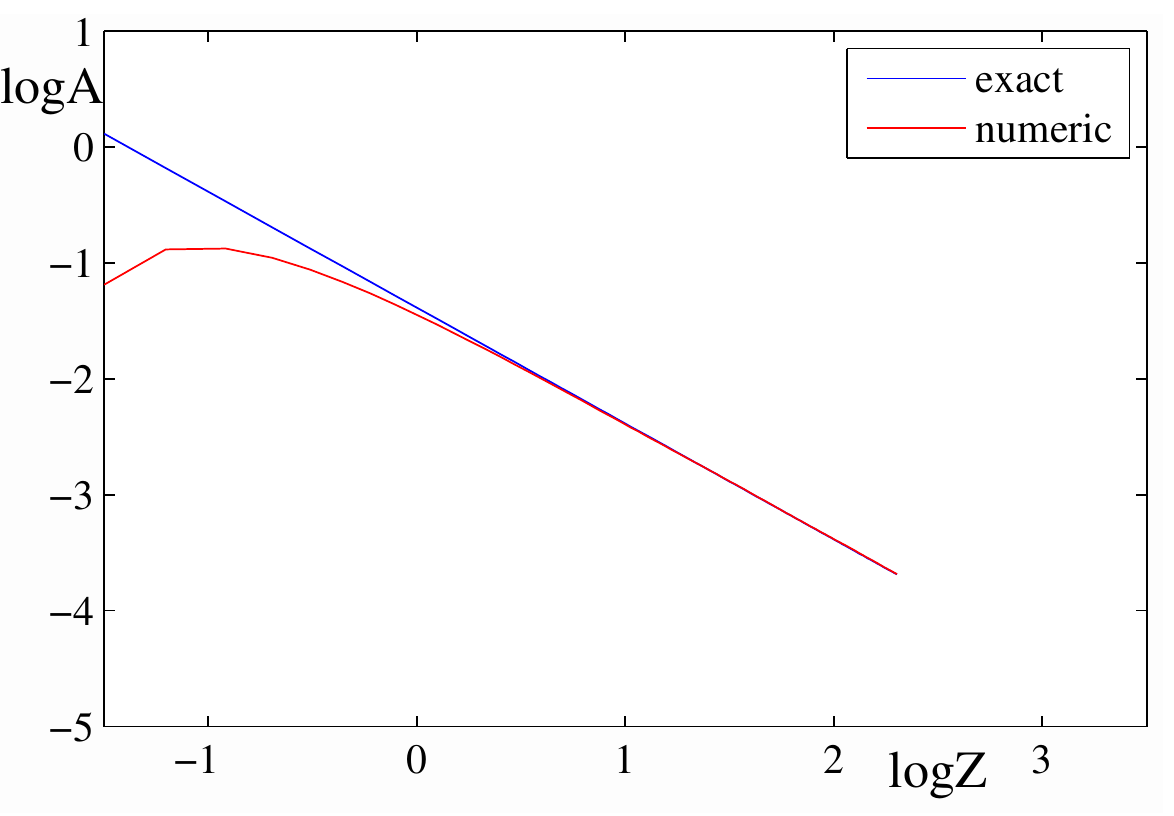}  \hspace{1cm}  \includegraphics[width=4cm]{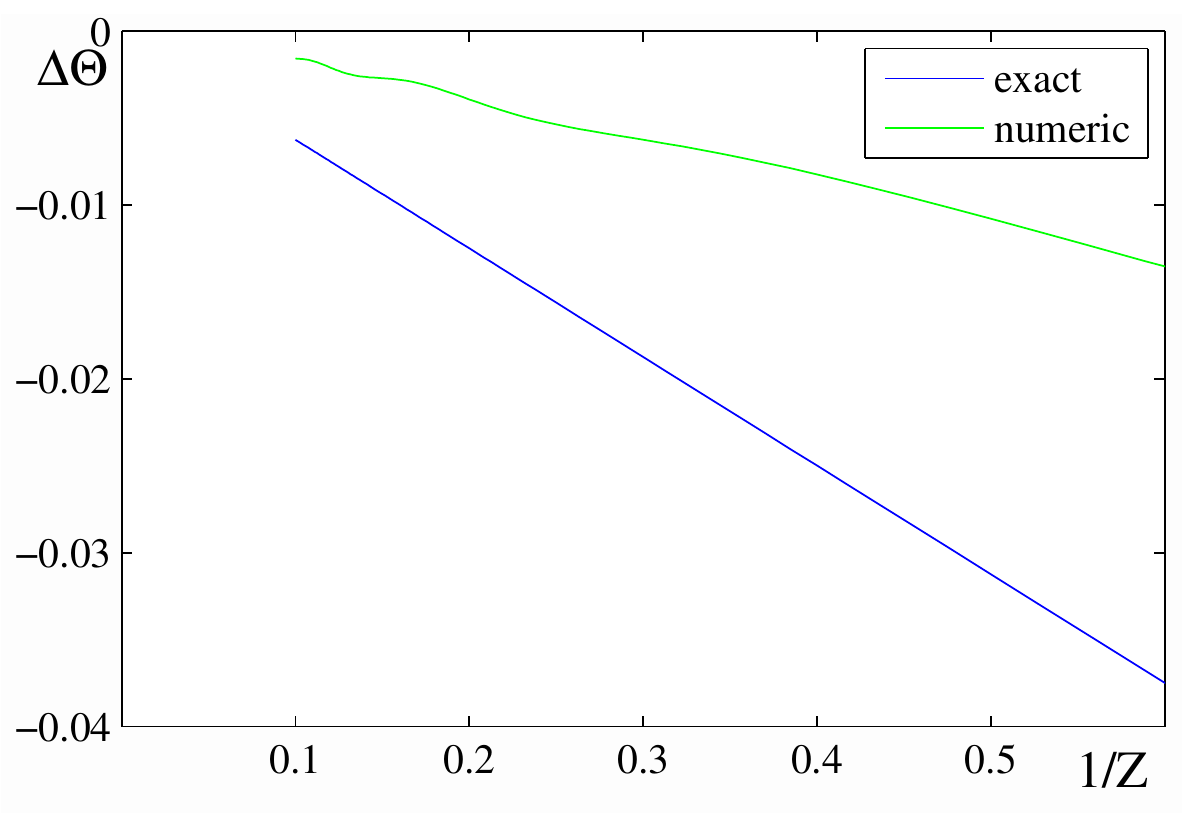}
\end{figure}

\begin{figure}
\includegraphics[width=4cm]{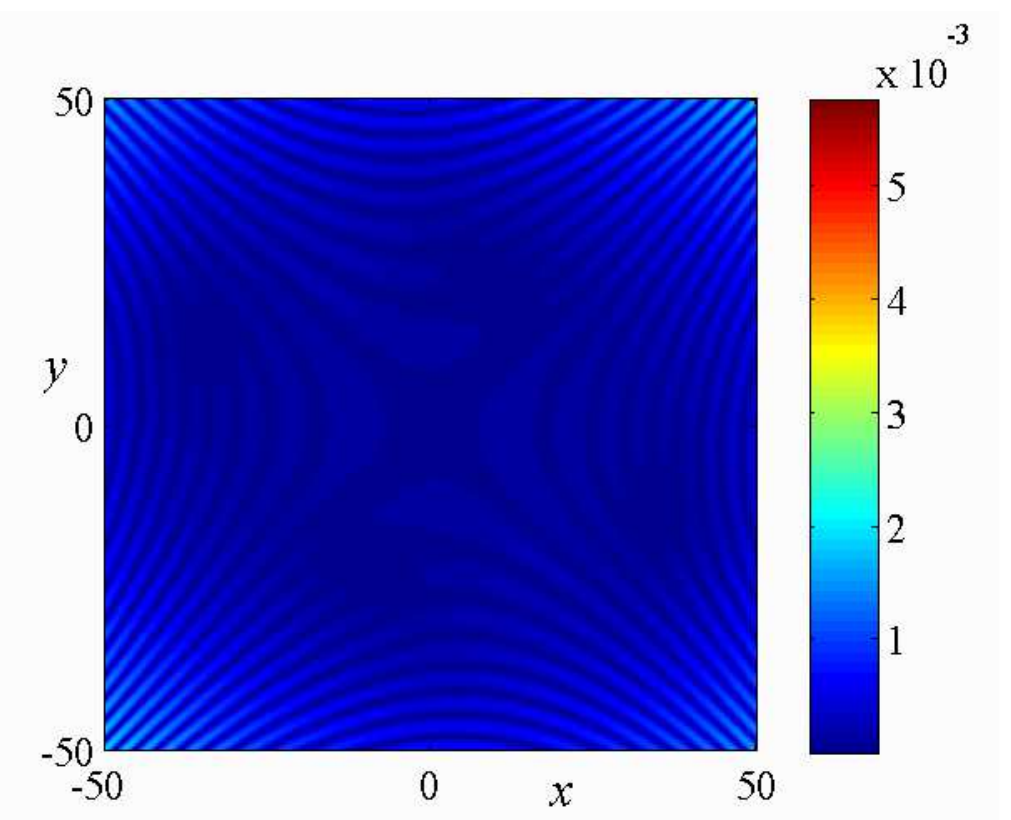}  \hspace{1cm}  \includegraphics[width=4cm]{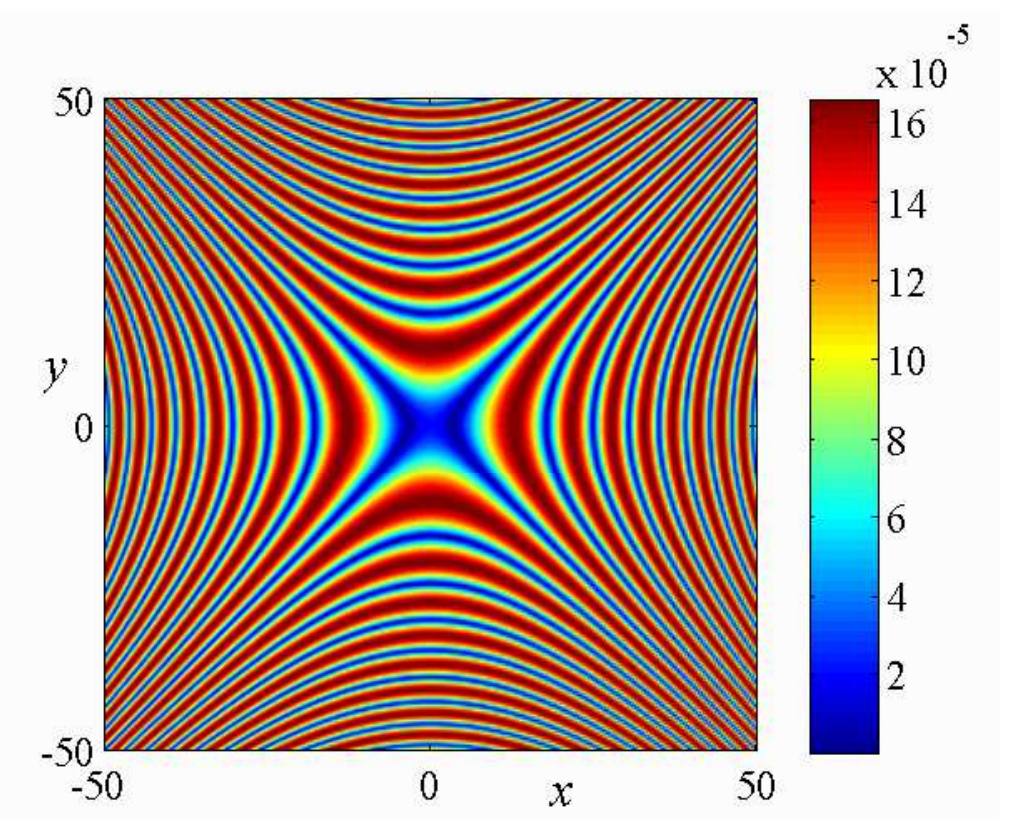} \hspace{1cm}  \includegraphics[width=4cm]{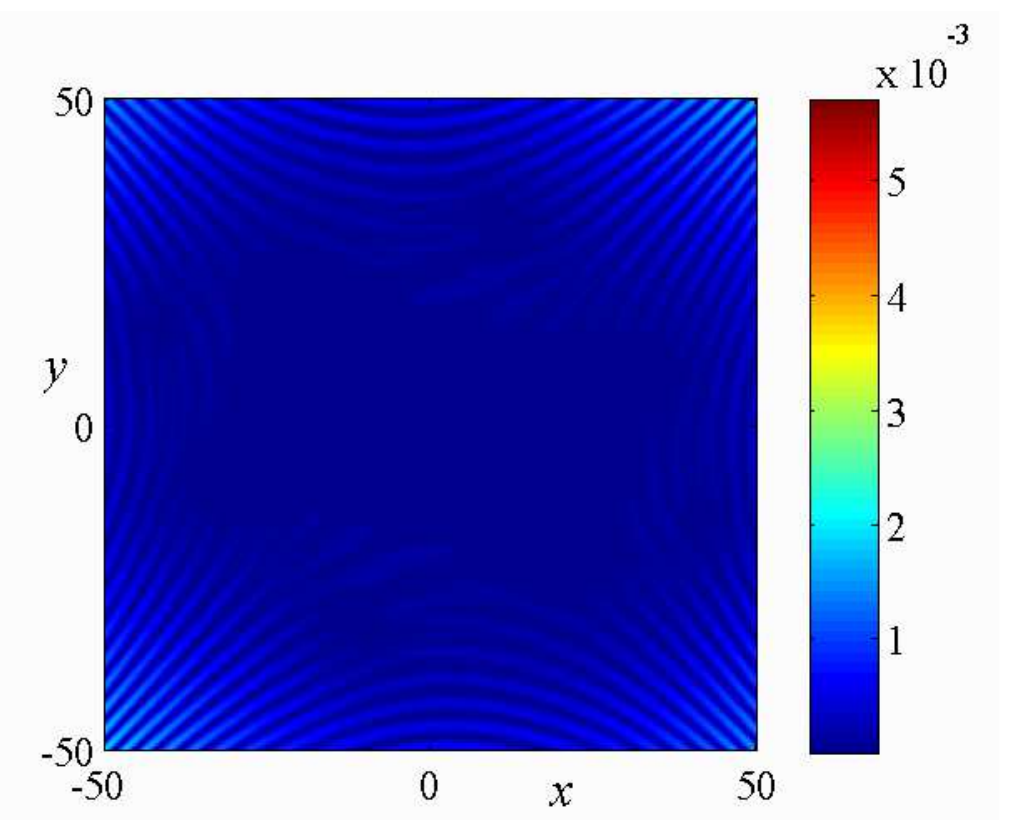}

\caption{Top: Initial condition $(x+iy)^4e^{-x^2-y^2}$: Numerical solution, Exact similarity solution and absolute value of their difference at $Z=24$.  Bottom: Log-amplitude vs. $\log Z$, $\Delta\theta = \theta-\theta_0-s$ vs. $1/Z$.}

\includegraphics[width=4cm]{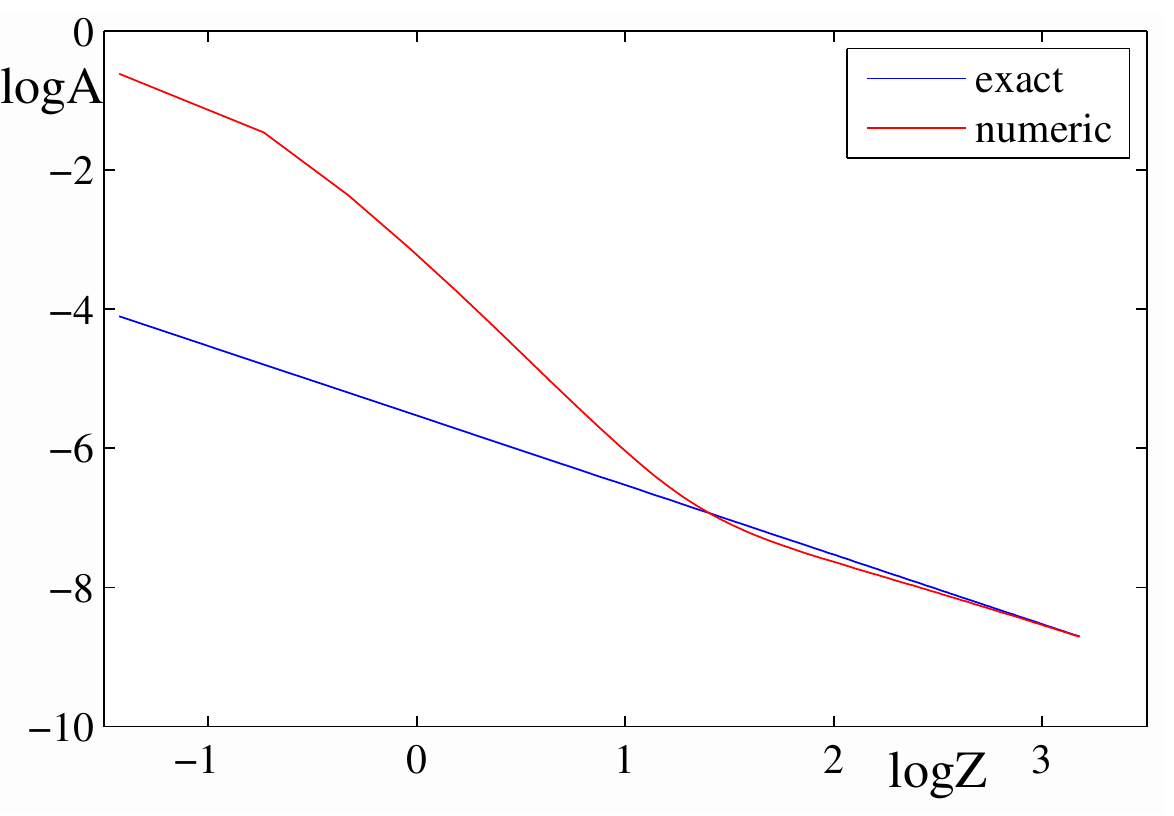} \hspace{1cm}  \includegraphics[width=4cm]{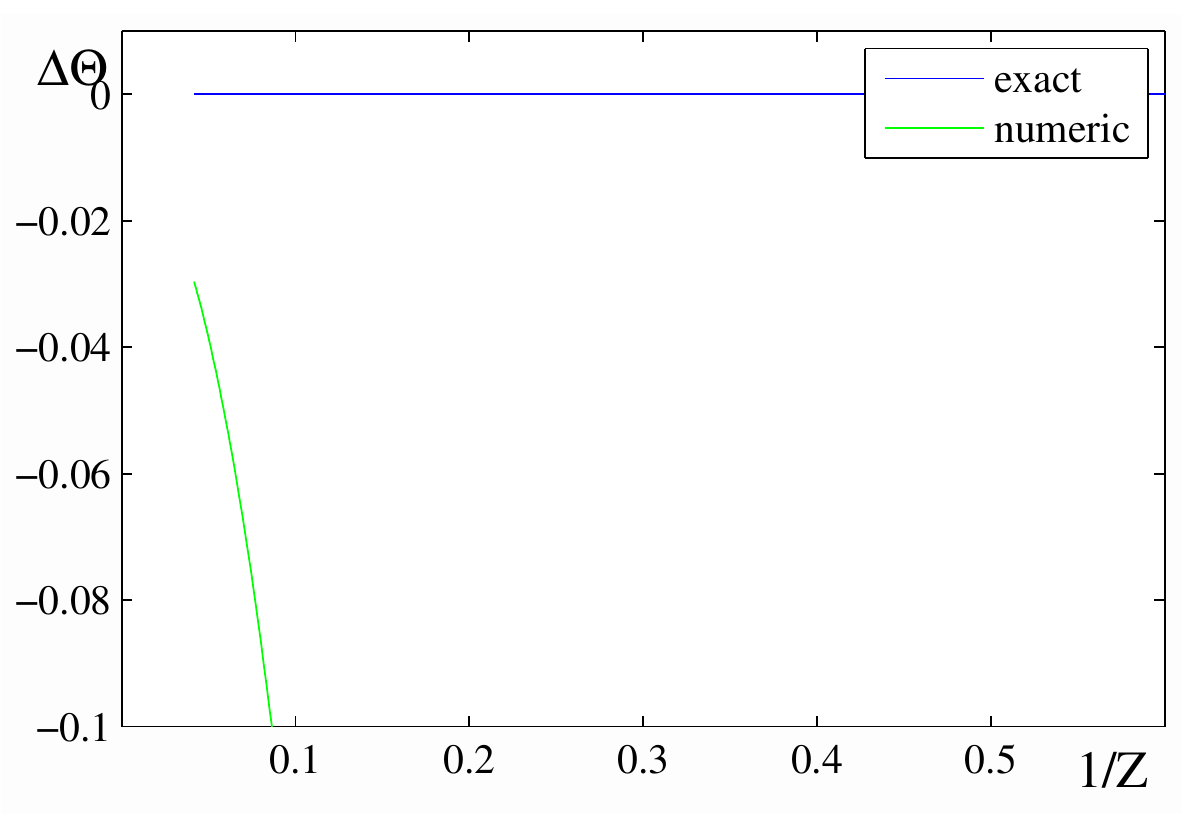}
\end{figure}

\newpage
\clearpage

\begin{figure}
\includegraphics[width=4cm]{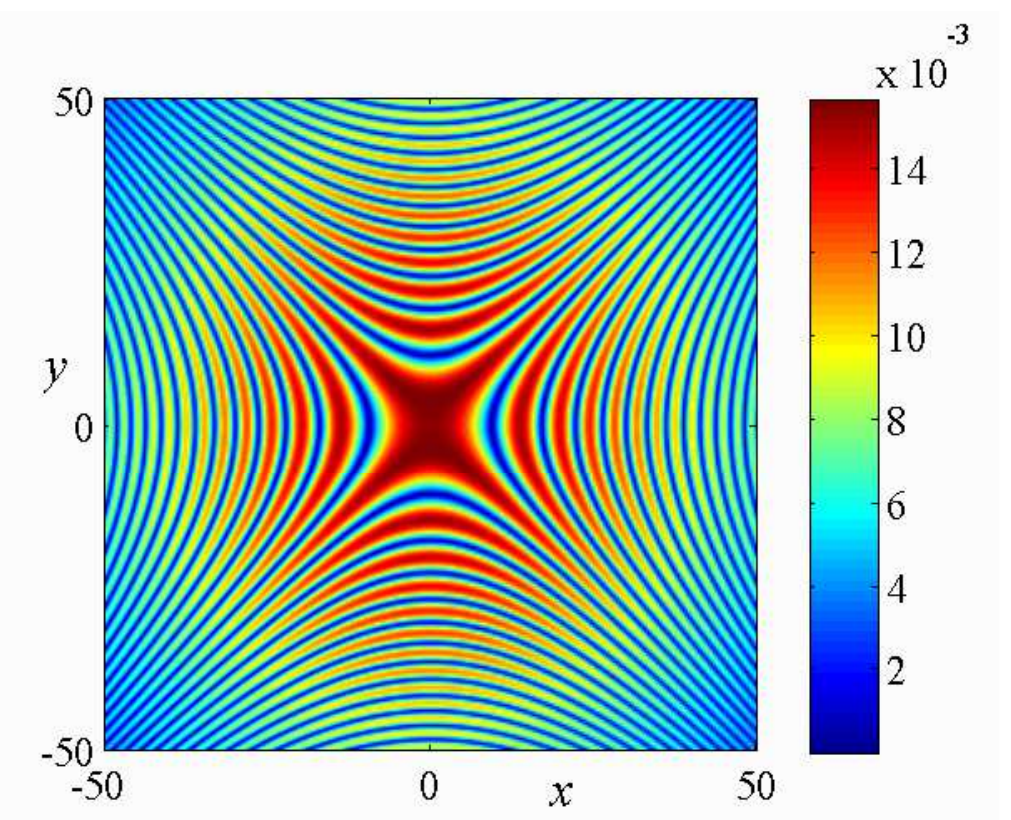}  \hspace{1cm}  \includegraphics[width=4cm]{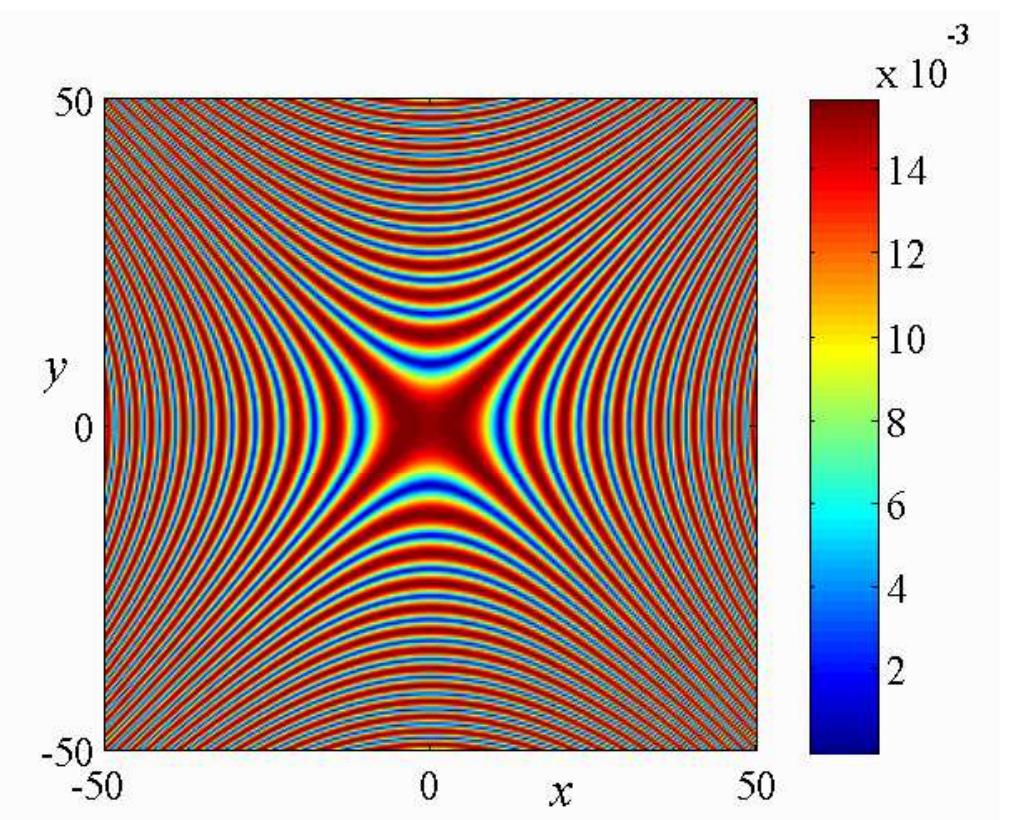} \hspace{1cm}  \includegraphics[width=4cm]{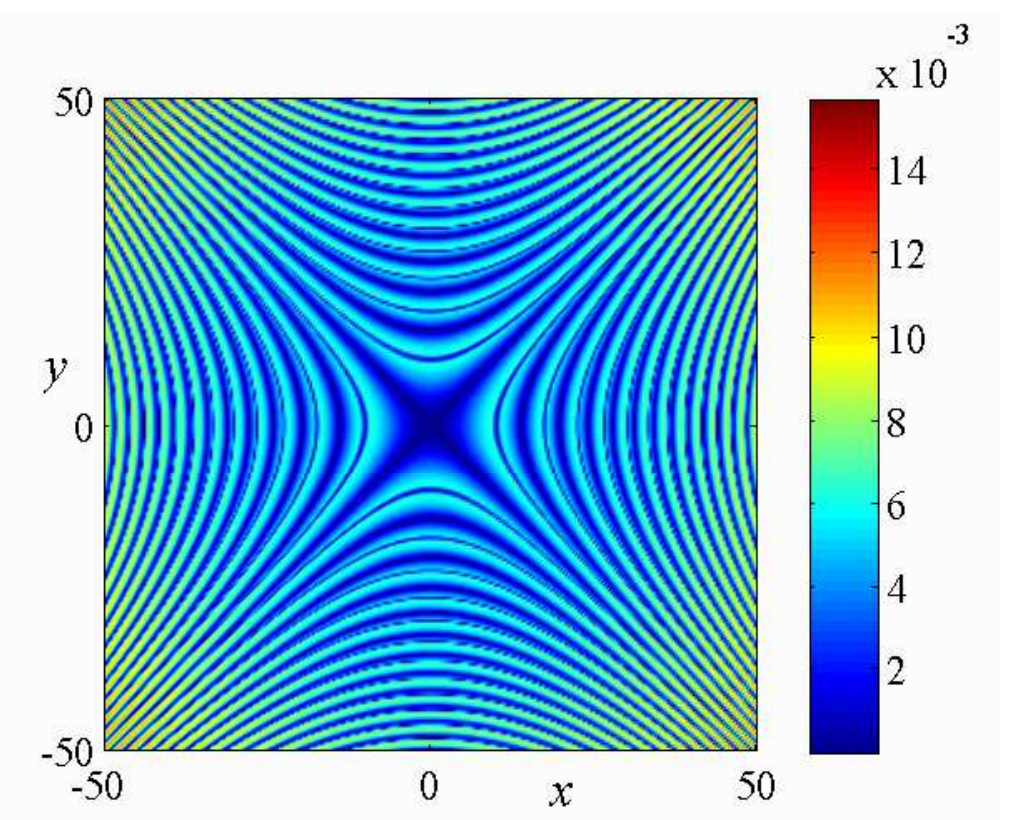}

\caption{Top: Initial condition $e^{-x^2-y^2}(1+0.1\text{randn})$, noise at every 16th gridpt.: Numerical solution, Exact similarity solution and absolute value of their difference at $Z=16$.  Bottom: Log-amplitude vs. $\log Z$, $\Delta\theta = \theta-\theta_0-s$ vs. $1/Z$.}

\includegraphics[width=4cm]{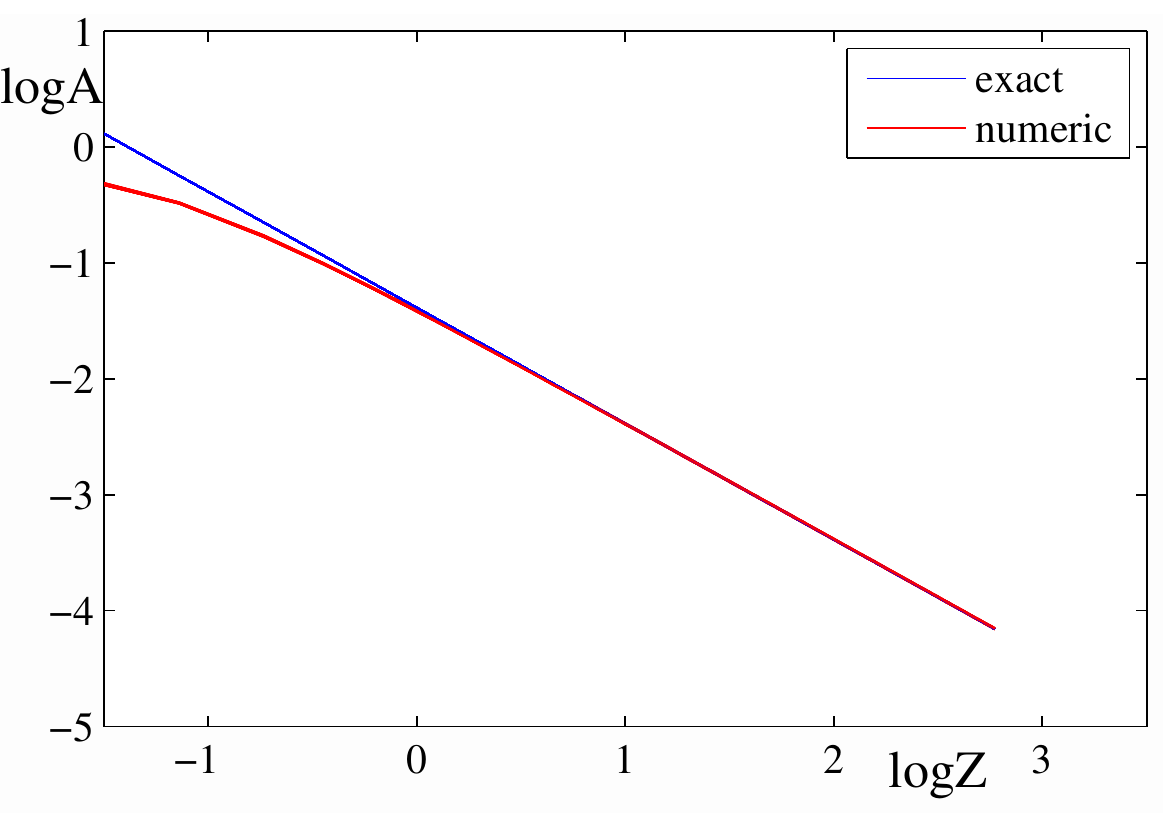}  \hspace{1cm}  \includegraphics[width=4cm]{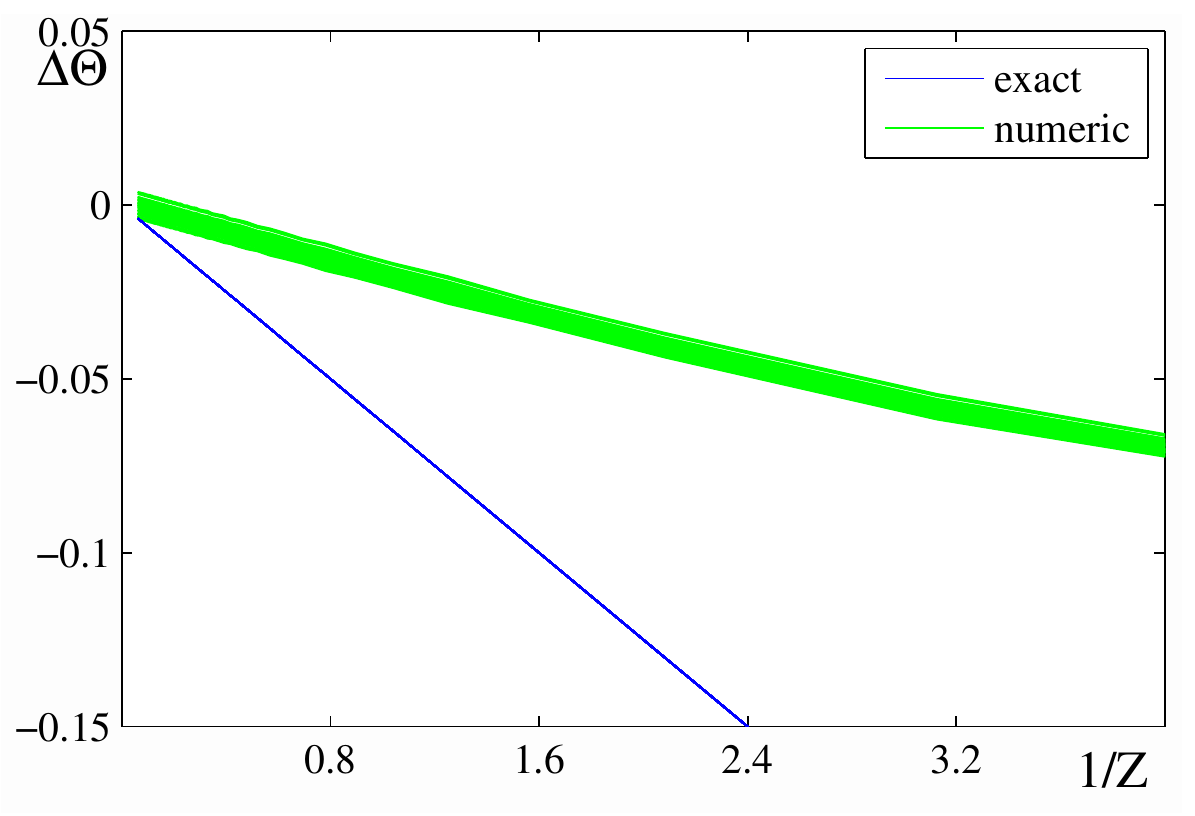}

\end{figure}

\begin{figure}
\includegraphics[width=4cm]{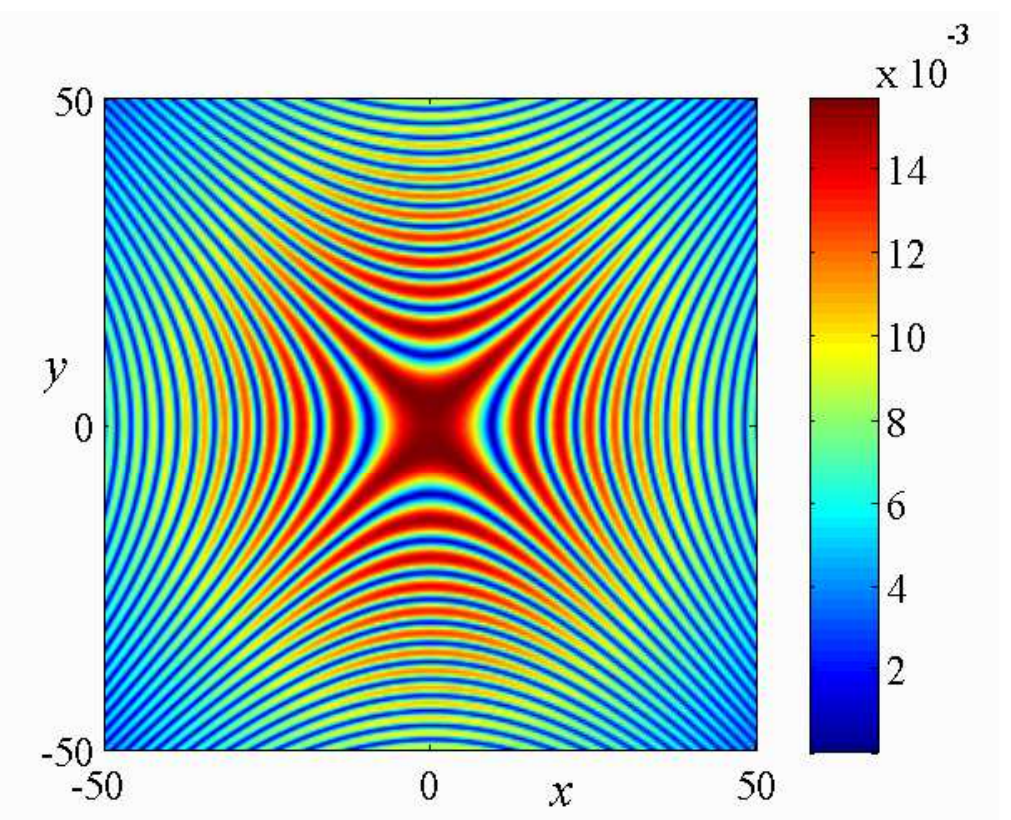}  \hspace{1cm}  \includegraphics[width=4cm]{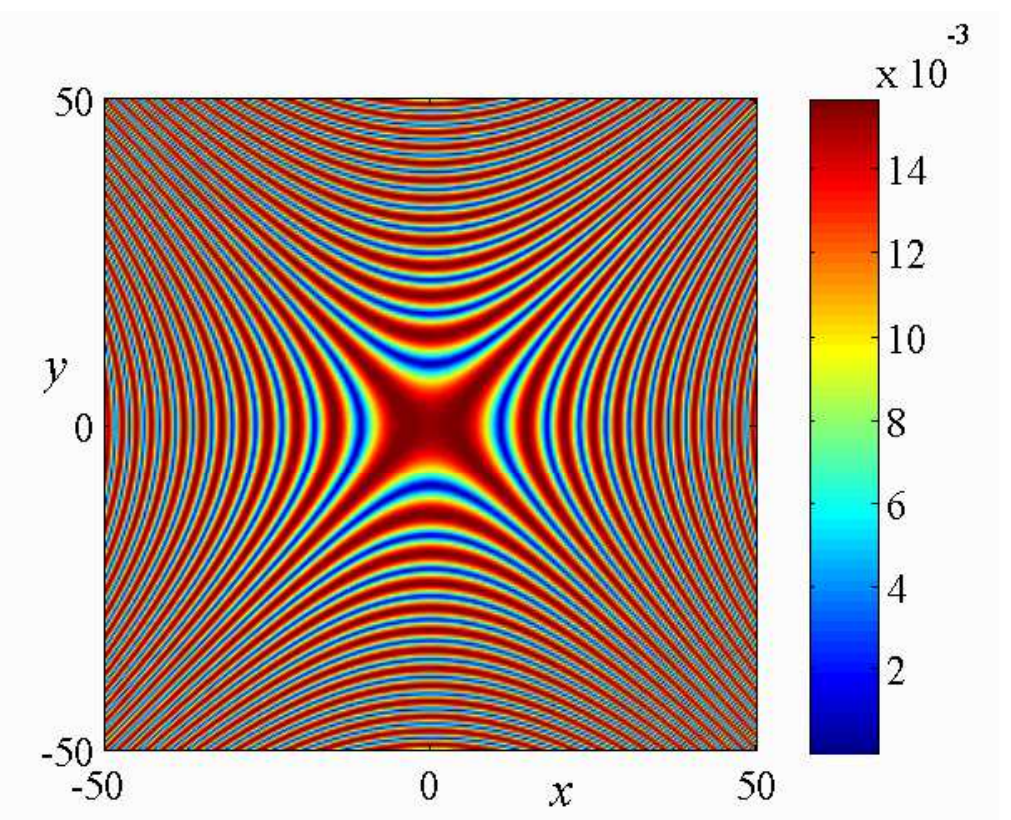} \hspace{1cm}  \includegraphics[width=4cm]{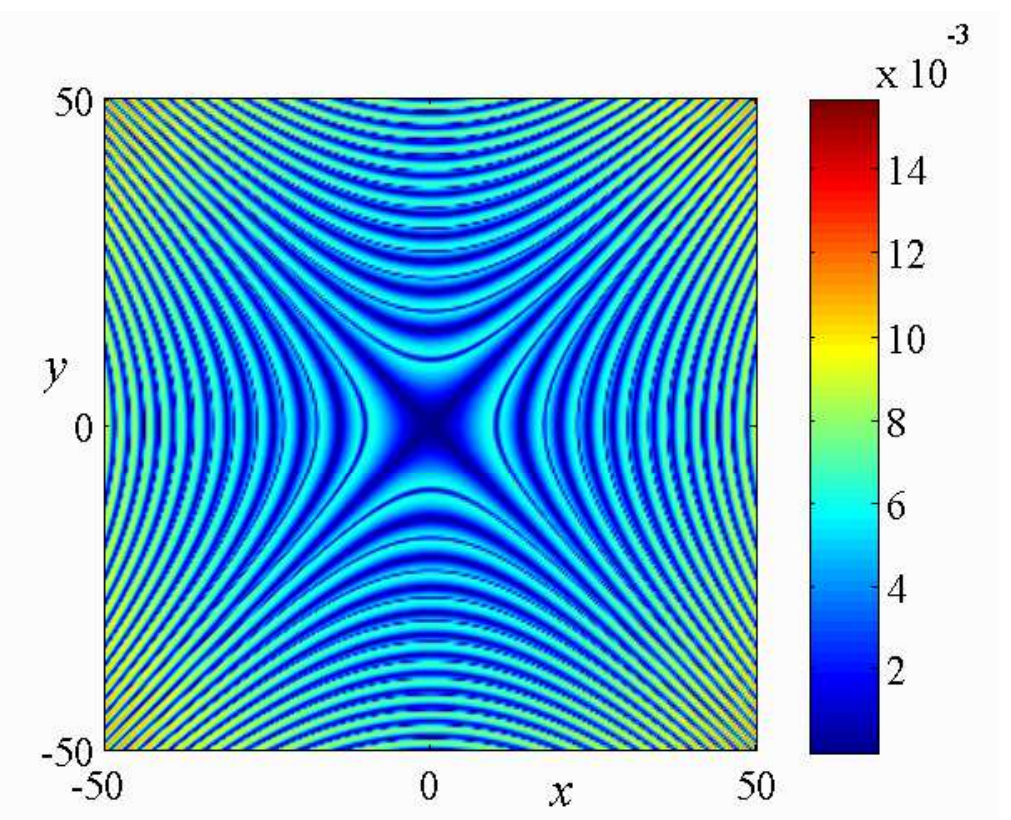}

\caption{Top: Initial condition $e^{-x^2-y^2}(1+0.1\text{randn})$, noise at every 8th gridpt.: Numerical solution, Exact similarity solution and absolute value of their difference at $Z=16$.  Bottom: Log-amplitude vs. $\log Z$, $\Delta\theta = \theta-\theta_0-s$ vs. $1/Z$.}

\includegraphics[width=4cm]{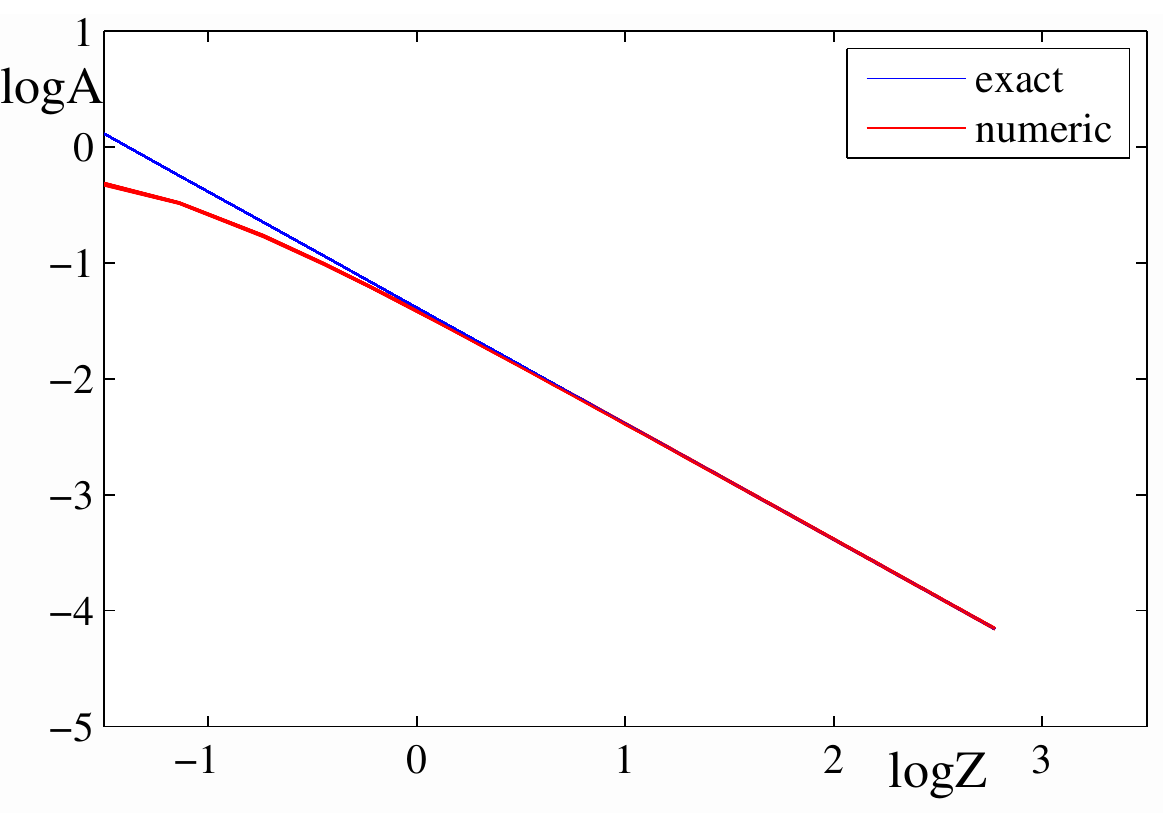}  \hspace{1cm}  \includegraphics[width=4cm]{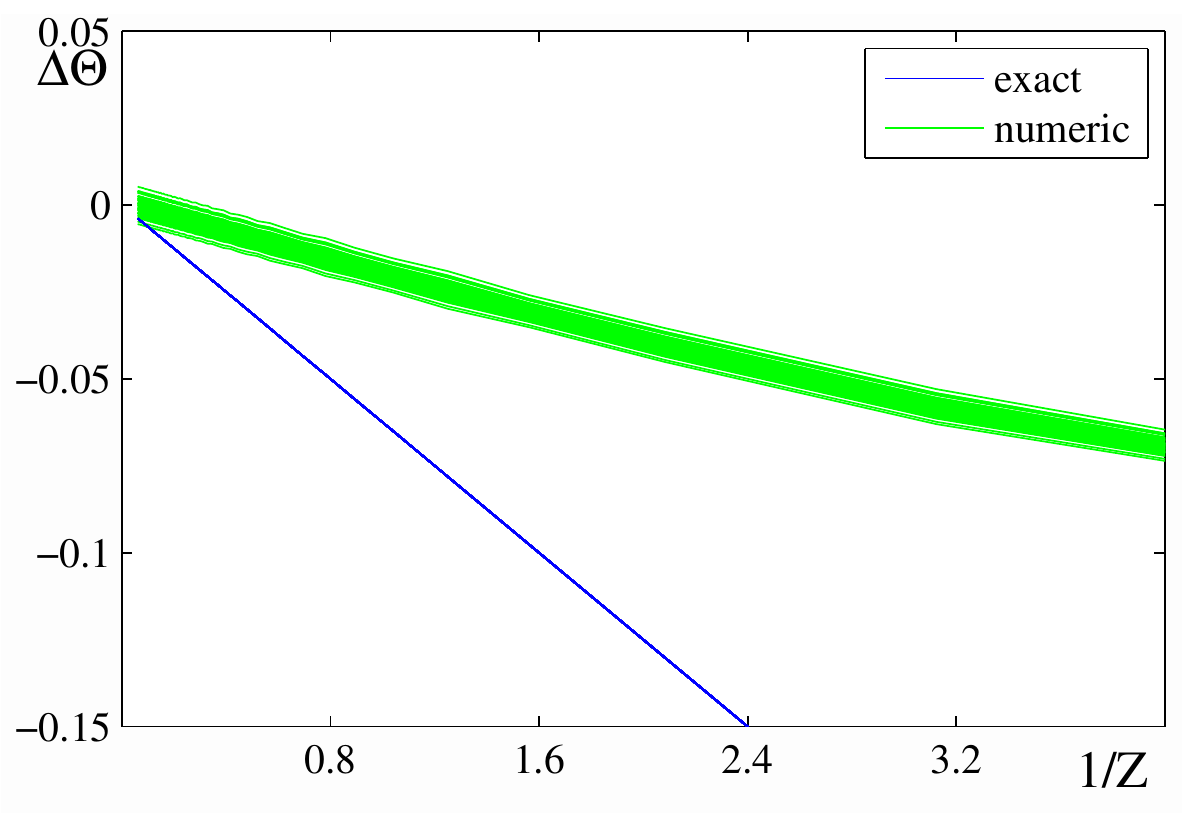}

\end{figure}

\newpage

\begin{figure}
\includegraphics[width=4cm]{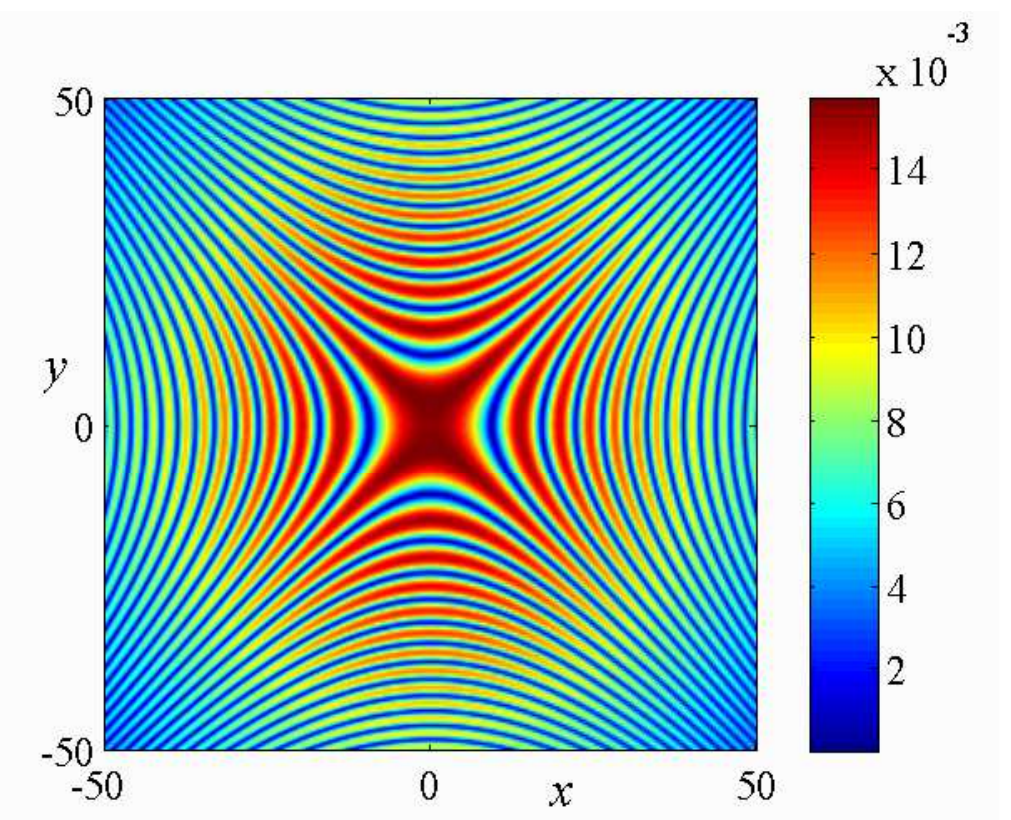}  \hspace{1cm}  \includegraphics[width=4cm]{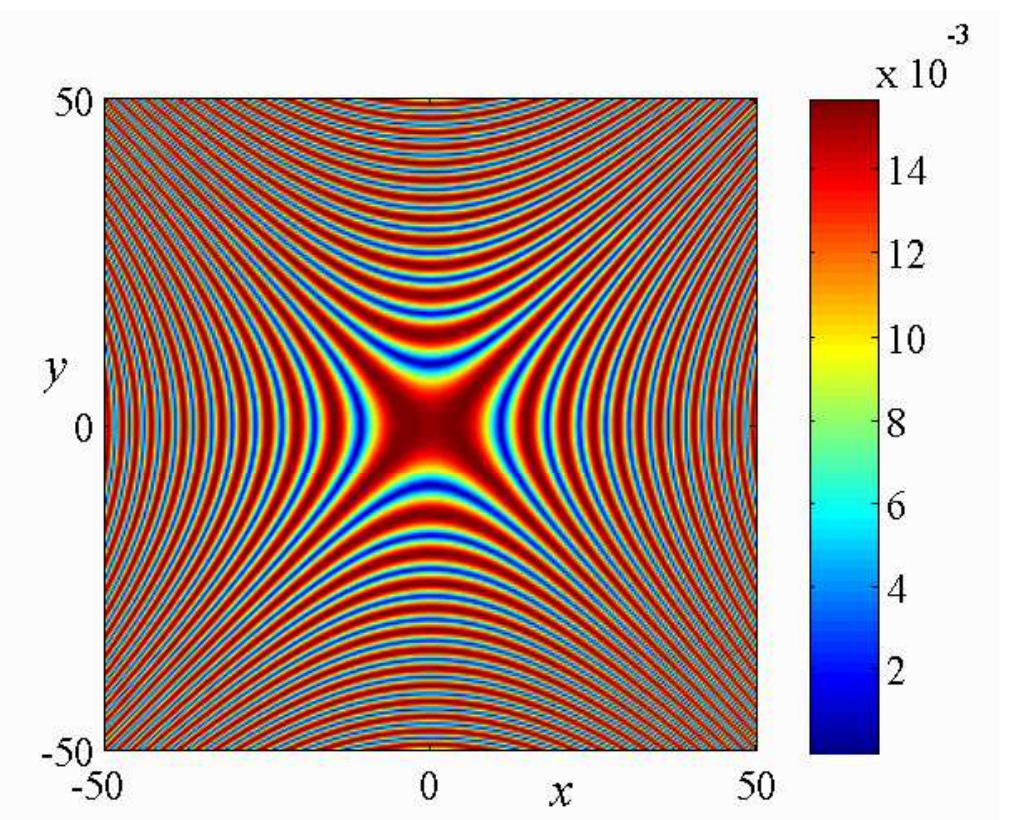} \hspace{1cm}  \includegraphics[width=4cm]{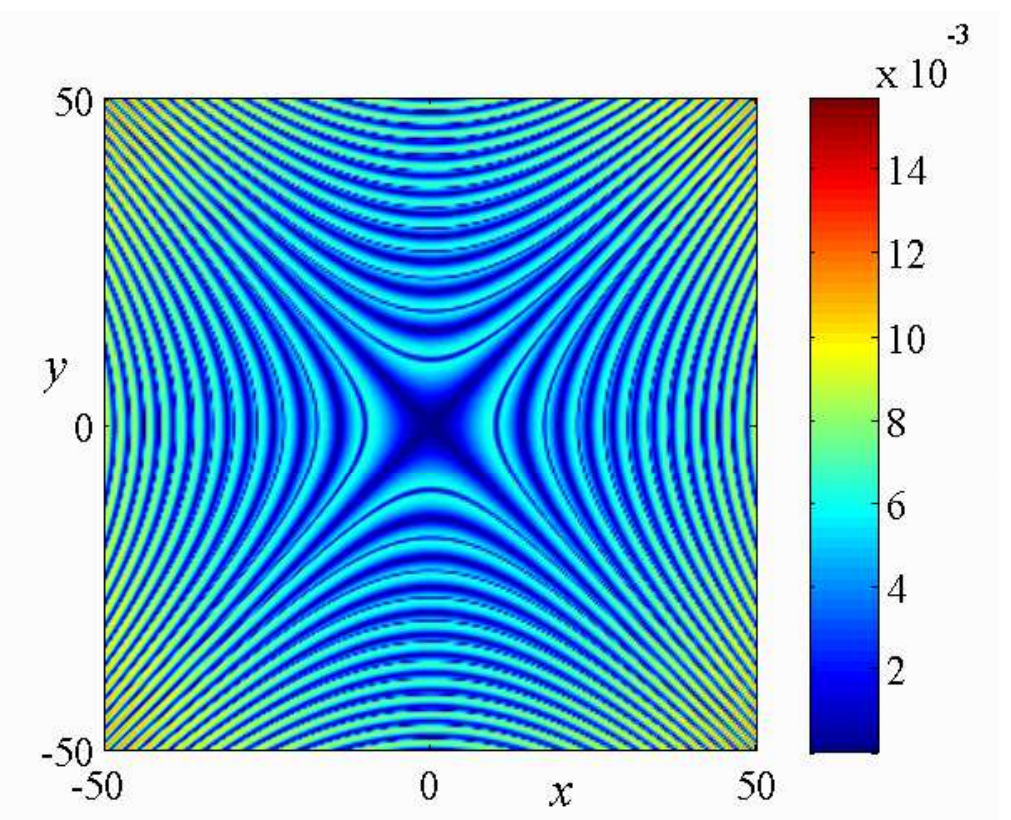}

\caption{Top: Initial condition $e^{-x^2-y^2}(1+0.1\text{randn})$, noise at every gridpt.: Numerical solution, Exact similarity solution and absolute value of their difference at $Z=16$.  Bottom: Log-amplitude vs. $\log Z$, $\Delta\theta = \theta-\theta_0-s$ vs. $1/Z$.}

\includegraphics[width=4cm]{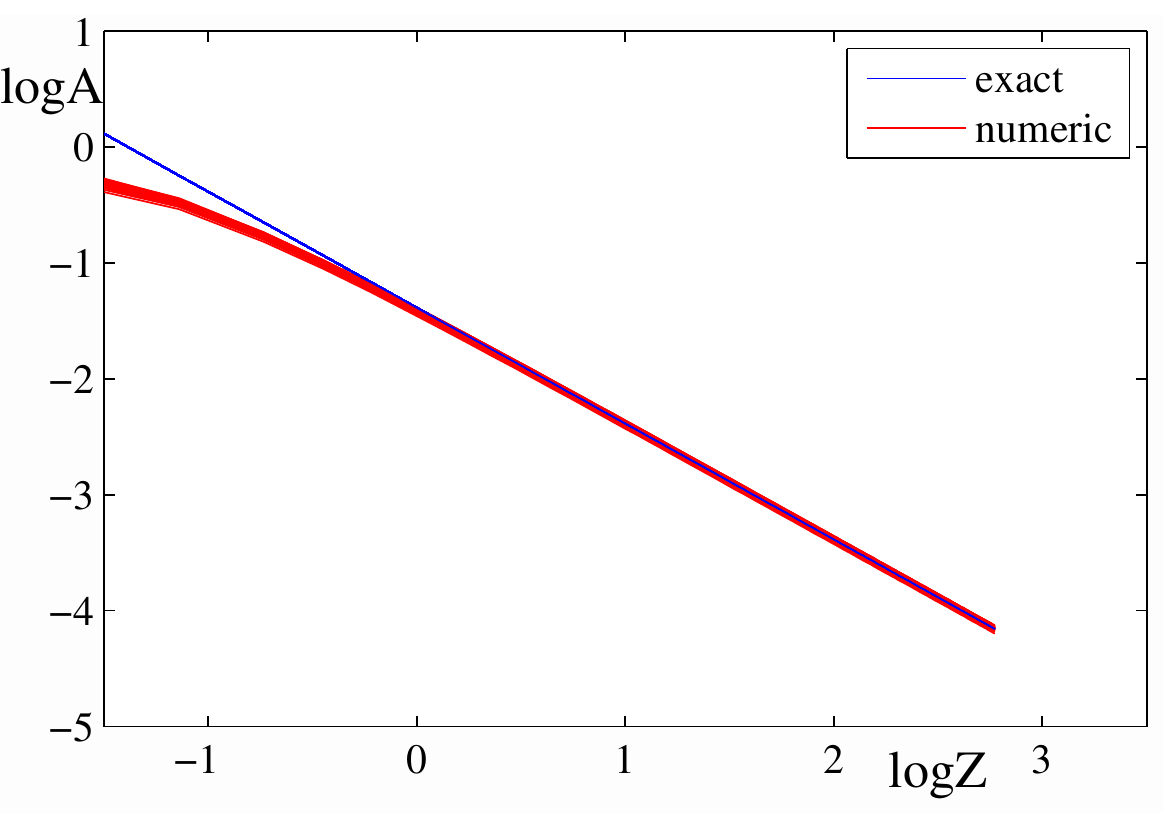}  \hspace{1cm}  \includegraphics[width=4cm]{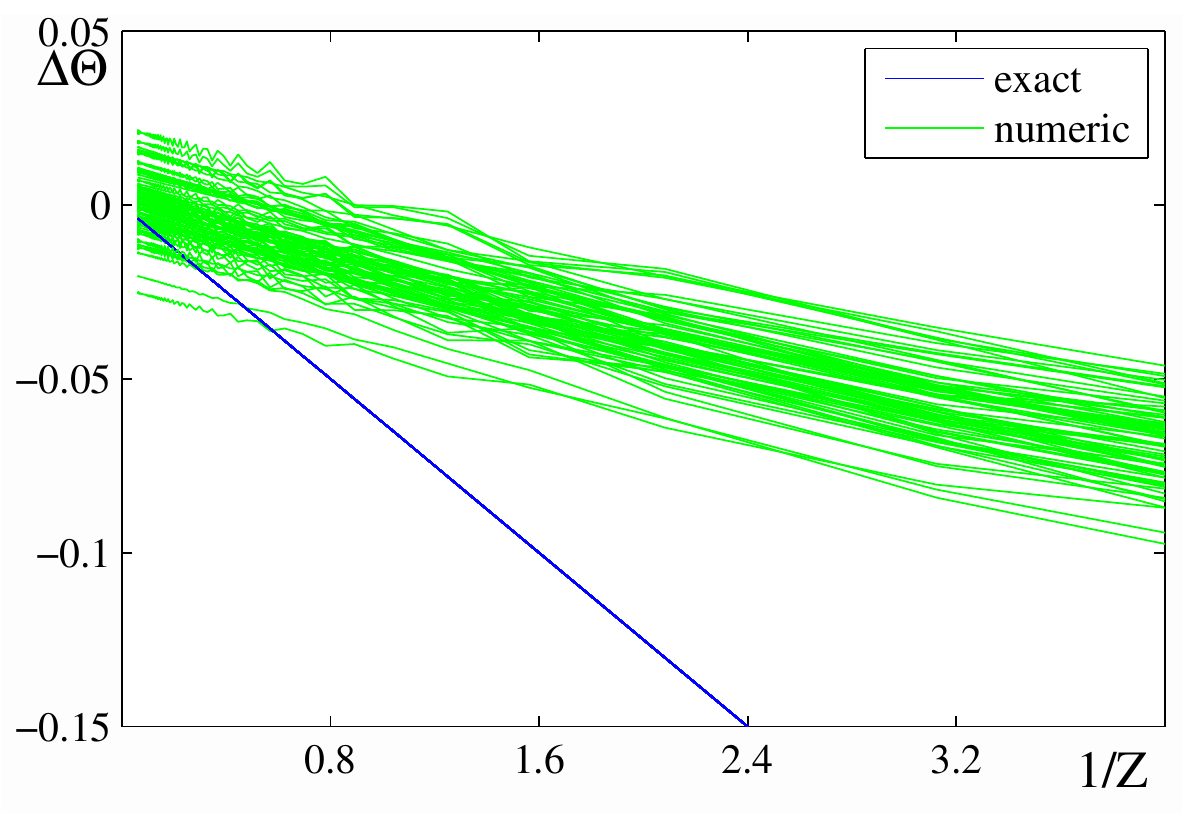}

\end{figure}

\begin{figure}
\includegraphics[width=4cm]{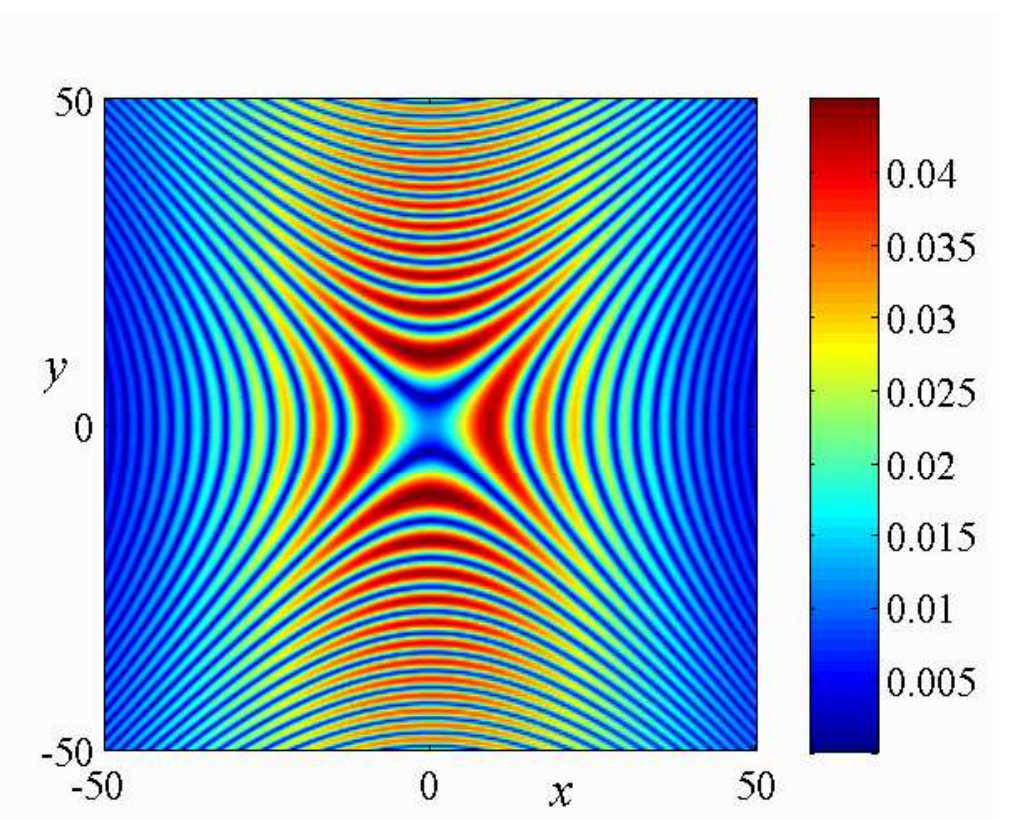}  \hspace{1cm}  \includegraphics[width=4cm]{Gauss3Ex-eps-converted-to.pdf} \hspace{1cm}  \includegraphics[width=4cm]{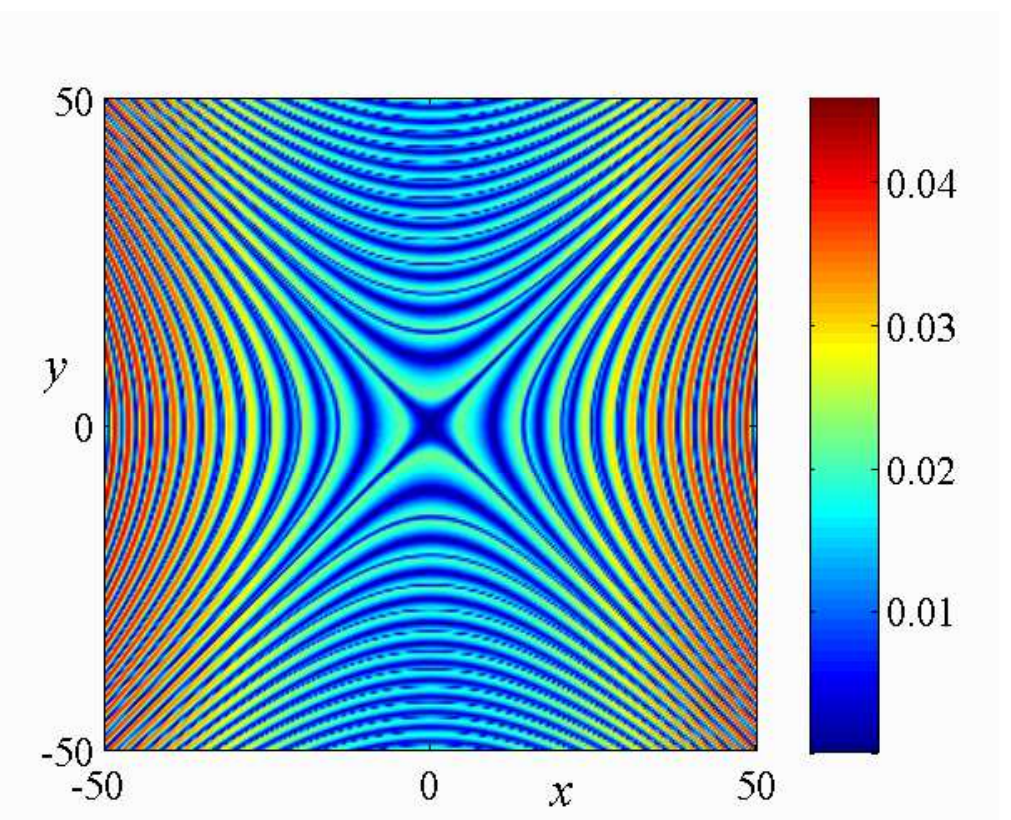}

\caption{Top: Initial condition $3e^{-x^2-y^2}(1+0.1\text{randn})$, noise at every 8th gridpt.: Numerical solution, Exact similarity solution and absolute value of their difference at $Z=16$.  Bottom: Log-amplitude vs. $\log Z$, $\Delta\theta = \theta-\theta_0-s$ vs. $1/Z$.}

\includegraphics[width=4cm]{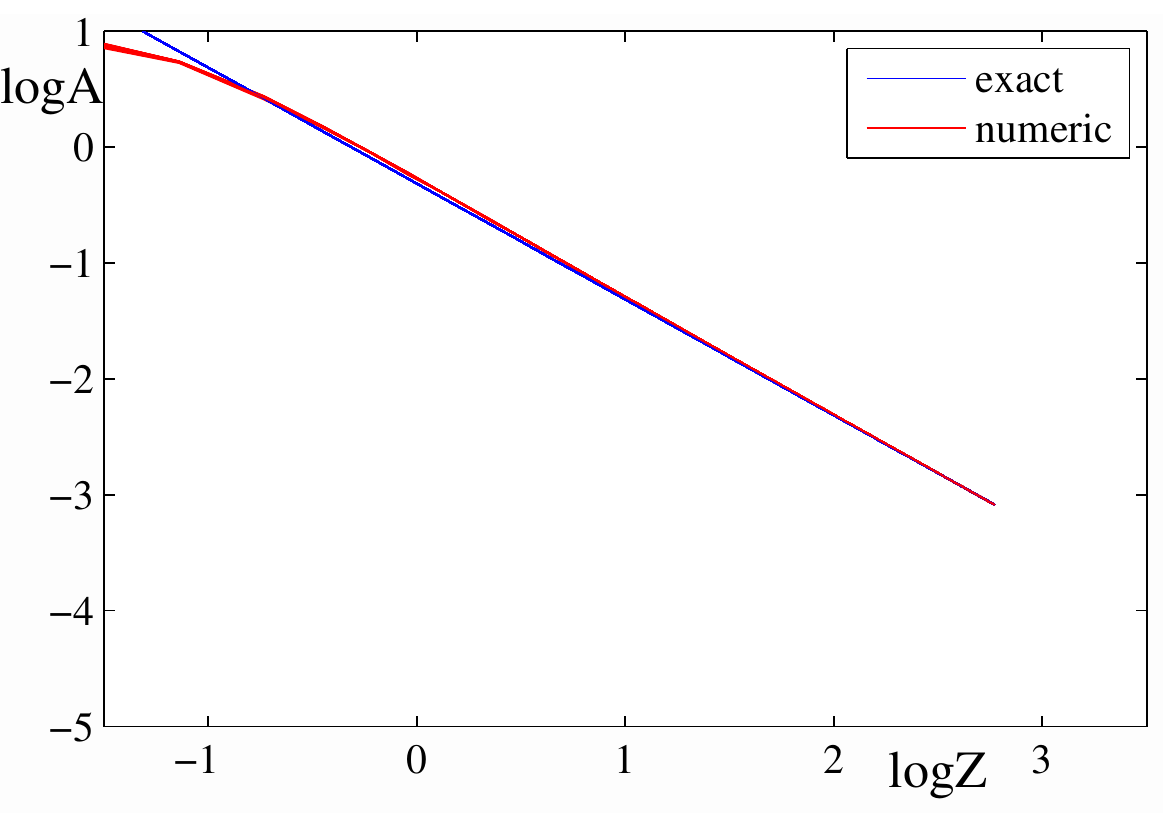}  \hspace{1cm}  \includegraphics[width=4cm]{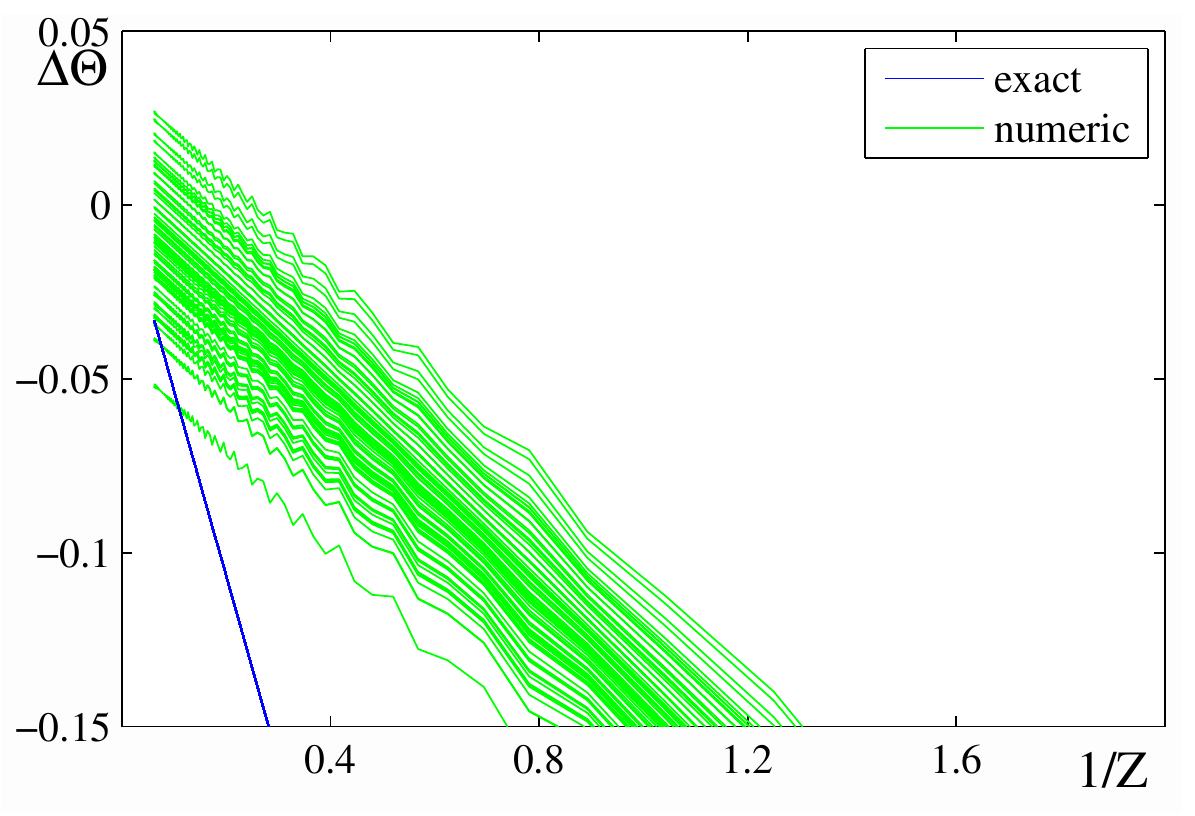}
\end{figure}

\newpage

\begin{figure}
\includegraphics[width=4cm]{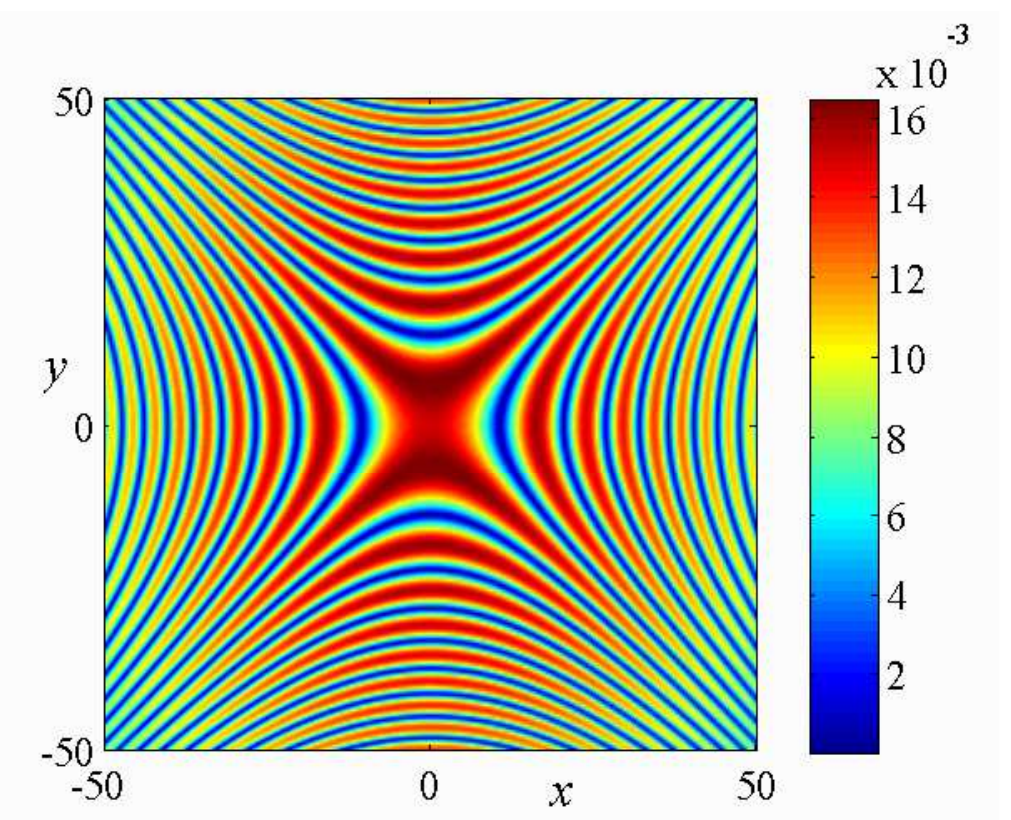}   \hspace{1cm}  \includegraphics[width=4cm]{SechT24Ex-eps-converted-to.pdf}  \hspace{1cm}  \includegraphics[width=4cm]{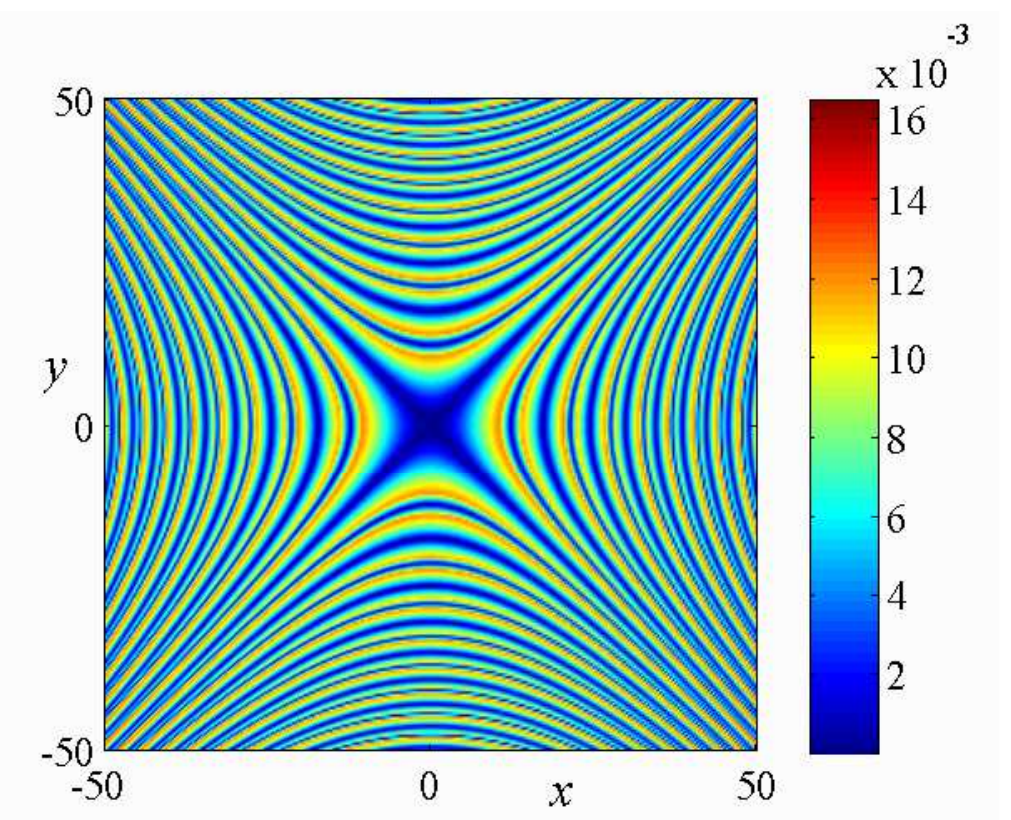}

\caption{Top:  Initial condition $\text{sech}(x^2+y^2)(1+0.1\text{randn})$, noise at every 16th gridpt.: Numerical solution, Exact similarity solution and absolute value of their difference at $Z=24$.  Bottom: Log-amplitude vs. $\log Z$, $\Delta\theta = \theta-\theta_0-s$ vs. $1/Z$.}

\includegraphics[width=4cm]{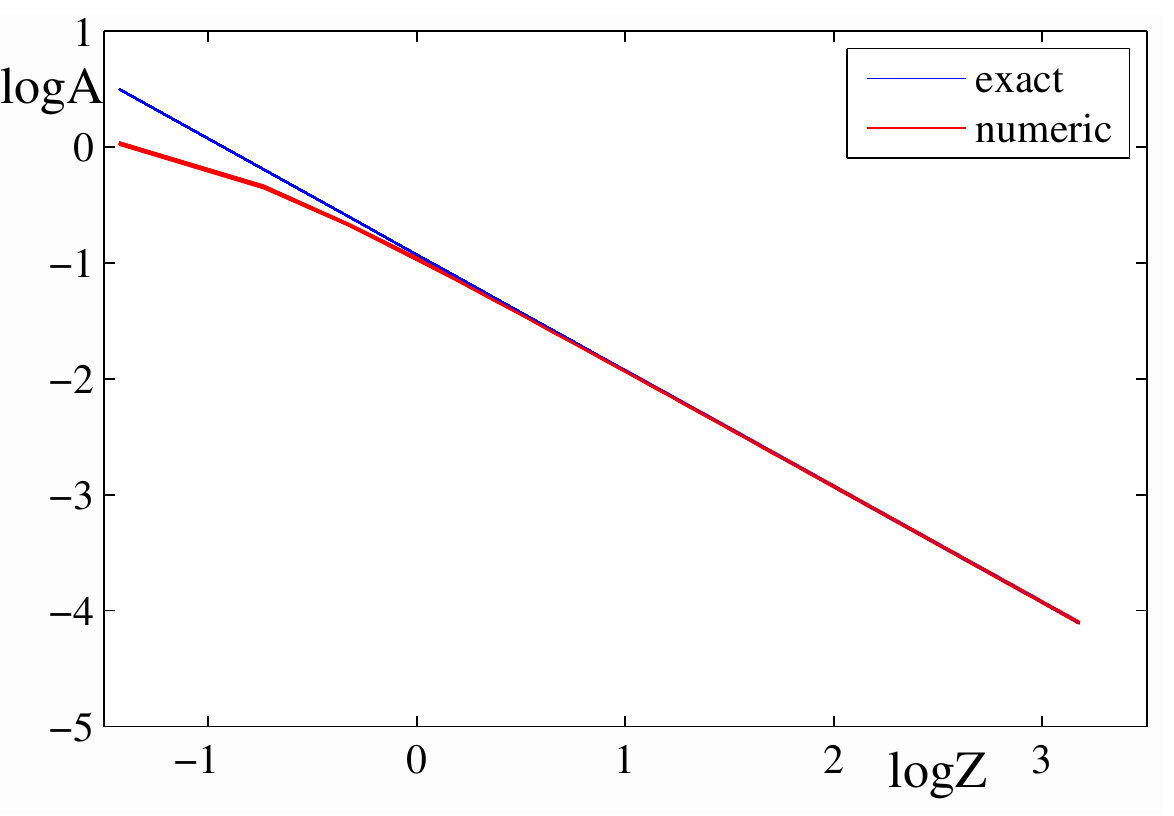}  \hspace{1cm}  \includegraphics[width=4cm]{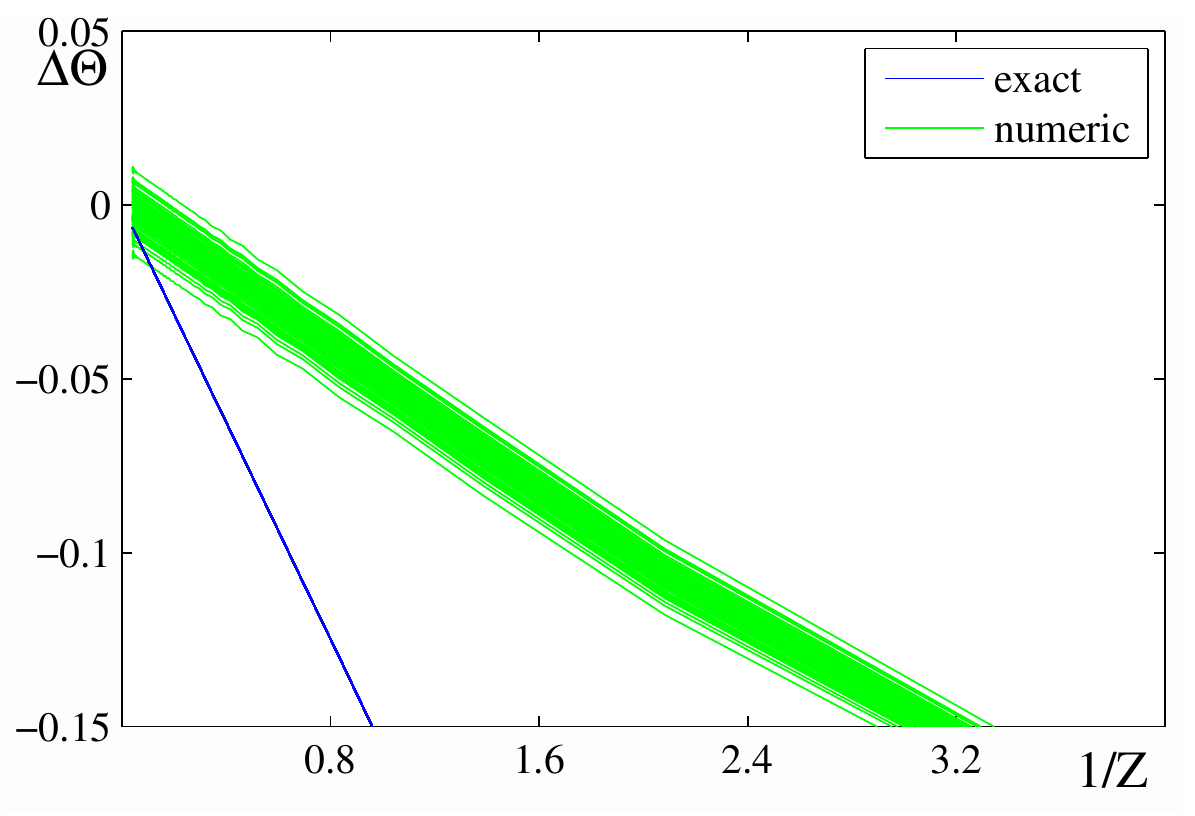}
\end{figure}

\begin{figure}
\includegraphics[width=4cm]{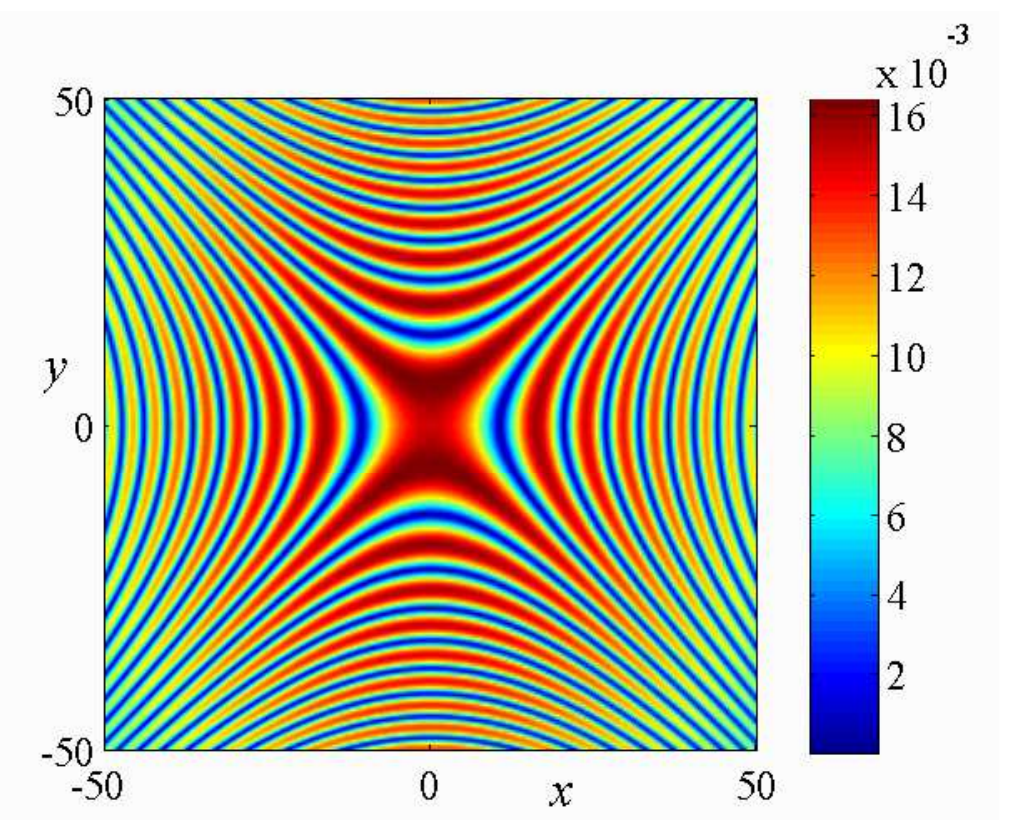}   \hspace{1cm}  \includegraphics[width=4cm]{SechT24Ex-eps-converted-to.pdf}  \hspace{1cm}  \includegraphics[width=4cm]{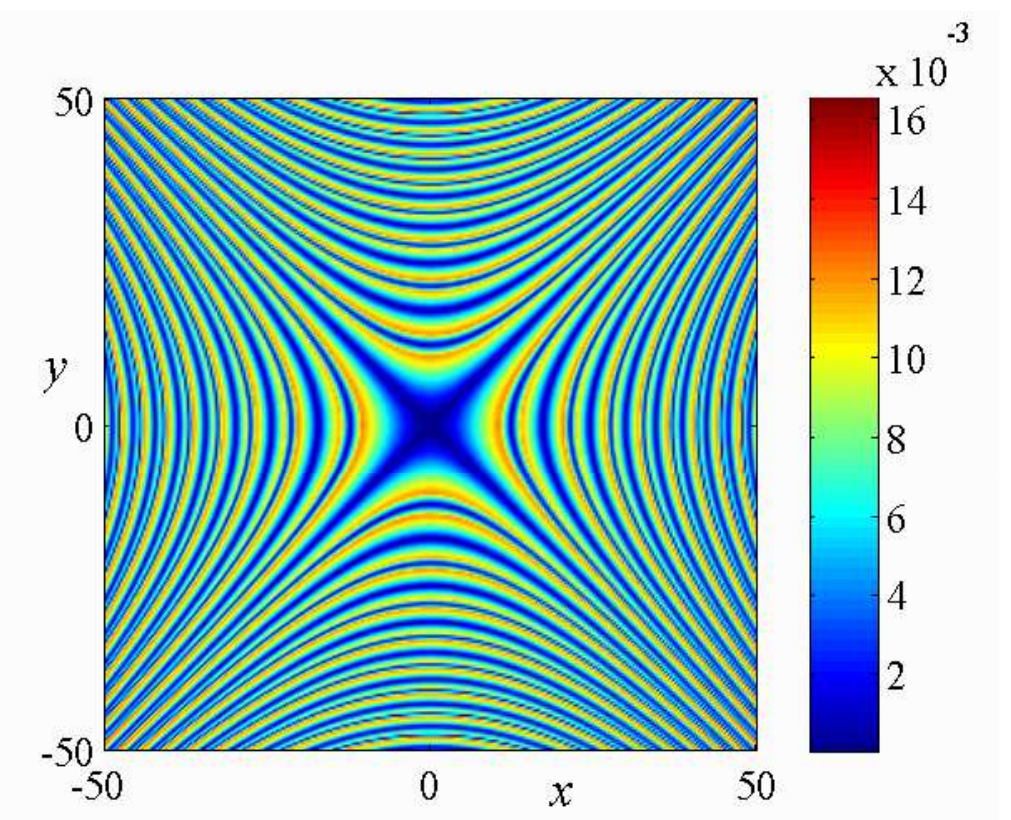}

\caption{Top:  Initial condition $\text{sech}(x^2+y^2)(1+0.1\text{randn})$, noise at every gridpt.: Numerical solution, Exact similarity solution and absolute value of their difference at $Z=24$.  Bottom: Log-amplitude vs. $\log Z$, $\Delta\theta = \theta-\theta_0-s$ vs. $1/Z$.}

\includegraphics[width=4cm]{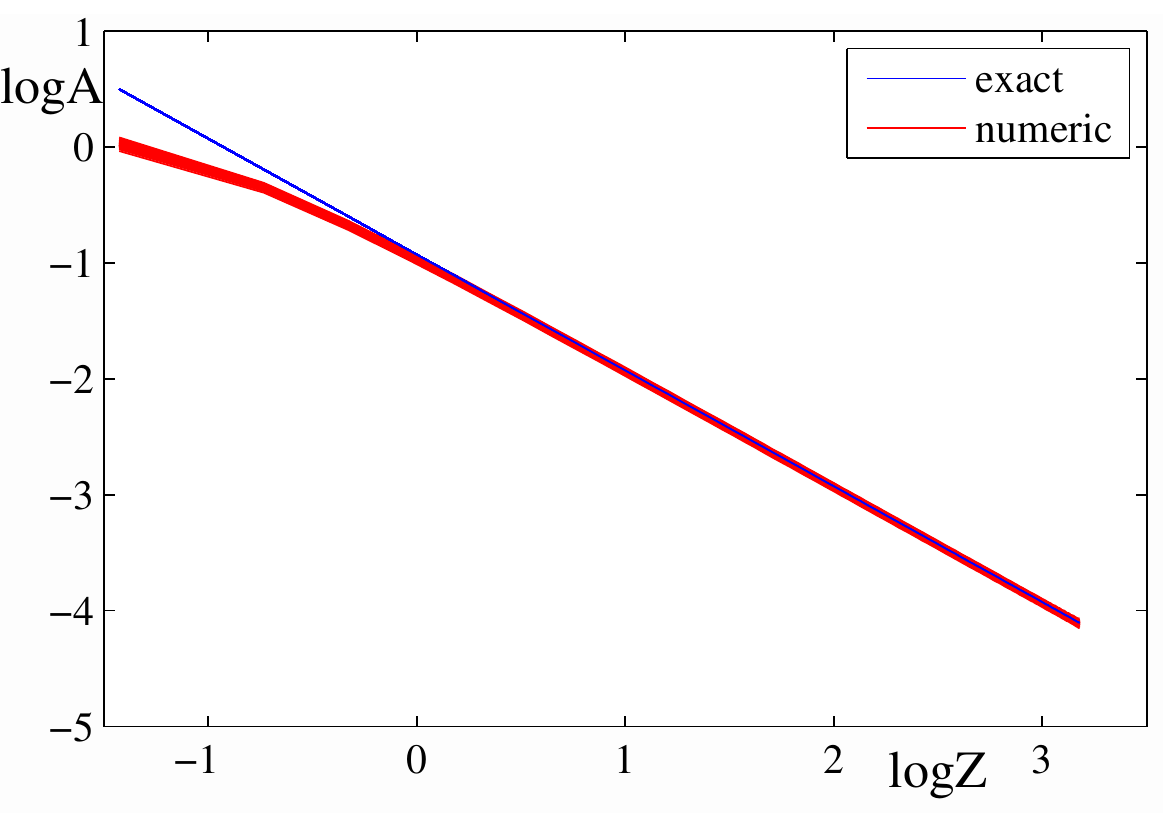}  \hspace{1cm}  \includegraphics[width=4cm]{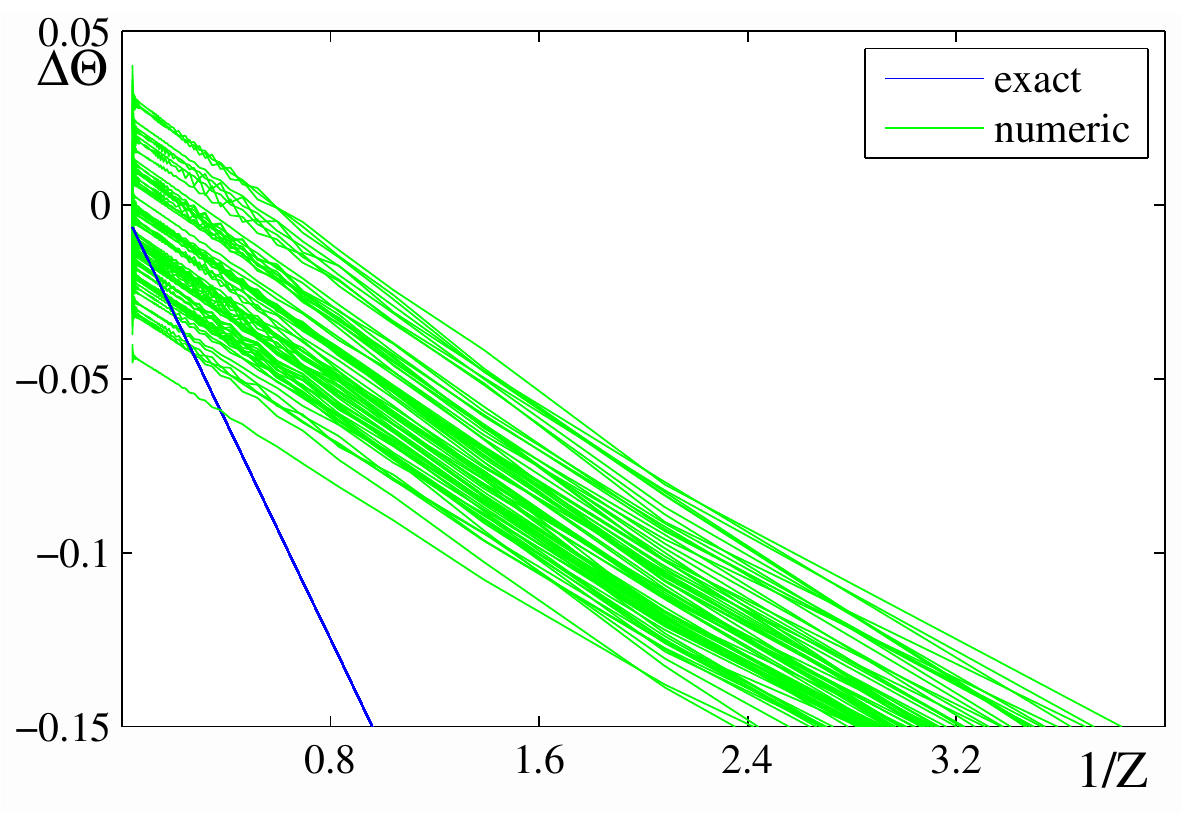}
\end{figure}

\end{document}